\documentclass[a4paper,11pt]{article}

\pdfoutput=1 

\usepackage{jcappub} 
\usepackage{comment}

\usepackage{subcaption}

\usepackage[T1]{fontenc} 

\makeatletter
\newcommand\makebig[2]{%
  \@xp\newcommand\@xp*\csname#1\endcsname{\bBigg@{#2}}%
  \@xp\newcommand\@xp*\csname#1l\endcsname{\@xp\mathopen\csname#1\endcsname}%
  \@xp\newcommand\@xp*\csname#1r\endcsname{\@xp\mathclose\csname#1\endcsname}%
}
\makeatother 

        \usepackage{amsfonts}
        \usepackage{amsmath}
	\usepackage{float}
\usepackage{array}
\usepackage{booktabs}
        \usepackage{colortbl}
        \usepackage{fancyhdr}
        \usepackage{textcomp}
\definecolor{dullpurple}{rgb}{0.431,0.188,0.534}
\definecolor{darkgreen}{rgb}{0.133,0.545,0.133}
\definecolor{dullred}{rgb}{0.706,0.208,0.192}

\usepackage{bm}
\usepackage{latexsym}
\usepackage{colortbl}
\usepackage{multirow}
\usepackage{array}
\usepackage{booktabs}
\usepackage{graphicx}
        \definecolor{dullpurple}{rgb}{0.431,0.188,0.534}
\hypersetup{urlcolor=dullpurple}
\hypersetup{citecolor=dullpurple}

\newcolumntype{Q}{>{$\displaystyle}l<{$}}
\newcolumntype{q}{>{\columncolor[gray]{0.9}$\displaystyle}l<{$}}
\newcolumntype{R}{>{$\displaystyle}r<{$}}
\newcolumntype{S}{>{$\displaystyle}c<{$}}
\newcolumntype{s}{>{\columncolor[gray]{0.9}$\displaystyle}c<{$}}
\newcolumntype{T}{>{\columncolor[gray]{0.9}}c<{}}

\newsavebox{\tableA}
\newsavebox{\tableB}

\newsavebox{\boxplot}
\newsavebox{\boxplota}

\makebig{biggg} {3.5}
\makebig{Biggg} {4.0}
\makebig{bigggg}{4.5}
\makebig{Bigggg}{5.5}

\title{\boldmath Entangled states as a probe of early universe history: a Higgs case study}

\author[a,1]{Rose~Baunach~\note{Corresponding author.}}

\affiliation[a]{Center for Quantum Mathematics and Physics and Department of Physics and Astronomy, University of California at Davis, One Shields Ave, Davis, CA 95616, U.S.A.}

\emailAdd{baunach@ucdavis.edu}

\abstract{I explore whether distinguishing features of phase transitions and/or the inflationary energy scale can be imprinted on cosmological observables due to entanglement during inflation, given a spectator scalar field with a Higgs-like potential. As a consequence of this analysis, I also present results that illustrate the variety of features a Higgs-like spectator can imprint on the primordial power spectrum due to entanglement, as well as how easy it might be to distinguish such spectra from other similar scalar field results at the level of CMB residuals. I utilize the technical framework for dynamically generated entangled states developed in~\cite{Adil:2022rgt,Baunach:2021yvu} to obtain my results. }

\begin{document}
\maketitle
\flushbottom

\section{Introduction}

\label{sec:intro}
We currently believe the anisotropies in the Cosmic Microwave Background radiation (CMB) are due to a period of inflationary expansion~\cite{Guth:1980zm, Kazanas:1980tx, Linde:1981mu, Albrecht:1982wi, Guth:1982ec, Bardeen:1983qw} that stretches quantum fluctuations in the early universe from micro to macro scales and sources the formation of cosmic structure. While the simplest single-field inflation model predictions are in strong agreement with our current data, they rely on the key assumption that quantum fluctuations during inflation are generated by a particular state---the Bunch-Davies (BD) state~\cite{Bunch:1978yq}. Even though this choice is well-motivated theoretically, it remains an interesting question whether current data can rule out other well-motivated choices of states. As previously demonstrated in~\cite{Baunach:2021yvu, Adil:2022rgt}, entanglement is naturally and inevitably dynamically generated during inflation given the presence of a “rolling” spectator scalar field---and the resulting entangled state will yield a primordial power spectrum with potentially measurable deviations compared
to the canonical BD result. 

In~\cite{Adil:2022rgt}, a Monte Carlo analysis was performed to investigate just how strongly Planck data prefers the standard BD vacuum state, versus an entangled one.  The results of that work demonstrated that most entangled states generated by the technical formalism in~\cite{Adil:2022rgt, Baunach:2021yvu} are consistent with Planck data\footnote{Note that as discussed in~\cite{Adil:2022rgt}, the analysis in that paper was limited technically from fully saturating the
Planck constraints, and one would have to expand the framework utilized in~\cite{Adil:2022rgt} to allow a greater variety of entangled states in order to more systematically determine which states are preferred by the data. However, while the results in~\cite{Adil:2022rgt} cannot claim that the data strongly prefers an entangled state over BD, the analysis did reveal significant entanglement was consistent with Planck data for some model parameters.}---which begs the question, what else might you be able to infer from signatures of entangled states in the CMB? For this work, I was motivated to explore whether an entangled state might imprint distinguishing features of the inflationary energy scale and/or cosmological phase transitions on CMB observables---and if such signatures could be used to observationally distinguish between alternate early universe scenarios various research programs might propose.


The inflationary energy scale is of high theoretical interest but is poorly observationally constrained.  In theoretical cosmology we are often conditioned to assume inflation occurs at very high energies (e.g. for GUT motivations, etc), but this has not been observationally proven. The upper limit for the inflationary energy scale, parameterized by the value of the Hubble parameter during inflation (in quasi de Sitter space), $H_{ds}$, is constrained by the tensor-to-scalar ratio, $r$. The current bound of $r<0.032$~\cite{Tristram:2021tvh} for single field inflationary models gives an approximate upper limit of $H_{ds} < O(10^{13} \; \mathrm{GeV})$.  However, the only universal lower bound on the inflationary energy scale is the requirement that big bang nucleosynthesis (BBN) successfully occur at $O(1 \; \mathrm{MeV})$.  (There are additional constraints due to reheating and baryogenesis scenarios, but these are model dependent.)  So there are actually a large set of energies at which inflation might occur that are currently within experimental bounds.

Another piece of the early universe narrative that is similarly interesting but poorly constrained is the timing of the electroweak phase transition. Current state of the art lattice calculations place the temperature of this phase transition for the SM Higgs at about 159.5 GeV~\cite{PhysRevD.93.025003}. This places the EW phase transition before BBN, which occurs at temperatures $O(1 \; \mathrm{MeV})$~\cite{Mukhanov:2005sc}, but beyond that things are less clearly defined.  One interesting question to consider is whether the EW phase transition might occur during or even before inflation. Given the experimental uncertainty in the inflationary energy scale, this is within the scope of possibility. If one could determine whether the EW phase transition took place before, during, or after inflation, that would have important implications for theories of reheating, baryogenesis, leptogeneis, and other early universe processes.  For example, if the Higgs develops a vacuum expectation value before the onset of reheating, that would imply the gauge bosons are no longer massless (due to the Higgs mechanism) during that era, which could alter the channels the inflaton degrees of freedom decay into during reheating, as well as the efficiency of such decays (e.g.~\cite{Freese:2017ace}).
Similarly, knowing the timing of the EW phase transition relative to inflation would automatically place a stronger bound on the inflationary energy scale (which, as discussed above, is currently a poorly constrained quantity experimentally). Lastly, the SM Higgs is not the only field postulated to undergo a phase transition in the early universe. For example, there are various models of the dark sector that contain a `dark Higgs,' and narrowing down the era of its phase transition would also have strong implications for model building in the dark sector (and would also potentially place constraints on the inflationary energy scale).

This project is a first step towards answering such questions.  I have utilized a model containing the inflaton and a spectator scalar field with a Higgs-like potential to investigate whether signatures of phase transitions (such as the EW transition) and/or the inflationary energy scale might imprint themselves on cosmological observables due to entanglement. Since this is the first instance the technical formalism in~\cite{Baunach:2021yvu,Adil:2022rgt}
has been employed to answer such questions, my aim was to map out general trends and possibilities in a `case study' analysis, without attempting to exhaustively prove every possible nuance. 

I organize this paper as follows. In section~\ref{sec:entReview}, I review the 
technical framework for entangled states as derived in~\cite{Baunach:2021yvu,Adil:2022rgt}.  The reader who is already familiar with~\cite{Baunach:2021yvu,Adil:2022rgt}, and in particular~\cite{Adil:2022rgt}, should feel free to skip this section.  Then, in section~\ref{sec:Higgs} I begin by reviewing standard lore for symmetry breaking and restoration in the early universe. I then lay out the definitions and assumptions made for this work in section~\ref{sec:Assume}.  Section~\ref{sec:PowSpec} presents a summary of my explorations investigating the variety of signatures a Higgs-like spectator can imprint on the primordial power spectrum due to entanglement.  In addition to pointing out the variety of solutions, this section also identifies which signals have the potential to be observationally constrained, based on the thresholds phenomenologically identified in previous work~\cite{Adil:2022rgt}. 

Section~\ref{sec:CMB} presents my results investigating whether entanglement with a Higgs-like spectator is distinguishable from entanglement with other scalar spectators at the level of Planck data, given a similar level of entanglement for all spectators. Then in section~\ref{sec:Sensitivity} I explore whether the observational imprints of entanglement are fine grained enough to distinguish if the Higgs-like potential is symmetry broken or symmetry restored (with a view of corroborating phase transition narratives). I also explore the uniqueness of the signals identified for a given inflationary energy scale---or whether degeneracies impede utilizing entanglement as an independent probe of that parameter.  In section~\ref{sec:Conclude} I conclude and provide some discussion on what lessons the case study analysis in this work provides for future research. There are also a few technical results from the formalism reviewed in section~\ref{sec:entReview} regulated to appendix~\ref{app:Review}, which the reader is directed to as appropriate.

\section{\label{sec:entReview}Review of entangled states formalism}

In this section I review the technical formalism developed in~\cite{Adil:2022rgt,Baunach:2021yvu} for entangled states---describing entanglement between the inflaton perturbations and those of a spectator scalar field.  For further details of this formalism beyond what is reviewed here, please see~\cite{Adil:2022rgt,Baunach:2021yvu}.

\subsection{\label{sec:Twopoint}Entangled two-point function}

Here I review the steps necessary to obtain the entangled two-point function---and thus the scalar inflationary power spectrum---in Schr\"odinger picture QFT.

As discussed in~\cite{Baunach:2021yvu,Adil:2022rgt}, one begins with the action for two scalar fields
\begin{equation}
    S = \frac{1}{2}\int d^{4}x \sqrt{-g} \big( \mathcal{R} -(\partial \Phi)^{2} -(\partial \Sigma)^{2} - 2 V(\Phi) - 2 V(\Sigma) \big)
\end{equation}
given $V(\Phi,\Sigma) = V(\Phi) + V(\Sigma)$, with
\begin{align}
    \Phi(t,\vec{x}) & = \phi(t) + \delta\phi(t,\vec{x}) \notag \\
    \Sigma(t,\vec{x}) & = \sigma(t) + \chi(t,\vec{x})
\end{align}
describing the background and perturbations of the inflaton field ($\Phi$) and spectator scalar field ($\Sigma$). Next, employ the ADM formalism~\cite{Arnowitt:1962hi} by re-writing the metric as:
\begin{equation}
    ds^2 = -N^{2}dt^{2} + h_{ij}(dx^{i} + N^{i} dt)(dx^{j} + N^{j} dt)
\end{equation}
with 
\begin{equation}
    h_{ij} = a^{2}e^{2\zeta}\delta_{ij}
\end{equation}
where $a$ is the scale factor, $\zeta$ is the co-moving curvature perturbation, and $\delta_{ij}$ is the usual Kronecker delta. Making the gauge choice of $\delta\phi = 0$, one effectively re-expresses the scalar degrees of freedom given by the inflaton perturbations, $\delta\phi$, in terms of the scalar metric perturbations, $\zeta$. After solving the ADM constraint equations in the usual way, much integration by parts, and transforming to conformal time, one eventually obtains the following action to second order in perturbations:
\begin{equation}
\label{eq:kspaceaction}
S=\int d\eta\ \int \frac{d^3 k}{(2\pi)^3}\ {\mathcal L}_k
\end{equation}
where
\begin{equation}
    \label{eq:kspacelagrangian}
    {\mathcal L}_k = \frac{1}{2} \vec{X}^{T \prime}_{\vec{k}}\ {\mathcal O}\ \vec{X}^{\prime}_{-\vec{k}}+\vec{X}^{T \prime}_{\vec{k}} \ {\mathcal M}_A\ \vec{X}_{-\vec{k}}-\frac{1}{2} \vec{X}^T_{\vec{k}}\ \Omega_k^2\ \vec{X}_{-\vec{k}}, 
\end{equation}
in which primes denote conformal time derivatives. The field variables are
\begin{equation}
    \label{eq:fieldvector}
     \vec{X}_{\vec{k}}= \begin{pmatrix}
        v_{\vec{k}}\\ 
        \theta_{\vec{k}}
    \end{pmatrix},
\end{equation}
with the following further redefinitions
\begin{equation}
\label{eq:fieldredef}
 v_{\vec{k}}=
   z \  
    \zeta_{\vec{k}} \quad \textrm{and} \quad \theta_{\vec{k}} = a \chi_{\vec{k}}
\end{equation}
given $z(\eta)=\sqrt{2 M_{p}^2\ \epsilon\ a^2(\eta)}$, $\epsilon$ measuring deviations from pure de Sitter space, and $M_{p}$ being the reduced Planck mass.

The matrices in eq.~\eqref{eq:kspacelagrangian}, ${\mathcal O}$, ${\mathcal M}$, and the symmetric $\Omega_k^2$, to lowest-order in slow-roll, 
are given by:
\begin{subequations}\label{eq:actionmatrices}
\begin{align}
\label{eq:actionmatrices:1}
{\mathcal O} & = \begin{pmatrix}
	1  & \ -\tanh\alpha\\
	-\tanh \alpha &\ 1
	\end{pmatrix}
\\
\label{eq:actionmatrices:2}
        {\mathcal M}_A & = \dfrac{\mathcal{H}}{2} \biggg[ 
        \left(3-\epsilon + \frac{\eta_{\rm sl}}{2} \right) \tanh\alpha + 
        \dfrac{a^2\  \partial_{\sigma} V}{\mathcal{H}^2\sqrt{2 M_{p}^2\epsilon}}
        \biggg] \begin{pmatrix}
	0 &\ -1\\
	1 &\ 0
	\end{pmatrix}
\\
\label{eq:actionmatrices:3}
 \Omega_k^2 & =\begin{pmatrix}
	 k^2 - \dfrac{z''}{z} & & & \Omega^2_{k\ 12}\\
	\Omega^2_{k\ 12} & & &\Omega^2_{k \ 22} \\ 
	\end{pmatrix},
\end{align}
\end{subequations}
with
\begin{subequations}
\begin{align}
\Omega^2_{k \ 12}  &\equiv
-\tanh\alpha \Bigg[
k^2 +a^2\partial_{\sigma}^2 V/2 +\mathcal{H}^2 \left(1 + \frac{5\eta_{\rm sl}}{4} \right)
\Bigg] 
-\mathcal{H}^2\Bigg(1 + \epsilon + \frac{\eta_{\rm sl}}{2} \Bigg)
\dfrac{a^2\  \partial_{\sigma} V}{\mathcal{H}^2\sqrt{2 M_{p}^2\epsilon}} \\
\Omega^2_{k \ 22}  &\equiv  k^2 +a^2 \partial^2_{\sigma}V
 -\dfrac{a''}{a} -2\epsilon (\epsilon -3) \mathcal{H}^2 \, \tanh^2 \alpha 
 +4\epsilon \mathcal{H}^2 \tanh \alpha 
 \biggg( \dfrac{a^2\  \partial_{\sigma} V}{\mathcal{H}^2\sqrt{2 M_{p}^2\epsilon}} \biggg).
\end{align}
\end{subequations}
As done in~\cite{Adil:2022rgt}, $\tanh \alpha$ is defined as
\begin{equation}
 \tanh \alpha \equiv \dfrac{\sigma'}{\mathcal{H}\sqrt{2 M_P^2\epsilon }} \ ,  
 \label{eq:alpha_def}
\end{equation}
with the Hubble parameter $ {\mathcal H}$ in conformal time being 
\begin{equation}
{\mathcal H}\equiv \frac{a^{\prime}}{a}
\end{equation}
whereby the slow-roll parameter $\epsilon$ is given in conformal time by ${\mathcal H}^{\prime}= (1-\epsilon){\mathcal H}^2$ and $\eta_{sl}$ denotes the second slow roll parameter $\eta_{sl} \equiv \epsilon'/\mathcal{H} \epsilon$.

The corresponding Hamiltonian density can then be calculated via
\begin{equation}
    \label{eq:kspacehamiltonian1}
    {\mathcal H}_k=\vec{\Pi}_{\vec{k}}^T \vec{X}^{\prime}_{-\vec{k}}-{\mathcal L}_k,
\end{equation}
and one obtains
\begin{equation}
    \label{eq:kspacehamiltonian2}
    {\mathcal H}_k = \frac{1}{2} \vec{\Pi}^T_{\vec{k}}\ {\mathcal O}^{-1}\  \vec{\Pi}_{-\vec{k}} + \vec{X}^T_{\vec{k}}\ {\mathcal M}_A^T {\mathcal O}^{-1}\ \vec{\Pi}_{-\vec{k}} 
      + \frac{1}{2} \vec{X}^T_{\vec{k}} \left(\Omega_k^2 + {\mathcal M}_A^T  {\mathcal O}^{-1} {\mathcal M}_A\right)\vec{X}_{-\vec{k}},
\end{equation}
in which $\vec{\Pi}_{\pm \vec{k}}$ is the momentum operator conjugate to $\vec{X}_{\pm\vec{k}}$. 

The next step is to solve the functional Schr\"odinger equation for each k-mode\footnote{Since the Hamiltonian density has no interactions mixing different k modes, this is fine to do.}, via:
\begin{equation} \label{eq:Schrodinger}
i \frac{\partial \Psi_{k}}{\partial \eta} = \mathcal{H}_{k} \Psi_{k}
\end{equation}
taking the wavefunctional describing the perturbations in the inflaton and spectator fields to be:
\begin{align}
\Psi_{\vec{k}} \left[ v_{\vec{k}} ,  \theta_{\vec{k}} ; \eta \right] & = \mathcal{N}_k(\eta)\  \mathrm{\exp}
\bigg[
-\dfrac{1}{2} \bigg( A_{k}(\eta)v_{\vec{k}}v_{-\vec{k}} + B_{k}(\eta)\theta_{\vec{k}}\theta_{-\vec{k}} 
  + C_{k}(\eta) \big(v_{\vec{k}}\theta_{-\vec{k}} + \theta_{\vec{k}}v_{-\vec{k}} \big) \bigg) \bigg]\ ,
\label{eq:wavefunctional}
\end{align}
with $\mathcal{N}_k(\eta)$ normalizing $\Psi_{\vec{k}}$. Solving the Schr\"odinger equation then generates coupled differential equations for the time evolution kernels $A_{k}(\eta)$, $B_{k}(\eta)$, and $C_{k}(\eta)$, where $C_{k}$ encodes the entanglement between the field fluctuations\footnote{If $C_{k} = 0$ at some time, that means there is no entanglement at that time and the two sectors are decoupled---i.e. $\Psi_{\vec{k}}$ is reduced to a product state.}. Together with the zero mode equation for the spectator field:
\begin{equation}
    \label{eq:zeromode}
    \sigma''(\eta) + 2 \mathcal{H}\sigma'(\eta) +a^{2}(\eta)\partial_{\sigma}^{2}V(\sigma) = 0\ ,
\end{equation}
the solutions of these differential equations specify the time evolution of the two-field vacuum state given by eq.~\eqref{eq:wavefunctional}. (The differential equations for $A_k$, $B_k$, and $C_k$ are listed in appendix~\ref{app:Kernel}, since they were previously derived in~\cite{Adil:2022rgt} and the main analysis in this paper uses a perturbative version of these equations, as discussed in section~\ref{sec:Perturb}.) One can then calculate the two-point function for $v_{\vec{k}}=z \ \zeta_{\vec{k}}$, given by~\cite{Adil:2022rgt,Albrecht:2014aga}:
\begin{align}
\label{eq:v2pt}
\langle v_{\vec{k}} v_{\vec{k}^{\prime}} \rangle &=(2\pi)^3 \delta^{(3)}\left(\vec{k}+\vec{k}^{\prime}\right)\left( \frac{B_{k R}}{2\left(A_{k R} B_{k R}-C_{k R}^2\right)}\right) \notag \\
& \equiv (2\pi)^3 \delta^{(3)}\left(\vec{k}+\vec{k}^{\prime}\right) P_{v}(k) \ ,
\end{align}
where `R' denotes the real part of the kernel.  This is related to the standard dimensionless inflationary power spectrum of curvature perturbations via:
\begin{equation}
\label{eq:dimlessPS}
\Delta^2_s=\frac{k^3}{2\pi^2} P_{\zeta}(k) = \frac{k^3}{2\pi^2} \frac{1}{z^2} P_{v}(k) \ .
\end{equation}
For more details on the steps contained in this section, please see~\cite{Adil:2022rgt,Baunach:2021yvu,Albrecht:2014aga} where these quantities are derived and explored in more detail.

\subsection{\label{sec:Perturb}Perturbative approach}

As in~\cite{Adil:2022rgt}, I will ultimately make use of a perturbative approach in this paper to systematically explore the lowest order corrections to the inflationary power spectrum due to entanglement. Expanding the kernels as follows\footnote{The expansion for $C_{k}$ begins at first order in $\lambda$, because $C_{k} = 0$ is the standard single field limit.  Also, there are no first order terms in $A_{k}$ and $B_{k}$ because there is nothing in the equations to source them if $C_{k}$ is zero initially (which I take to be the case in this paper, as discussed subsequently), as demonstrated in~\cite{Adil:2022rgt}.}:
\begin{subequations}\label{eq:kernels_expansion}
\begin{align}
A_k & = A_k^{(0)} +\lambda^2 A_k^{(2)} + ...
\\
B_k & = B_k^{(0)} +\lambda^2 B_k^{(2)} + ...
\\
C_k & = \lambda C_k^{(1)} + ...
\end{align}
\end{subequations}
where the zeroth order terms are the standard Bunch-Davies vacuum (no entanglement) solutions, one can then express the scalar power spectrum to lowest order in $\lambda$ as:
\begin{equation} \label{eq:psPerturb}
    \Delta^2_s = \Delta^2_{s,BD}  \left[1 + \lambda^{2} \left(\frac{-A_{k R}^{(2)}}{A_{k R}^{(0)}} +\frac{(C_{k R}^{(1)})^2}{A_{k R}^{(0)} B_{k R}^{(0)}} \right) \right] 
\end{equation}
where 
\begin{equation} \label{eq:psBD}
\Delta^2_{s,BD} = \frac{k^3}{2\pi^2} \frac{1}{z^2} \frac{1}{2 A_{k R}^{(0)}}\ .
\end{equation}
But what is $\lambda$?  As discussed in~\cite{Adil:2022rgt}, there are two relevant quantities that show up in the Lagrangian:
\begin{subequations}
\begin{align}
\lambda_1 & \equiv \tanh\alpha \equiv \dfrac{\sigma'}{\mathcal{H}\sqrt{2 M_P^2\epsilon }}
\\
\lambda_2 & \equiv\dfrac{a^2\partial_\sigma V}{\mathcal{H}^2\sqrt{2 M_{\rm Pl}^2 \epsilon}} \  .
\end{align}
\end{subequations}
The $\lambda_{i}$ are already required to be small to ensure the spectator field is subdominant to the inflaton during the course of inflation (i.e. one stays within the quasi-single field limit). As done in~\cite{Adil:2022rgt}, I take the expansion parameter to be 
\begin{equation}
\label{eq:lambdadef}
    \lambda = \mathrm{max}\{\lambda_{i}(\eta): i=1,2\} \quad \eta_0 \leq \eta \leq \eta_{end}
\end{equation}
where $\eta_0$ is the time at which entangled evolution `begins' and $\eta_{end}$ is theoretically the end of inflation (but computationally some explicit late time where all observationally relevant modes are far past the horizon\footnote{As discussed in~\cite{Adil:2022rgt}, there is no guarantee that $\zeta_{\vec{k}} = \frac{v_{\vec{k}}}{z}$ remains constant outside the horizon, compared to the standard BD result, so the entangled power spectrum given by eq.~\eqref{eq:psPerturb} must be evaluated explicitly at late times.}). As shown in~\cite{Adil:2022rgt}, this choice of $\lambda$ ensures increasing $\lambda$ is directly correlated to increasing the fractional change in $\Delta_s^2$ (due to entanglement) relative to the BD case. 

Given these choices, one can expand the full kernel equations (see appendix~\ref{app:Kernel}) using the expansion specified by eq.~\eqref{eq:kernels_expansion} to obtain the equations needed to compute the power spectrum in  eq.~\eqref{eq:psPerturb}.  One obtains:
\begin{subequations}\label{eq:dimless_kern_expand}
\begin{align}
\label{eq:dimlessA0}
i \partial_{\tau}A_{q}^{(0)} & = (A_q^{(0)})^2 - \left[ \left(\frac{q}{1-\epsilon}\right)^2 - \frac{\left(\nu_{f}^2 -\frac{1}{4}\right)}{\tau^2} \right]  
\\
i \partial_{\tau}A_{q}^{(2)} & = 2 A_q^{(0)}A_q^{(2)} \notag \\
& {} {} \quad + \Bigg[ \tilde{\lambda}_1 A_q^{(0)} + C_q^{(1)} 
 - \frac{i}{2(1-\epsilon)\tau} \left[ \left(3-\epsilon+\frac{\eta_{sl}}{2}\right) \tilde{\lambda}_1 + \tilde{\lambda}_2 \right] \Bigg]^{2} \label{eq:dimlessA2}
\\
\label{eq:dimlessB0}
i \partial_{\tau}B_{q}^{(0)} & = (B_q^{(0)})^2 - \left[ \left(\frac{q}{1-\epsilon}\right)^2 - \frac{\left(\nu_{g}^2 -\frac{1}{4}\right)}{\tau^2} \right] 
\\
i \partial_{\tau}C_{q}^{(1)} & = C_{q}^{(1)} \left(A_{q}^{(0)} + B_{q}^{(0)} \right) + \tilde{\lambda}_1 A_{q}^{(0)}B_{q}^{(0)} \notag \\
& {} {} \quad + \frac{i}{2(1-\epsilon)\tau} \left[ \left(3-\epsilon+\frac{\eta_{sl}}{2}\right) \tilde{\lambda}_1 + \tilde{\lambda}_2 \right] (A_{q}^{(0)} - B_{q}^{(0)} ) \notag \\
& {} {} \quad + \tilde{\lambda}_1 \left[ \left(\frac{q}{1-\epsilon}\right)^2 +\frac{\mu^{2} \partial_{s}^{2}V(s)}{2 (1-\epsilon)^{2} \tau^{2}} +\frac{1 +\frac{5}{4}\eta_{sl}}{(1-\epsilon)^{2}\tau^{2}} \right] 
+ \tilde{\lambda}_2 \left[\frac{1 + \epsilon + \frac{\eta_{sl}}{2}}{(1-\epsilon)^{2}\tau^{2}} \right]  \label{eq:dimlessC1}
\end{align}
\end{subequations}
where $\tilde{\lambda}_{1,2} = \frac{\lambda_{1,2}}{\lambda}$ is an algebraic simplification, and the following re-definitions are made to solve the kernel equations in terms of dimensionless quantities (for ease of numerical computations):
\begin{subequations} \label{eq:dimless}
\begin{align}
&\tau = -\frac{\eta}{\eta_0},\ q= \frac{k}{k_0} = -(1-\epsilon) k\eta_0 \label{eq:dimlessa}\\
&\sigma = s M_{p} ,\ \  V(\sigma) = \Lambda^{4} V(s) , \ \ \mu^{2} = \frac{\Lambda^4}{H_{ds}^{2} M_{p}^{2}} \\
&\lambda_1  = \frac{(1-\epsilon)}{\sqrt{2 \epsilon}} (-\tau) \partial_{\tau}s
\\
&\lambda_2  =  \frac{\mu^{2}}{\sqrt{2 \epsilon}} \ \partial_{s}V(s) \\
&A_k(\eta)  = \frac{{A}_q(\tau) }{\left(-\eta_0\right)} , \ \  B_k(\eta) = \frac{{B}_q(\tau) }{\left(-\eta_0\right)} ,\  \ C_k(\eta) = \frac{{C}_q(\tau) }{\left(-\eta_0\right)}  \\
&\mathcal{H}(\eta)  = \frac{\mathcal{H}(\tau) }{\left(-\eta_0\right)} , \ \  \mathcal{H}(\tau) = \frac{-1}{(1-\epsilon)\tau} \ .
\end{align}
\end{subequations}
Note that dimensionless conformal time, $\tau$, runs from -1 to 0, and that the form of eq.~\eqref{eq:psPerturb} will be the same in terms of dimensionless variables, since the factors of $\eta_0$ will cancel in the ratios of kernels according to the definitions in eq.~\eqref{eq:dimless}. Also, since the zeroth order equations for $A^{(0)}$ and $B^{(0)}$ are just the Schr\"odinger picture version of the Mukhanov-Sasaki equation, the quantities $\nu_f$ and $\nu_g$ are defined in the usual way (here in terms of dimensionless variables):
\begin{subequations}
\begin{align}
\label{eq:Hankelnuf}
    &\nu_{f}  = \frac{3}{2} + \epsilon + \frac{\eta_{sl}}{2} \\
   &\nu_{g}(s)  = \sqrt{\frac{9}{4} + 3\epsilon - \frac{\mu^{2}\partial_{s}^{2}V(s)}{(1-\epsilon)^{2}}} \ , \label{eq:Hankelnug}
\end{align}
\end{subequations}
where I highlight to the reader that
\begin{equation} \label{eq:Meff}
    \mu^{2}\partial_{s}^{2}V(s) = \frac{\partial_{\sigma}^{2}V(\sigma)}{H^{2}_{ds}} = \frac{M_{eff}^{2}(\sigma)}{H^{2}_{ds}} \ . 
\end{equation}
Note that if the spectator is a free massive scalar with a quadratic potential, $M_{eff}^{2}$ will be independent of $\sigma$ and therefore $\nu_g$ will be a constant.  It is one of the features of this work, discussed in section~\ref{sec:Higgs}, that I allow $\nu_g$ to vary---due to the variation of $M_{eff}^{2}$---for a Higgs-like potential.

Also, the dimensionless form of the equation describing the evolution of the zero mode, eq.~\eqref{eq:zeromode}, becomes:
\begin{equation}
    \label{eq:dimlessBackground}
    s''(\tau) - \frac{2}{\tau(1-\epsilon)} s'(\tau) + \frac{\mu^{2} \partial_{s}V(s)}{\tau^{2}(1-\epsilon)^{2}} = 0 \ .
\end{equation}

Finally, as previously derived in~\cite{Adil:2022rgt}, the initial conditions for eq.~\eqref{eq:dimless_kern_expand} are:
\begin{subequations}\label{eq:realimA0B0}
\begin{align}
\label{eq:A0real}
A_{q R}^{(0)}(\tau_0 = -1) =& \frac{2}{ \pi \left |H^{(2)}_{\nu_{f}}(\frac{q}{1-\epsilon})\right|^2} \\
\label{eq:A0im}
A_{q I}^{(0)}(\tau_0 = -1) =& \frac{1}{2} \Bigg[ 1 - 2 \nu_{f}
 + x \left . \left[ \frac{H^{(1)}_{\nu_{f} -1}(x)}{H^{(1)}_{\nu_{f}}(x)} + \frac{H^{(2)}_{\nu_{f}-1} (x)}{H^{(2)}_{\nu_{f}}(x)} \right] \right|_ {x=\frac{q}{(1-\epsilon)}} \Bigg] \\
\label{eq:B0real}
B_{q R}^{(0)}(\tau_0 = -1) =& \frac{2}{ \pi \left |H^{(2)}_{\nu_{g}}(\frac{q}{1-\epsilon})\right|^2} \\
\label{eq:B0im}
B_{q I}^{(0)}(\tau_0 = -1) =& \frac{1}{2} \Bigg[ 1 - 2 \nu_{g}  
 + x \left . \left[ \frac{H^{(1)}_{\nu_{g} -1}(x)}{H^{(1)}_{\nu_{g}}(x)} + \frac{H^{(2)}_{\nu_{g}-1} (x)}{H^{(2)}_{\nu_{g}}(x)} \right] \right |_ {x=\frac{q}{(1-\epsilon)}} \Bigg]  \\
\label{eq:A2real}
A_{q R}^{(2)}(\tau_0 = -1) =& -\tilde{\lambda}_{1,0}^{2}A_{q R}^{(0)}(\tau_0 = -1) \\
\label{eq:A2im}
A_{q I}^{(2)}(\tau_0 = -1) =& -\tilde{\lambda}_{1,0}^{2}A_{q I}^{(0)}(\tau_0 = -1) 
-\frac{1}{2(1-\epsilon)} \left[\left(3 - \epsilon + \frac{\eta_{sl}}{2} \right) \tilde{\lambda}_{1,0}^{2} + \tilde{\lambda}_{1,0}\tilde{\lambda}_{2,0} \right] \notag \\
& {} \quad +\left[ \frac{\eta_{sl}\tilde{\lambda}_{1,0}^{2}}{2(1-\epsilon)} + \tilde{\lambda}_{1,0}^{2} + \frac{2\tilde{\lambda}_{1,0}^{2}}{(1-\epsilon)} +\frac{\tilde{\lambda}_{1,0}\tilde{\lambda}_{2,0}}{(1-\epsilon)} \right]
\end{align}
\end{subequations}
where $\tilde{\lambda}_{1,0}$ denotes that $\tilde{\lambda}_{1}$ should be evaluated at $\tau_0 = -1$, and a term $\mathcal{O}(\eta_{sl}\epsilon)$ has been dropped from $A_{q I}^{(2)}$. As done in~\cite{Adil:2022rgt}, the initial conditions for the kernel equations are constructed such that the corresponding modes are the standard Bunch–Davies ones at the initial time $\eta_0$ (corresponding to $\tau_0 = -1$) at which entanglement begins to be dynamically generated. Appendix~\ref{app:ICs} reviews the derivation of these initial conditions, as previously presented in [1].

In this section, I have made the choice---as was also done in~\cite{Adil:2022rgt,Baunach:2021yvu}---to assume there is no entanglement at some initial time $\eta_0$ (corresponding to $\tau = -1$) and to instead study the dynamical generation of entanglement over the course of inflation and its impact on cosmological observables.  If I had instead chosen $C_{k}(\eta_0) \neq 0$, as was done in some earlier work~\cite{Albrecht:2014aga,Bolis:2016vas,Albrecht:2018hoh}, I would have had to consider linear terms in $\lambda$ in the expansions for $A_k$ and $B_k$ in eq.~\eqref{eq:kernels_expansion} (as discussed in~\cite{Adil:2022rgt}), which would lead to different results for the primordial power spectrum.

To compare my results with Planck data, I use the following form of the perturbative primordial power spectrum in eq.~\eqref{eq:psPerturb}:
\begin{equation} \label{eq:psCMB}
    \Delta^2_s = A_{s} \Big(\frac{k}{k_{\rm piv}}\Big)^{n_s - 1}  \left[1 + \lambda^{2} \left(\frac{-A_{q R}^{(2)}}{A_{q R}^{(0)}} +\frac{(C_{q R}^{(1)})^2}{A_{q R}^{(0)} B_{q R}^{(0)}} \right) \right] 
\end{equation}
where I have replaced the theoretical value for $\Delta^{2}_{s,BD}$ with the usual observational parameterization in terms of $A_s$ and $n_s$ (with $k_{\rm piv} = 0.05 \; \rm Mpc^{-1}$~\cite{2020Planck}), and the entanglement quantities in the square bracket have been expressed in terms of dimensionless parameters. (However, as noted earlier in this section, the expression in the square brackets in terms of dimensionless variables is equivalent to the one in eq.~\eqref{eq:psPerturb}, since all dimensionful kernels have mass dimension 1.) Also note that I take the relationship between $\epsilon$, $\eta_{sl}$, and $n_s$ to be $\eta_{sl} = 1 - 2 \epsilon - n_{s}$.

Lastly, as in previous 
work~\cite{Baunach:2021yvu,Adil:2022rgt}, I will often make use of a phenomenological parameter called $k_{\rm ent}$ when comparing the results of entangled power spectra with data. $k_{\rm ent}$ is defined as follows. One can see from eq.~\eqref{eq:dimless} that shifting the initial time $\eta_{0}$ (where entanglement begins) is equivalent to shifting the scale that leaves the horizon at $\eta_0$ (the largest observable scale that will show evidence of entanglement). Therefore, one can chose to parameterize the onset of entanglement via this distinctive scale, called $k_{\rm ent}$ in this analysis.~\footnote{From the discussion above, one can see that $k_{\rm ent}$ is the phenomenological equivalent to $k_0$, defined as $k_0 \propto \frac{1}{\eta_0}$ in eq.~\eqref{eq:dimlessa}. If one chooses to fix $k_0 = 10 ^{-6} \textrm{Mpc}^{-1}$, corresponding to the largest CMB observable scale, then $k_{\rm ent} = xk_0$ with x some constant factor. However, to keep theoretical and phenomenological variables distinct, I have chosen to define $k_{\rm ent}$ without reference to $k_0$ in eq.~\eqref{eq:kent}.} And since the entangled equations are solved using dimensionless time---i.e., the dimensionless quantity in square brackets in eq.~\eqref{eq:psCMB} will be the same no matter what $\eta_0$ is---one can post-process power spectra to include the effects of shifting the onset of entanglement with a straightforward $k$ shift.  To do this, simply make the conversion
\begin{equation}
    \label{eq:kent}
    k \rightarrow k \left( \frac{k_{\rm ent}}{10^{-6} \mathrm{Mpc}^{-1}} \right) 
\end{equation}
in eq.~\eqref{eq:psCMB}. I take $ {k_{\rm ent}}\slash {10^{-6}} \geq 1$ in this work.

This completes a review of the technical framework.  The next section discusses my motivations for the current work and how the technical framework developed in this section will be used to explore them. 

\section{\label{sec:Higgs}Higgs-like spectator}
In this work, I make use of a spectator scalar field with a Higgs-like potential to consider whether signatures of phase transitions and/or the inflationary energy scale might imprint themselves on cosmological observables.  In this section I begin by discussing symmetry breaking and restoration in the early universe and the questions related to those concepts that motivated this work.  I then outline what choices and approximations were made for this case study analysis. I also present results showcasing the rich phenomenology of entangled primordial power spectra accessible to a Higgs-like spectator.


\subsection{\label{sec:Symm}Symmetry breaking and symmetry restoration}
Consider a real scalar field described by
\begin{align}
\mathcal{L} & = \frac{1}{2} \partial_{\mu}\phi \partial^{\mu}\phi - V(\phi) \notag \\
    V(\phi) & = -\frac{1}{2} m^{2} \phi^{2} + \frac{1}{4} \lambda \phi^{4} \ .
\end{align}
Formally, this potential has two degenerate stable minima, characterized by a reflection symmetry $\phi \leftrightarrow -\phi$ in the Lagrangian density. To construct a viable quantum theory, one expands the Lagrangian density about one of these stable minima, rather than $\phi = 0$, which is unstable. This requires a choice of vacuum state, and making this choice `breaks' the reflection symmetry. (Obviously this is a simplified example of symmetry breaking compared to the Higgs mechanism in the Standard Model.)

\begin{figure}[!h]
\centering
\includegraphics[width=0.52\linewidth]{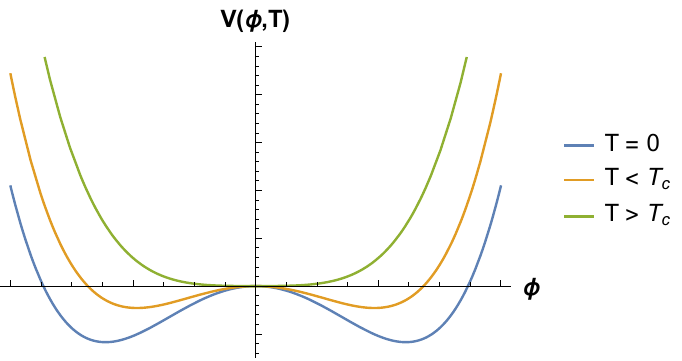} 
\vspace{-0.15cm}
\caption{The potential $V(\phi,T)$ in eq.~\eqref{eq:Vsimple} is plotted for a few values of $T$ above and below $T_c$---the temperature at which $\partial_{\phi}^{2}V(\phi,T)|_{\phi=0}$ changes sign---in order to graphically illustrate the difference between symmetry broken and symmetry restored behavior for this potential.  The x and y axis units are arbitrary.}
\label{fig:HiggsEx}
\end{figure}

However, continuing with this simple example for a moment, let us add another piece to the story. In standard lore~\cite{Kolb:1990vq, Mukhanov:2005sc} one comes across the phenomenon of `high-temperature symmetry restoration.' This arises from considering the fact that at nonzero temperature the $\phi$ field may no longer exist in isolation. For example, the radiation and matter densities in the early universe are non-negligible at various stages. To model such a background one can imagine the $\phi$ field in contact with a heat bath. The implications of this scenario can be rigorously dealt with in the language of thermal field theory (e.g.~\cite{Quiros:1999jp}), with a net effect of introducing a temperature dependent mass term in the potential, such that one might have
\begin{equation} \label{eq:Vsimple}
    V(\phi, T) = -\frac{1}{2} m^{2} \phi^{2} + \frac{1}{4} \lambda \phi^{4} + a \lambda T^{2} \phi^{2}
\end{equation}
in the case of our simplified model (where $a$ is some dimensionless $O(1)$ constant).  At a high enough temperature, the temperature dependent term will dominate and $\phi = 0$ will be the single stable minima of the potential---i.e. the symmetry has been `restored'.  However, as the temperature decreases the original mass term will begin to dominate and the degenerate minima will begin to appear, signalling symmetry breaking once again.  The simple potential in eq.~\eqref{eq:Vsimple} is plotted in figure~\ref{fig:HiggsEx} for a few values of T to illustrate this graphically. $T_{c}$ is the temperature at which $M_{eff}^{2}(\phi,T) = \partial_{\phi}^{2}V(\phi,T)$ evaluated at $\phi=0$ changes sign, such that $M_{eff}^{2}(0,T_{c}) = 0$ and $V(\phi,T_{c}) = \frac{1}{4} \lambda \phi^{4}$ (given the potential in eq.~\eqref{eq:Vsimple}).

For the SM Higgs, standard lore~\cite{Kolb:1990vq,Mukhanov:2005sc} does assume a similar behavior to our simplified model---that there exists some temperature in the early universe above which symmetry is `restored', and the origin of the SM Higgs potential becomes a stable minimum, with a $T_{c}$ value of about 159.5 GeV~\cite{PhysRevD.93.025003}. (Alternative scenarios, such as symmetry non-restoration---e.g.~\cite{Carena:2021onl,Meade:2018saz}---do of course exist.) The process by which the SM Higgs potential transitions from one type of behavior to another is an important component of the electroweak phase transition.

As discussed in the introduction, the timing of the EW phase transition and its proximity to inflation is interesting but poorly determined. The large experimental uncertainty in determining the inflationary energy scale implies that the EW phase transition might have occurred during or even before inflation. Such a scenario would have important implications for early universe processes such as baryogenesis and reheating. And these implications are not unique to the SM Higgs, but would also be relevant for Higgs-like model building in the dark sector. Thus, determining whether signatures of phase transitions (such as the EW transition) and/or the inflationary energy scale might imprint themselves on cosmological observables due to entanglement is a worthwhile goal. I utilize a spectator scalar field with a Higgs-like potential to investigate these questions, with specific choices detailed in the next section.

\subsection{\label{sec:Assume}Definitions and assumptions}

In the previous section, I stated my goal was to determine whether signatures of phase transitions (such as the EW phase transition), and/or the inflationary energy scale might imprint themselves on cosmological observables due to entanglement. If such signatures were visible, then one could use them to place further constraints on early universe history using CMB data.

To utilize the entangled states formalism described in section~\ref{sec:entReview} to investigate this goal, the spectator zero mode must be allowed to `roll'.~\footnote{Or the two sectors must start with nonzero initial entanglement, as was considered in earlier work~\cite{Albrecht:2014aga,Bolis:2016vas,Albrecht:2018hoh}.}  Entanglement will be sourced dynamically if the spectator zero mode has either a non-zero position or velocity~\cite{Adil:2022rgt,Baunach:2021yvu}.  To match this requirement onto a scenario that might resemble the EW phase transition (or a dark Higgs analog), I only consider second order and/or crossover-type phase transitions where the zero mode dynamically rolls to its new symmetry broken vacuum expectation value (rather than a first order transition where the symmetry breaking process is expected to proceed via tunneling~\cite{Kolb:1990vq,Mukhanov:2005sc}).  This does not greatly restrict the generality of my investigations, since current bounds forbid a first order phase transition for the SM Higgs due to its mass (i.e. it is too heavy to have a viable first order EW phase transition in the minimal SM~\cite{Mukhanov:2005sc}).

Before presenting results for entanglement with a Higgs-like spectator in the following section, I first define two quantities that play a key role in my investigations---the spectator potential and the energy scale of inflation.

\subsubsection{\label{sec:Potential}Higgs-like potential}

I utilize a Higgs-like spectator potential of the following form
\begin{equation}\label{eq:Higgspot}
    V(\sigma,T) = -\frac{1}{2} m_{h}^{2} \sigma^{2} + \frac{\lambda_h}{4}\sigma^{4} + \frac{\lambda_h}{8}T^{2}\sigma^{2}
\end{equation}
where $m_h = M_h/\sqrt{2}$, $M_h = 125 \; \mathrm{GeV}$, $\lambda_h = m_h^{2}/v^{2}$, and $v = 246 \; \mathrm{GeV}$, in analogy with the SM Higgs.  This is a very simple version of what is often called a `mean field approximation' in the literature (e.g.~\cite{Carena:2021onl}), where only leading order thermal corrections---obtained by expanding the thermal one-loop effective potential in powers of $M_{eff}^{2}/T^{2}$~\cite{Quiros:1999jp,Mukhanov:2005sc, Kolb:1990vq}---have been considered.~\footnote{Furthermore, I have ignored any constant (not $\sigma$ dependent) thermal contributions, because the entanglement equations only depend on $\partial_{\sigma}V(\sigma)$ and $\partial_{\sigma}^{2}V(\sigma)$.} 

Since this project was designed to be a first pass at answering the motivating questions discussed in section~\ref{sec:intro}, I feel justified utilizing such a simple potential.  If this `case study' yields interesting results, one could then go back and substitution the full temperature corrected and renormalized Higgs potential for more fine grained predictions.

The dimensionless version of eq.~\eqref{eq:Higgspot} is given by:
\begin{equation}\label{eq:HiggspotDimless}
    V(s,T) = -\frac{1}{2} \Big(\frac{m_{h}}{M_{p}}\Big)^{2} s^{2} + \frac{\lambda_h}{4}s^{4} + \frac{\lambda_h}{8}\Big(\frac{T}{M_p}\Big)^{2}s^{2}
\end{equation}
and the quantity $\mu^{2}\partial_{s}^{2}V(s,T)$ is given by:
\begin{equation} \label{eq:HiggsVdpr}
    \mu^{2}\partial_{s}^{2}V(s,T) = \frac{M_{p}^2}{H_{ds}^2} \Bigg[-\Big(\frac{m_{h}}{M_{p}}\Big)^{2} + 3 \lambda_{h} s^{2} + \frac{\lambda_{h}}{4} \Big(\frac{T}{M_{p}}\Big)^{2} \Bigg] \; .
\end{equation}
The fact that eq.~\eqref{eq:HiggsVdpr} is $s$ dependent even if $T$ is fixed will be partially responsible for driving some of the new entanglement features exhibited in the plots in subsequent sections---especially compared to the free massive scalar potential studied in~\cite{Baunach:2021yvu,Adil:2022rgt}---since this will cause the Hankel index for the spectator perturbations, $\nu_g$, to vary as the zero mode rolls through the potential.

Lastly, since this work is focused on exploring observational effects of entanglement during inflation, I take
\begin{equation} \label{eq:Tgh}
    T = T_{GH} = \frac{H_{ds}}{2 \pi} \; 
\end{equation}
for the rest of this paper.

\subsubsection{\label{sec:EnergyScale}Inflationary energy scale}
The other new ingredient in my analysis compared to previous work is considering the inflationary energy scale explicitly in my equations. Observationally, the scalar primordial power spectrum is parameterized as 
\begin{equation} \label{eq:psObs}
    \Delta_s^{2} = A_{s} \Bigg(\frac{k}{k_{piv}}\Bigg)^{n_s - 1} \ .
\end{equation}
At the pivot scale $\Delta_s^{2} = A_s$, and if one equates this with the standard scale invariant theoretical result
\begin{equation}
    \Delta_s^{2} = \frac{H_{ds}^{2}}{8 \pi^{2} M_{p}^{2}\epsilon}
\end{equation}
one can estimate a value of $\epsilon$ to input into the kernel evolution equations
\begin{equation} \label{eq:eps}
    \epsilon(H_{ds}) = \frac{H_{ds}^{2}}{8 \pi^{2} M_{p}^{2}A_s}
\end{equation}
given observational input for $A_s$ and a value of the inflationary energy scale, $H_{ds}$, that one would like to investigate (or equivalently $T_{GH}$, since they are just related by a factor of $2 \pi$).  These assumptions give the approximate equivalence relation:
\begin{equation}
    \epsilon \propto H_{ds}^{2} \propto T_{GH}^{2}
\end{equation} modulo important numerical conversion factors. The fact that these three quantities are linked in my simplified model will have some consequences for what questions I can and cannot answer with my subsequent analysis.

\vspace{-0.2cm}
\subsection{\label{sec:PowSpec}Primordial power spectrum phase space}
\begin{figure}[hp]
\centering
\subfloat[]{\label{fig:s1}\includegraphics[width=0.40\textwidth]{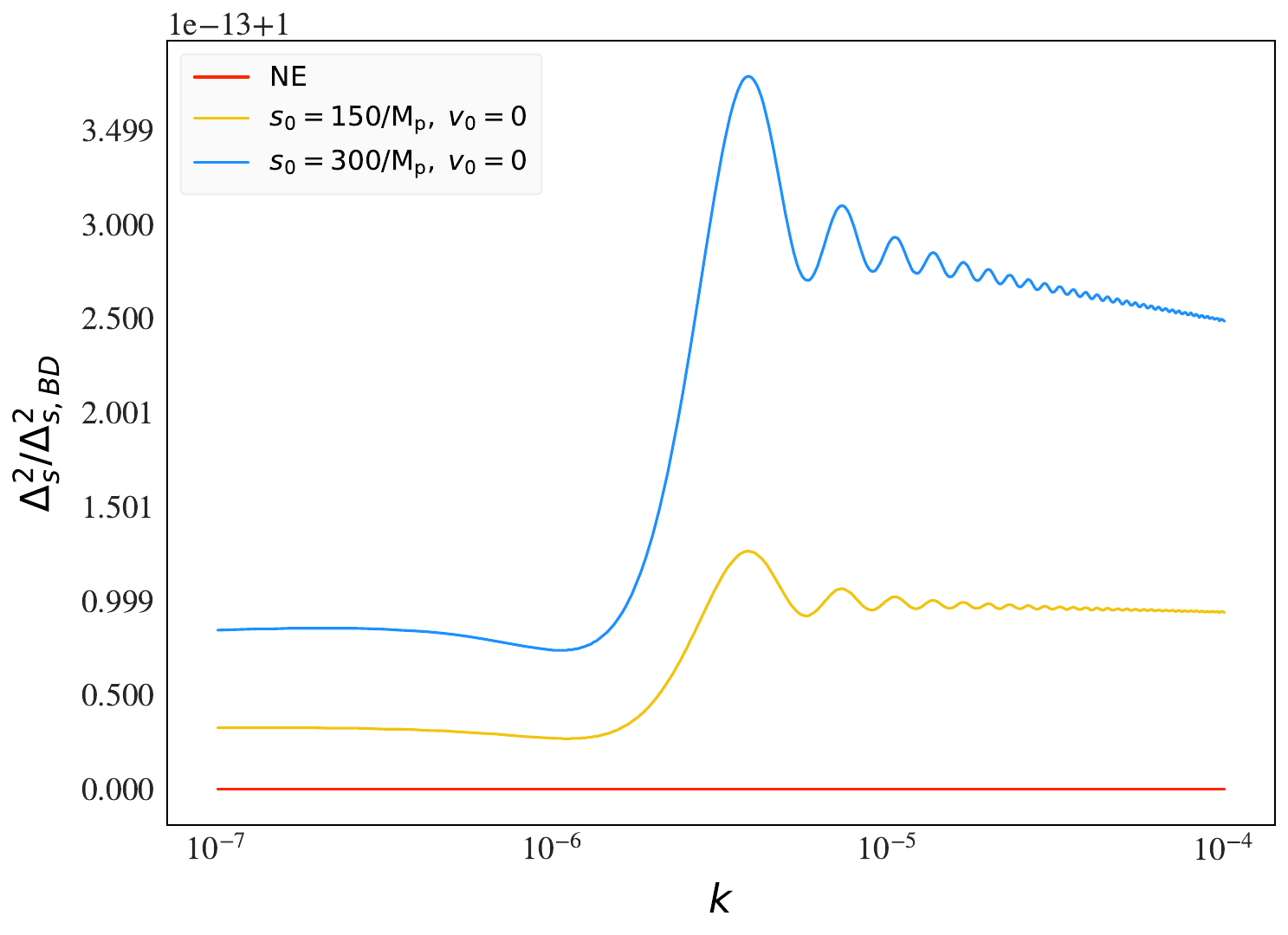}}\qquad
\subfloat[]{\label{fig:s2}\includegraphics[width=0.40\textwidth]{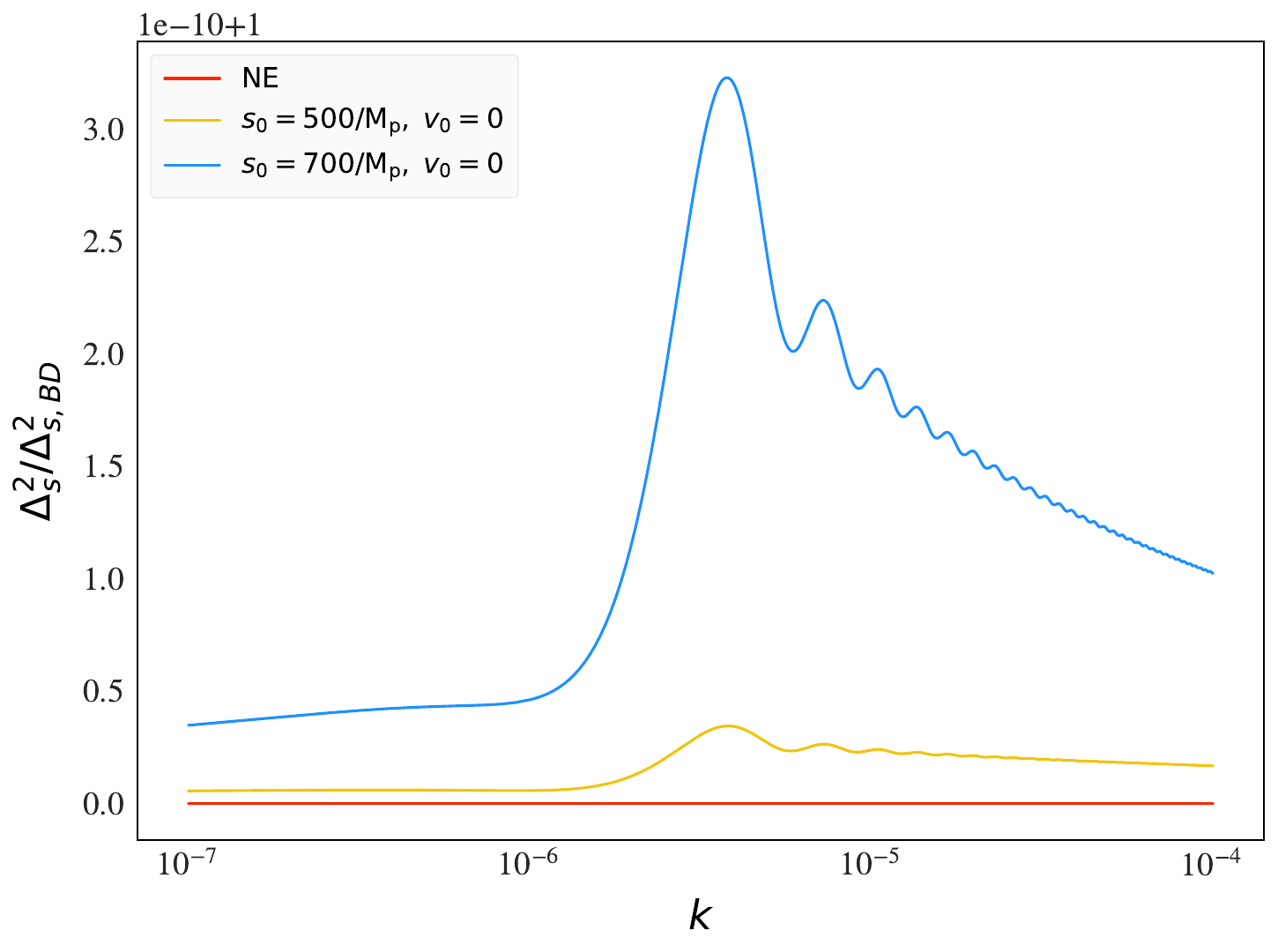}}\\
\subfloat[]{\label{fig:s3}\includegraphics[width=0.40\textwidth]{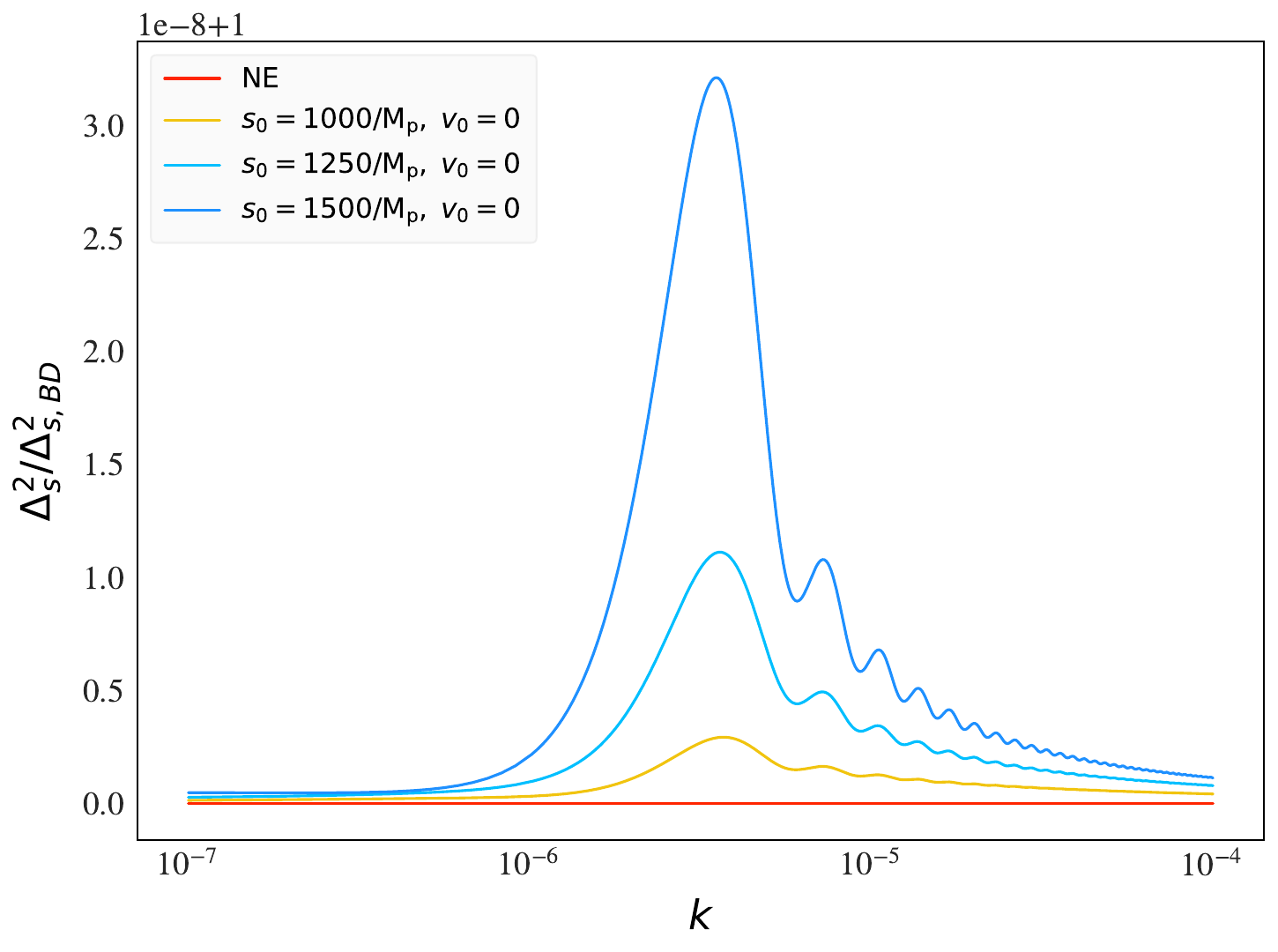}}\qquad
\subfloat[]{\label{fig:s4}\includegraphics[width=0.40\textwidth]{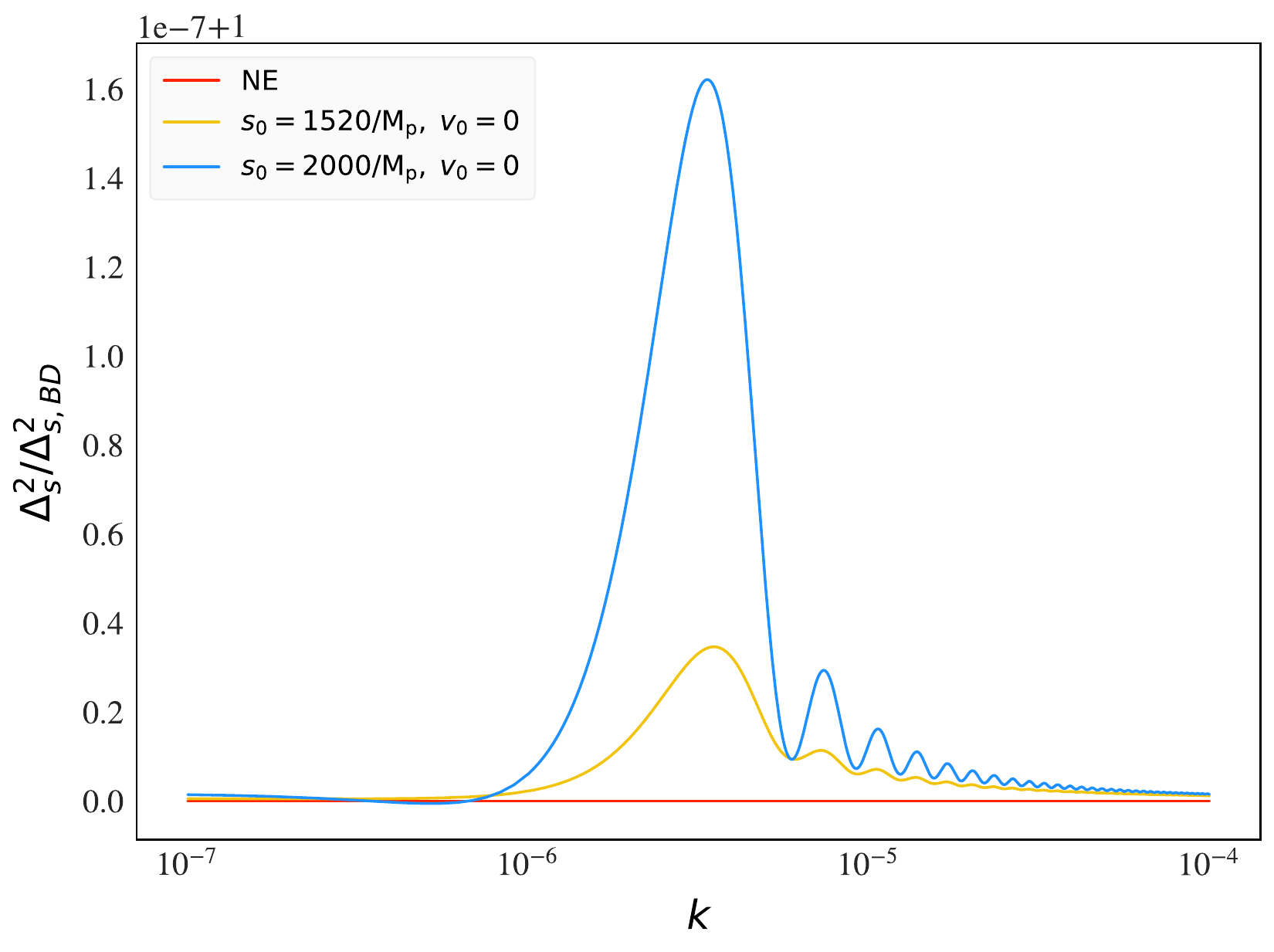}}\\
\subfloat[]{\label{fig:s5}\includegraphics[width=0.40\textwidth]{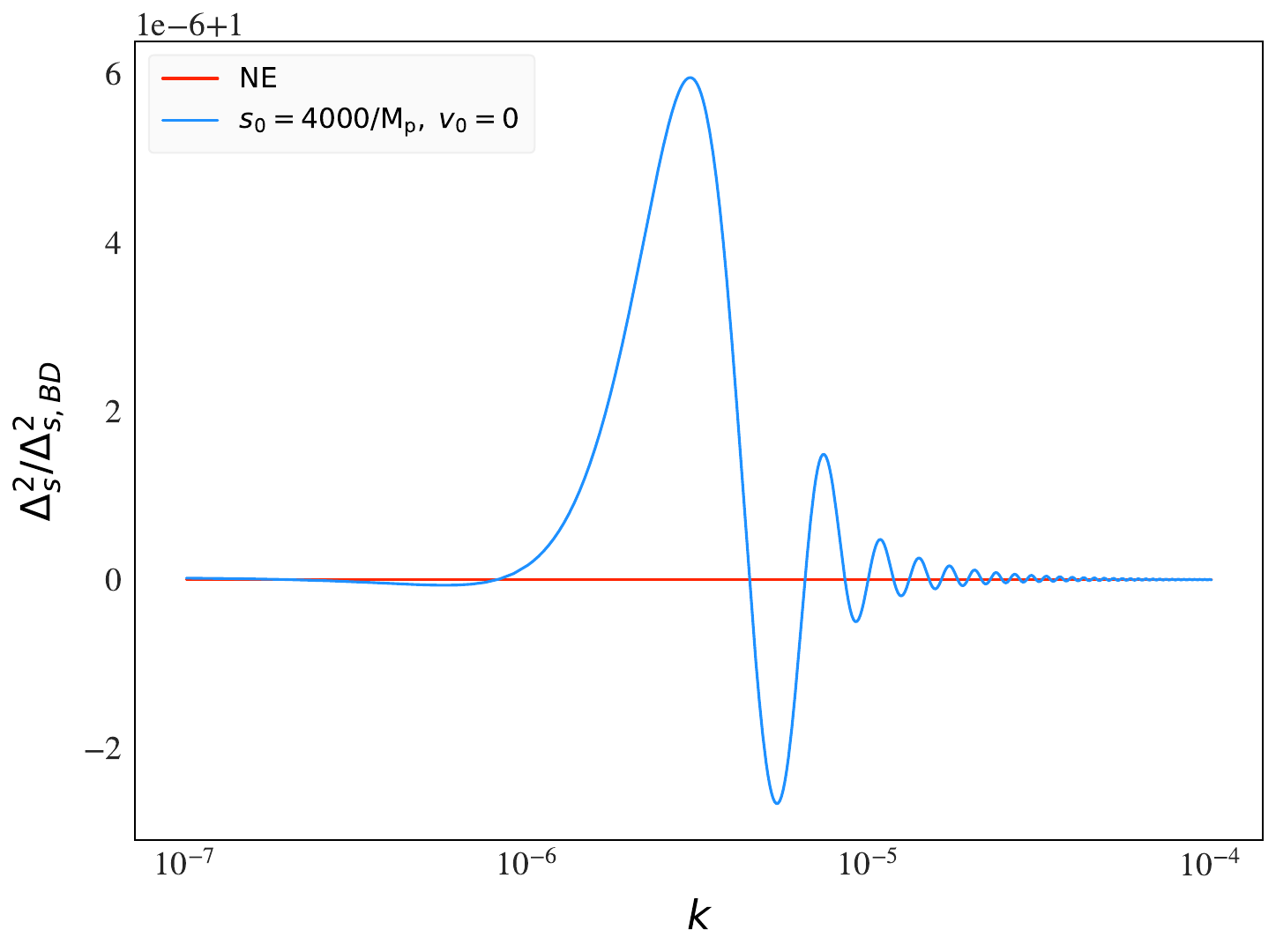}}\qquad
\subfloat[]{\label{fig:s6}\includegraphics[width=0.40\textwidth]{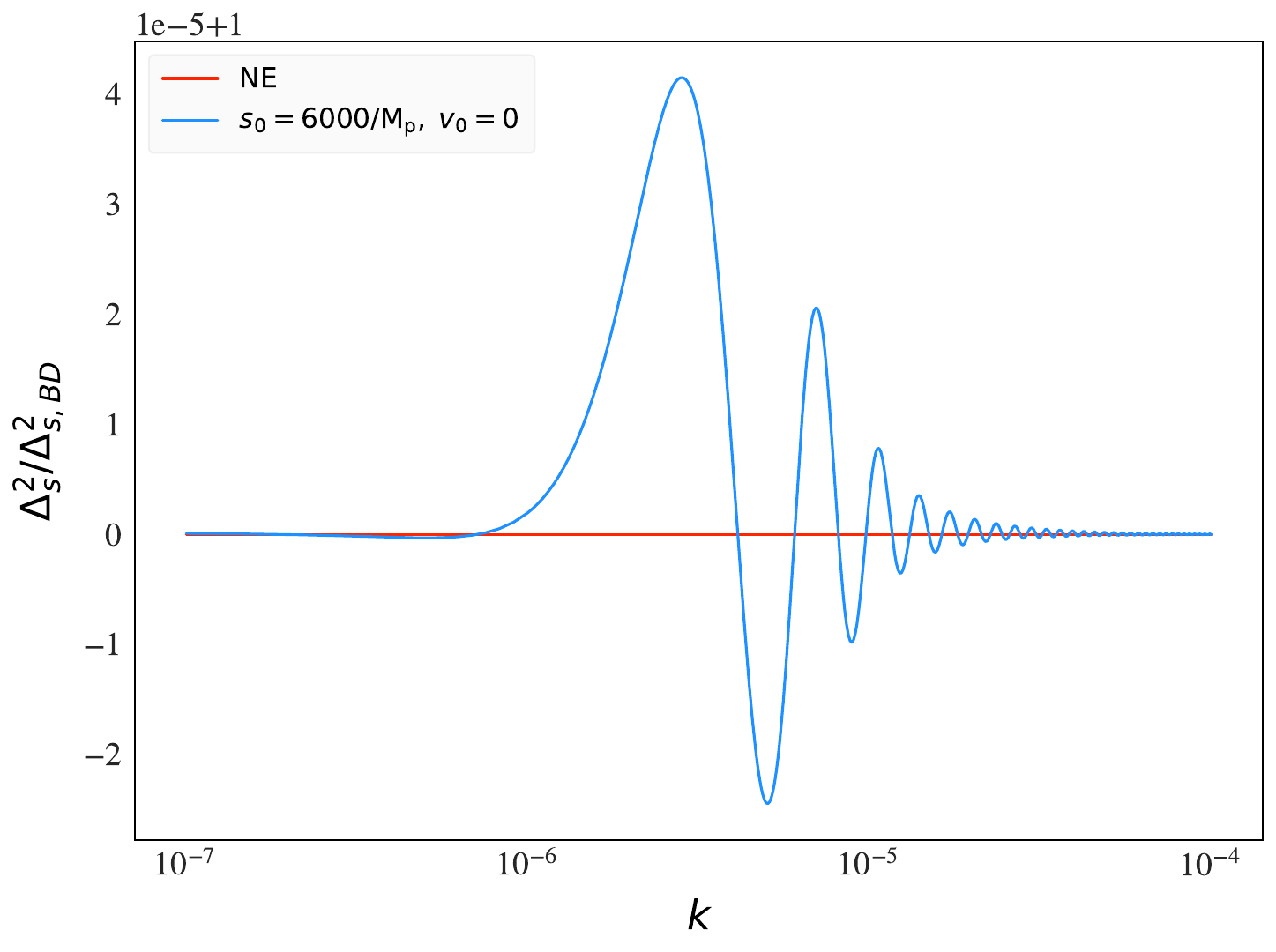}}\\
\subfloat[]{\label{fig:s7}\includegraphics[width=0.40\textwidth]{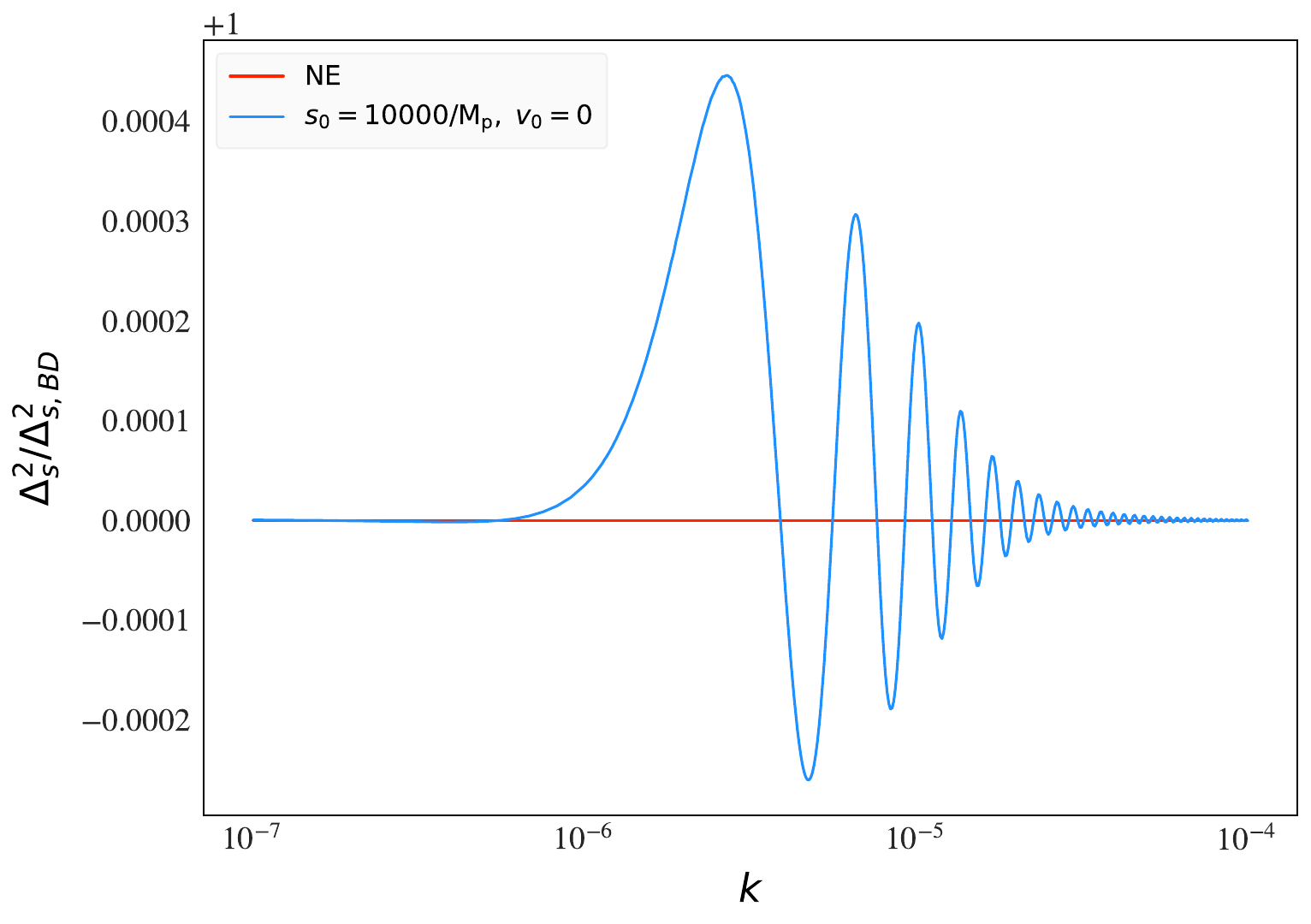}}\qquad
\subfloat[]{\label{fig:s8}\includegraphics[width=0.40\textwidth]{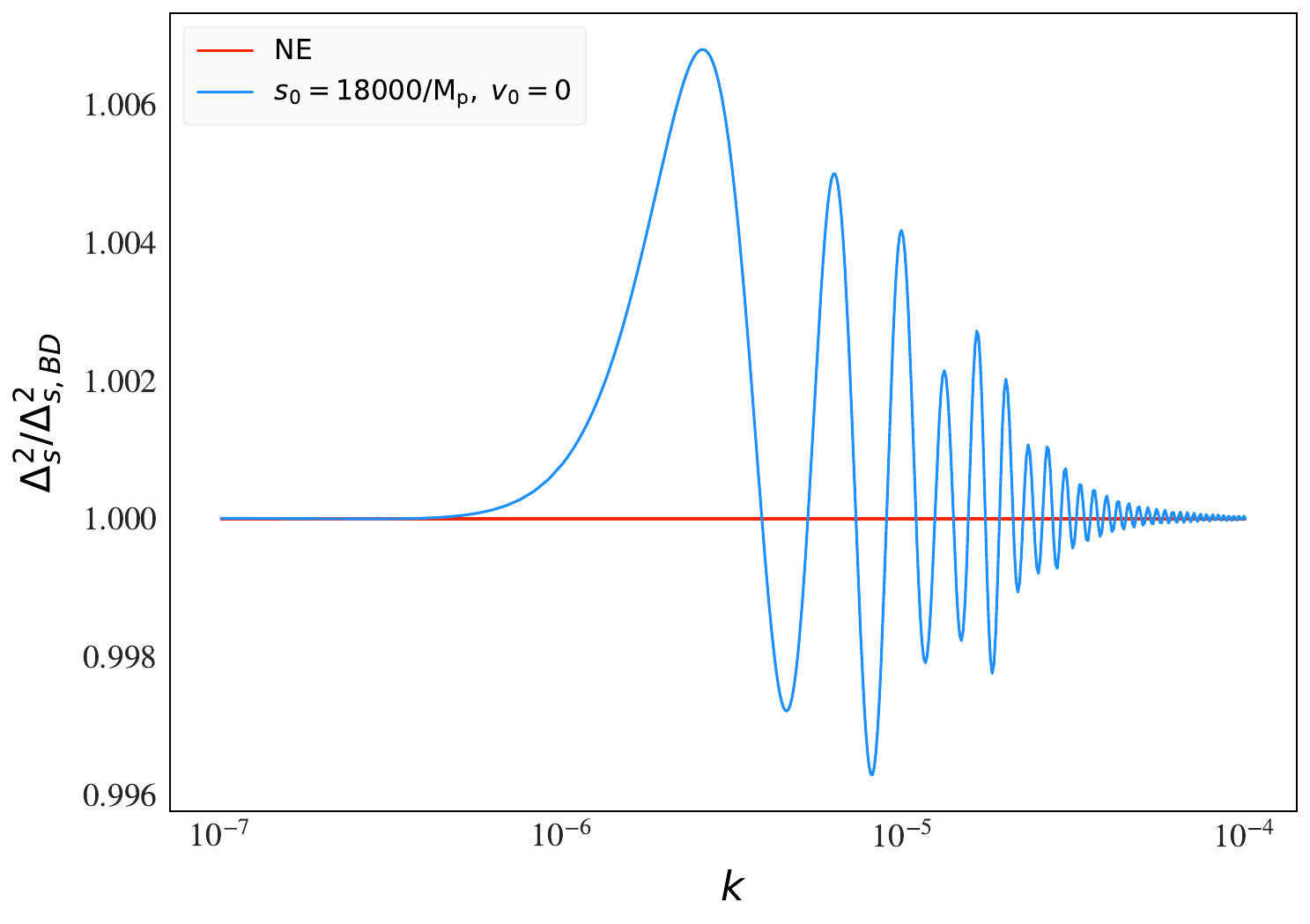}}
\vspace{-0.2cm}
\caption{ \label{fig:PS_100GevHiggs} A variety of results for fractional corrections to the scalar primordial power spectrum due to entanglement, $\frac{\Delta_s^{2}}{\Delta_{s,BD}^{2}}$, given a spectator scalar field with the Higgs-like potential defined in eq.~\eqref{eq:HiggspotDimless}. For all plots the spectator had a variety of initial positions and no initial velocity, with $T=T_{GH} = 100 \; \mathrm{GeV}$ (corresponding to an inflationary energy scale of $H_{ds} \approx 628 \; \mathrm{GeV}$ and $\epsilon = O(10^{-25})$). The non entangled (NE) case corresponds to $\frac{\Delta_s^{2}}{\Delta_{s,BD}^{2}} = 1$. $k_{\mathrm{ent}} = 10^{-6}$ for all plots (see eq.~\ref{eq:kent}).}
\end{figure}

\begin{figure}
    \centering
    \subfloat[$\sigma(\tau) = s(\tau) \times M_{p}$]{\label{fig:sigma}\includegraphics[width=0.47\textwidth]{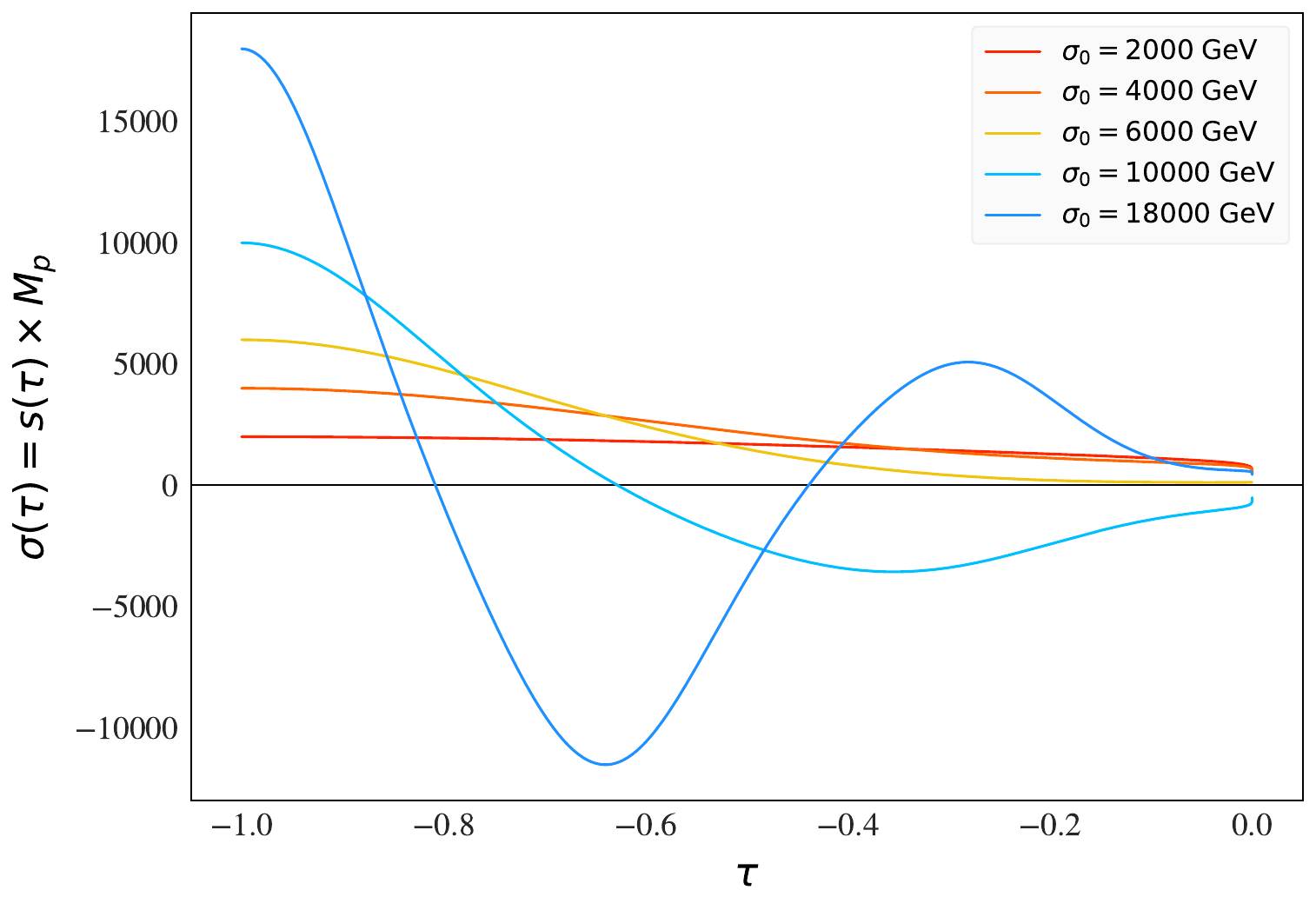}}\quad
\subfloat[$\nu_{g}^{2}\big(s(\tau)\big)$]{\label{fig:nug}\includegraphics[width=0.47\textwidth]{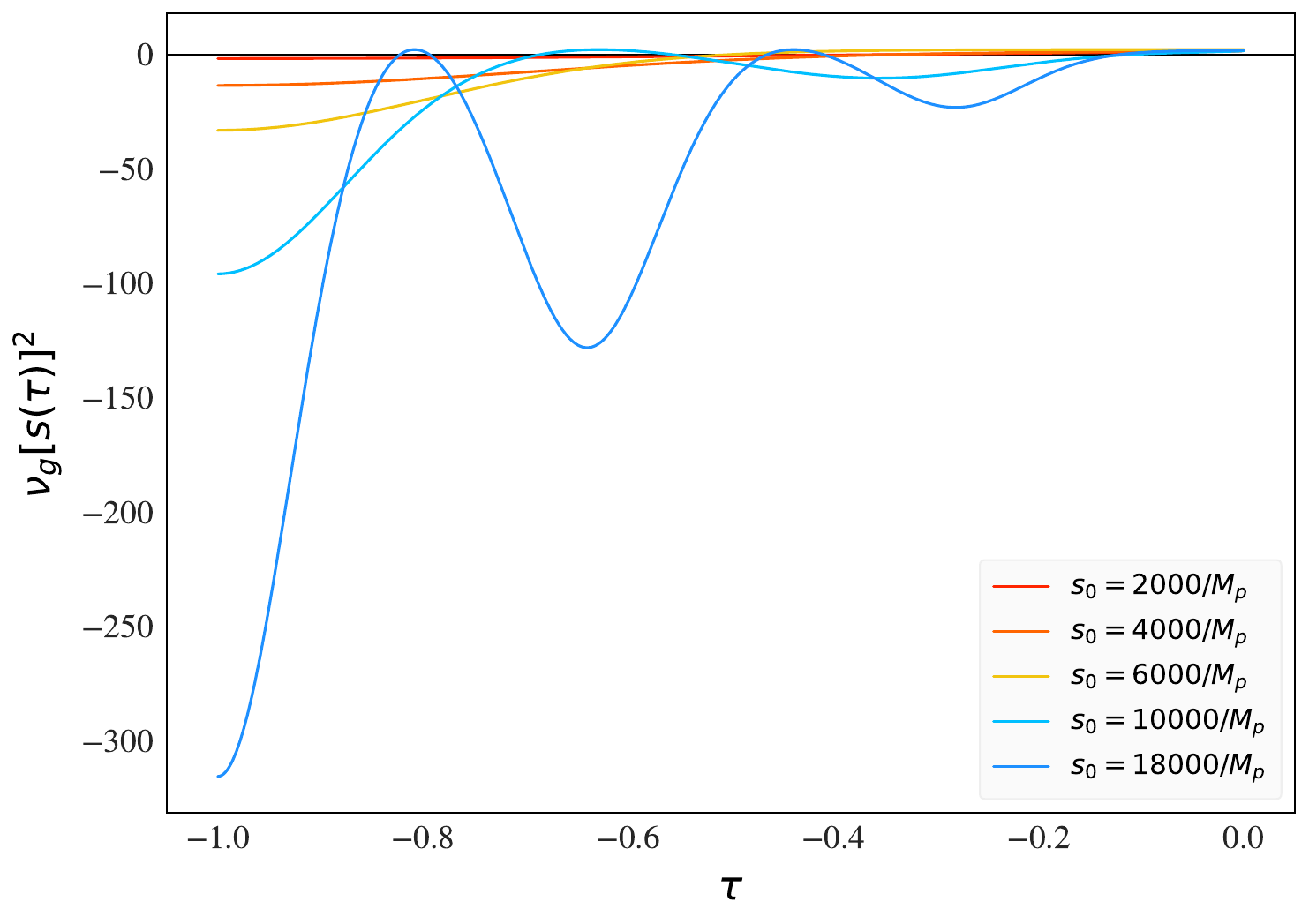}}\\
\caption{\label{fig:snug_100GevHiggs} Location of the zero mode, $\sigma = s M_{p}$, on the Higgs-like spectator potential of eq.~\eqref{eq:HiggspotDimless} and evolution of the quantity $\nu_g^{2}$---where $\nu_g$ is given by eqs.~\eqref{eq:Hankelnug} and~\eqref{eq:HiggsVdpr}---during inflation, corresponding to the primordial cases shown in figure~\ref{fig:PS_100GevHiggs}. As in figure~\ref{fig:PS_100GevHiggs}, $T=T_{GH} = 100 \; \mathrm{GeV}$, $H_{ds} \approx 628 \; \mathrm{GeV}$, and $\epsilon = O(10^{-25})$). Dimensionless conformal time $\tau$ is defined in eq.~\eqref{eq:dimless}.}
\end{figure}

Before investigating possible signatures of phase transitions and/or the inflationary energy scale, I first explored what types of primordial power spectrum features the Higgs-like spectator could produce, since this was a novel use of the formalism.
This also helped me narrow down which spectator initial conditions yield potentially observationally constrainable signals. A representative sample of my results is shown in figures~\ref{fig:PS_100GevHiggs} and~\ref{fig:snug_100GevHiggs}. 

Figure~\ref{fig:PS_100GevHiggs} plots the fractional change to the scalar primordial power spectrum due to entanglement, $\frac{\Delta_s^{2}}{\Delta_{s,BD}^{2}}$, with $\Delta_s^{2}$ defined in eq.~\eqref{eq:psPerturb} and the standard Bunch-Davies result given by eq.~\eqref{eq:psBD}. For all the results shown in figure~\ref{fig:PS_100GevHiggs} I took $T=T_{GH} = 100 \; \mathrm{GeV}$---corresponding to an inflationary energy scale of $H_{ds} \approx 628 \; \mathrm{GeV}$---which is below $T_c$ and therefore in the symmetry broken phase for the simple Higgs-like potential defined in eqs.~\eqref{eq:Higgspot}~-~\eqref{eq:HiggspotDimless}. As in previous work~\cite{Adil:2022rgt}, I set the initial velocity of the spectator zero mode to be zero, so that the variety of oscillatory features in the primordial power spectrum show in figure~\ref{fig:PS_100GevHiggs} are driven by the initial position of the spectator zero mode and its effective mass. However, in contrast to previous work, the effective mass is no longer constant since the second derivative of the Higgs-like potential depends on the position of the zero mode. Figure~\ref{fig:snug_100GevHiggs} shows the evolution of the zero mode location and of the parameter $\nu_g^{2}$---whose variation is due to variation in the spectator's effective mass, see eqs.~\eqref{eq:Hankelnug},~\eqref{eq:Meff}, and~\eqref{eq:HiggsVdpr}---during the course of inflation, given the initial conditions specified in figure~\ref{fig:PS_100GevHiggs}.

There are two relevant points to make about the results in figure~\ref{fig:PS_100GevHiggs}.  First, there is a interesting transition in the types of oscillatory behaviors that one observes. For smaller initial displacements from the origin, corrections due to entanglement are purely positive (i.e. $\frac{\Delta_s^{2}}{\Delta_{s,BD}^{2}} > 1$). However, this changes for larger displacements, and one begins to observe oscillations both above and below the zero point for such initial conditions. The transition point between these two behaviors seems to occur shortly after the initial value of the parameter $\nu_g$ equals zero (which corresponds to an initial displacement of $s_0 \approx 1520/M_p$ for the zero mode, given the parameters used for the numerical results in figure~\ref{fig:PS_100GevHiggs}).~\footnote{For larger initial displacements, the effective mass term in eq.~\eqref{eq:Hankelnug} is greater than the slow roll terms, which causes the parameter $\nu_g^{2}$ to become negative and $\nu_g$ itself to become imaginary. This causes numerical problems for calculating initial conditions for the kernel equations (since Hankel indices of imaginary order are not currently available in Python to the author's knowledge). To mitigate this issue, I generated initial conditions for Hankel indices of imaginary order in Mathematica, and then input these initial conditions into my Python code for the rest of the numerical analysis.  This workaround allowed me to study spectators with effective masses greater than $H_{ds}$, in contrast to the Monte Carlo analysis previously performed in~\cite{Adil:2022rgt}. However, such a workaround was only feasible since this is a `case study' analysis, as the resource intensiveness of this solution would not be conducive to Monte Carlo work.} One can see evidence of this by comparing figures~\ref{fig:s4} and~\ref{fig:s5}, for example.

The second relevant point deals with issues of observability and scale.  It's clear from figure~\ref{fig:PS_100GevHiggs} that utilizing a Higgs-like potential for the spectator scalar field can impart a variety of interesting oscillatory features on the primordial power spectrum due to entanglement. For example, some intriguing secondary structure in the oscillations of figure~\ref{fig:s7} and~\ref{fig:s8} occurs when the zero mode has enough energy to explore both sides of the potential (as shown in figure~\ref{fig:sigma}). However, if one looks at the y-axis of figure~\ref{fig:PS_100GevHiggs}, most of these features will never be observable. As demonstrated in previous work~\cite{Baunach:2021yvu,Adil:2022rgt}, corrections to the primordial power spectrum of $O(10^{-13})$ or even $O(10^{-4})$ are tiny and completely unobservable with current CMB data. So while the parameter space \textit{is} rich for this type of potential, only larger initial displacements have a hope of being observationally constrained, such as the near percent level corrections in figure~\ref{fig:s8}.

\subsubsection{\label{sec:interpret} Narrative for an observationally relevant result}

Despite the case study format of this work, it's worth taking a moment to imagine what theoretical scenario an observationally relevant result like figure~\ref{fig:s8} might correspond to. Let us start with the spectator potential and initial conditions.  The results in figure~\ref{fig:s8} utilize a Higgs-like potential in the symmetry broken phase ($T=T_{GH} = 100 \; \mathrm{GeV}   < T_{c}$ for my simplified model). They also require the zero mode of the field to begin its entangled evolution far away from the minimum (at an initial position of $\sigma = 18000 \; \mathrm{GeV}$). The zero mode is then allowed to roll during the course of inflation, and it has enough energy to explore both sides of the Higgs-like potential (as shown in figure~\ref{fig:sigma}). It then dynamically settles into one of the minima of the potential by the end of inflation as its kinetic energy decreases (a simple form of ``vacuum selection'').

For such a result to occur, there needs to be some event that precedes the entangled evolution and causes the spectator field to be high up on the potential at $\eta_0$.  This could be achieved by means of a bubble collison~\cite{Aguirre:2009ug}, a period of 
quantum fluctuations that would drive the location of the zero mode up the potential~\cite{Linde:2007fr, Guth:2007ng}, or some hitherto undiscovered mechanism.  (For example, the inflaton zero mode is typically assumed to begin inflation away from its potential minimum, and one could argue we do not yet have a satisfactory mechanism why this should be so.) Such an event or mechanism is not out of the question, but would require a deeper understanding of what precedes inflation in this scenario to fully motivate it. 

A related question is `what is the timing of the (electroweak) phase transition?' for the result in figure~\ref{fig:s8}. If one is interested in the time at which the Higgs-like potential develops a second minimum, then the (electroweak) phase transition precedes inflation in our narrative. However, if one is curious when vacuum selection has occurred, then this would happen at the end of inflation in the narrative corresponding to figure~\ref{fig:s8}. In either case, one would know that the phase transition occurred by the end of inflation, and not afterward, if a result such as figure~\ref{fig:s8} was found to be consistent with observational data.

Lastly, the results in figure~\ref{fig:s8} assume low-scale ($H_{ds} \approx 628 \; \mathrm{GeV}$) inflation as a theoretical input. While this may not be necessarily favored in the literature (i.e. GUT-type motivations), it is also not observationally ruled out (see the discussion in section~\ref{sec:intro}). It's also worth pointing out that I only assume the observationally (CMB) relevant part of inflation occurs at a low scale to obtain figure~\ref{fig:s8}. The era of entangled evolution that my equations probe might be proceeded by an era of higher scale inflation, 
or there may be some other pre-inflationary phase.

However, if a signal such as figure~\ref{fig:s8} has the potential to be observationally constrained, the question then becomes, how unique is it? At the lowest level, since we are fairly ignorant about the intricacies of the very early universe, can Planck data distinguish between the signal in figure~\ref{fig:s8} and other well-motivated scalar spectators with similar levels of entanglement? If it can, then can the data make more fine grained distinctions, such as determining whether a Higgs-like potential is symmetry broken or symmetry restored, at the level of CMB anisotropies? And finally, are such signals unique to a particular inflationary energy scale, or do degeneracies exist? I explore these questions in the next two sections.

\section{\label{sec:CMB}CMB-level potential differentiation} 
In this section I investigate whether, given an inflationary energy scale,  Planck data can distinguish between entanglement with a Higgs-like spectator compared to other spectator scalar fields with different potentials---given similar levels of entanglement for all spectators. This is a first step towards discerning whether entanglement can be used as a probe of phase transitions---since if differentiation between different scalar potentials proves impossible at the level of CMB anisotropies, it then seems unlikely the data would be able to distinguish between different behaviors of the same potential. I first define the other test cases to compare with the signal of figure~\ref{fig:s8}, and then present some initial results from $C_{\ell}^{TT}$ residuals.

\subsection{\label{sec:Imposter} Contrasting spectator scalars}
In the wild west of the early universe, there are, of course, a myriad of spectator scalars one could consider with a variety of theoretical motivations and UV completions.  The list I considered for this case study analysis therefore is definitely not exhaustive, and simply represents some straightforward examples that nevertheless had the potential to confuse the data.

The first contrasting possibility I considered was a free massive scalar with the same mass as the Higgs-like potential at $T=0$, i.e.
\begin{equation} \label{eq:FMSHiggs}
    V_{125 \; GeV} = \frac{1}{2} M_{h}^{2} \sigma^{2}, \;  \; \quad  M_h= 125 \; \mathrm{GeV} \; .
\end{equation}
The second possibility was also a free massive scalar, but it was engineered so that its mass was the same as the initial effective mass for the signal in figure~\ref{fig:s8}, i.e. 
\begin{align} \label{eq:11TeV}
V_{11 \; TeV} &= \frac{1}{2} M^{2} \sigma^{2}, \; \; \quad 
M = 11 \; \mathrm{TeV} \; s.t. \; \nu_{g,11\;TeV} = \nu_{g,i,Higgs}. 
\end{align}
This made it so the parameter $\nu_g$ was the same for both cases initially.  However, after the initial time the Higgs-like spectator's $\nu_g$ was allowed to evolve as the zero mode rolled, while $\nu_g$ for the 11 TeV free massive scalar remained constant throughout the entangled evolution (since the second derivative of the potential for a free massive scalar is constant).

Finally, the third contrasting case I considered was a spectator with a `strongly' symmetry non restored (SSNR) Higgs-like potential.  This potential could be used as a proxy for symmetry non-restoration in the early universe, or it could be a simple model of a `dark Higgs'-type scalar that is heavier than its standard model counterpart.  For this analysis I designed the SSNR potential as follows:
\begin{equation} \label{eq:SSNR}
    V_{SSNR} = -\frac{1}{2} (25m_{h})^{2} \sigma^{2} + \frac{\lambda_h}{4}\sigma^{4} + \frac{\lambda_h}{8}T^{2}\sigma^{2}
\end{equation}
with $m_h = M_h/\sqrt{2}$, $M_h = 125 \ \mathrm{GeV}$, $\lambda_h = m_h^{2}/v^{2}$, and $v = 246 \ \mathrm{GeV}$, as done in eq.~\eqref{eq:Higgspot}. In essence I increased the mass by a factor of 25, but kept the self coupling the same compared to the Higgs-like potential in eq.~\eqref{eq:Higgspot}, which deepens the potential wells and therefore raises the temperature at which the SSNR potential becomes `symmetry restored.'

\begin{figure}[h!]
    \centering
    \includegraphics[width=0.64\textwidth]{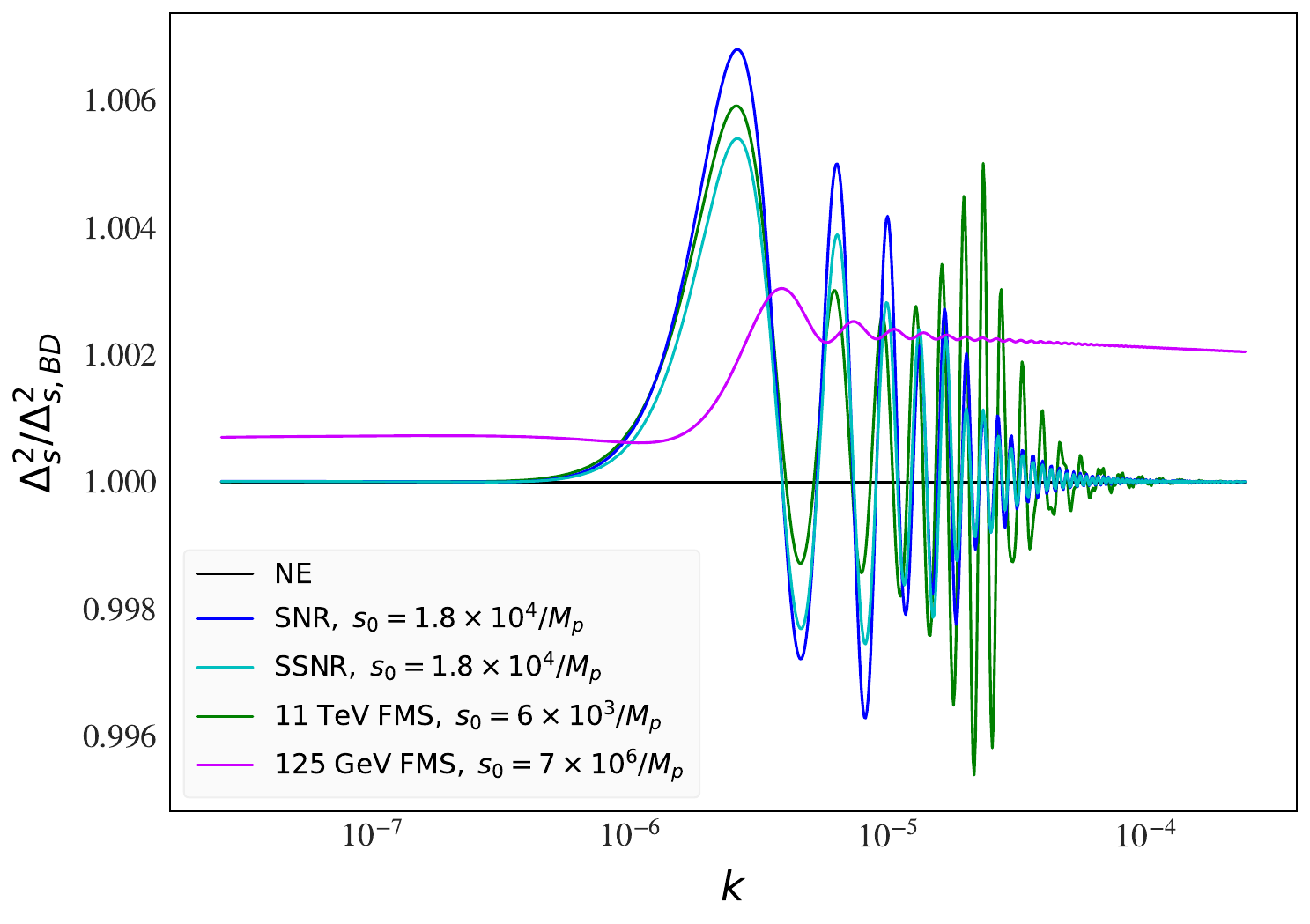}
    \caption{Fractional corrections to the scalar primordial power spectrum due to entanglement, $\frac{\Delta_s^{2}}{\Delta_{s,BD}^{2}}$, for spectator scalars utilizing the SNR Higgs-like spectator potential (eq.~\ref{eq:HiggspotDimless}, blue), the SSNR Higgs-like potential (eq.~\ref{eq:SSNR}, cyan), the 11 TeV free massive scalar potential (eq.~\ref{eq:11TeV}, green), and the 125 GeV free massive scalar potential (eq.~\ref{eq:FMSHiggs}, purple), for a similar level of entanglement (as discussed in the text). The non entangled (NE) case corresponds to $\frac{\Delta_s^{2}}{\Delta_{s,BD}^{2}} = 1$. $k_{\mathrm{ent}} = 10^{-6}$ (see eq.~\ref{eq:kent}).}
    \label{fig:ImposterPS}
\end{figure}
\vspace{-0.3cm}

Figure~\ref{fig:ImposterPS} plots corrections to the scalar primordial power spectrum due to entanglement for spectator scalars utilizing these three potentials together with the spectra from the Higgs-like spectator in figure~\ref{fig:s8}. For the potentials with explicit temperature dependence, $T=T_{GH} = 100 \; \mathrm{GeV}$ as in figure~\ref{fig:PS_100GevHiggs}, and all equations used the same value for $\epsilon$ (and therefore $H_{ds}$), as determined by eq.~\eqref{eq:eps}. The initial positions for each case are listed in the caption of figure~\ref{fig:ImposterPS}, and all four cases plotted in figure~\ref{fig:ImposterPS} had no initial velocity. These initial conditions were chosen so that each case corresponded to a roughly comparable amount of entanglement, parameterized by $\lambda$ (see the discussion surrounding eq.~\ref{eq:lambdadef}), as well as the degree to which $|\frac{\Delta_s^{2}}{\Delta_{s,BD}^{2}}| > 1$. For the cases shown in figure~\ref{fig:ImposterPS}, $\lambda = \lambda_{1}(\tau)|_{\tau = -1}$, where $\lambda_{1}$ is defined in eq.~\ref{eq:dimless} in terms of dimensionless variables.

At the level of the power spectrum, the free massive scalar of eq.~\eqref{eq:FMSHiggs}---plotted in purple in figure~\ref{fig:ImposterPS}---is the most visually different by eye. However, all the results plotted in figure~\ref{fig:ImposterPS} have visual `tells' that distinguish them. Utilizing the other free massive scalar potential, given by eq.~\eqref{eq:11TeV}, the resulting spectra (plotted in green) has similar behavior to the Higgs-like spectator at smaller k values, but has a prominent secondary oscillatory feature at slightly higher k that the other results do not share. Even spectators with the SNR vs SSNR Higgs-like potentials are distinguishable at this level---mostly with amplitude, but if one stares closely at the range just to the right of $k= 10^{-5} Mpc^{-1}$ in figure~\ref{fig:ImposterPS}, there is also some evidence of the two spectra exchanging which one has the greater amplitude for a given oscillation. In the next section I will present results investigating whether a form of these visual distinctions persist at the level of CMB anisotropies.


\vspace{-0.2cm}
\subsection{\label{sec:Clres} CMB residuals}

\begin{figure}[h!]
    \centering
    \includegraphics[width=0.58\textwidth]{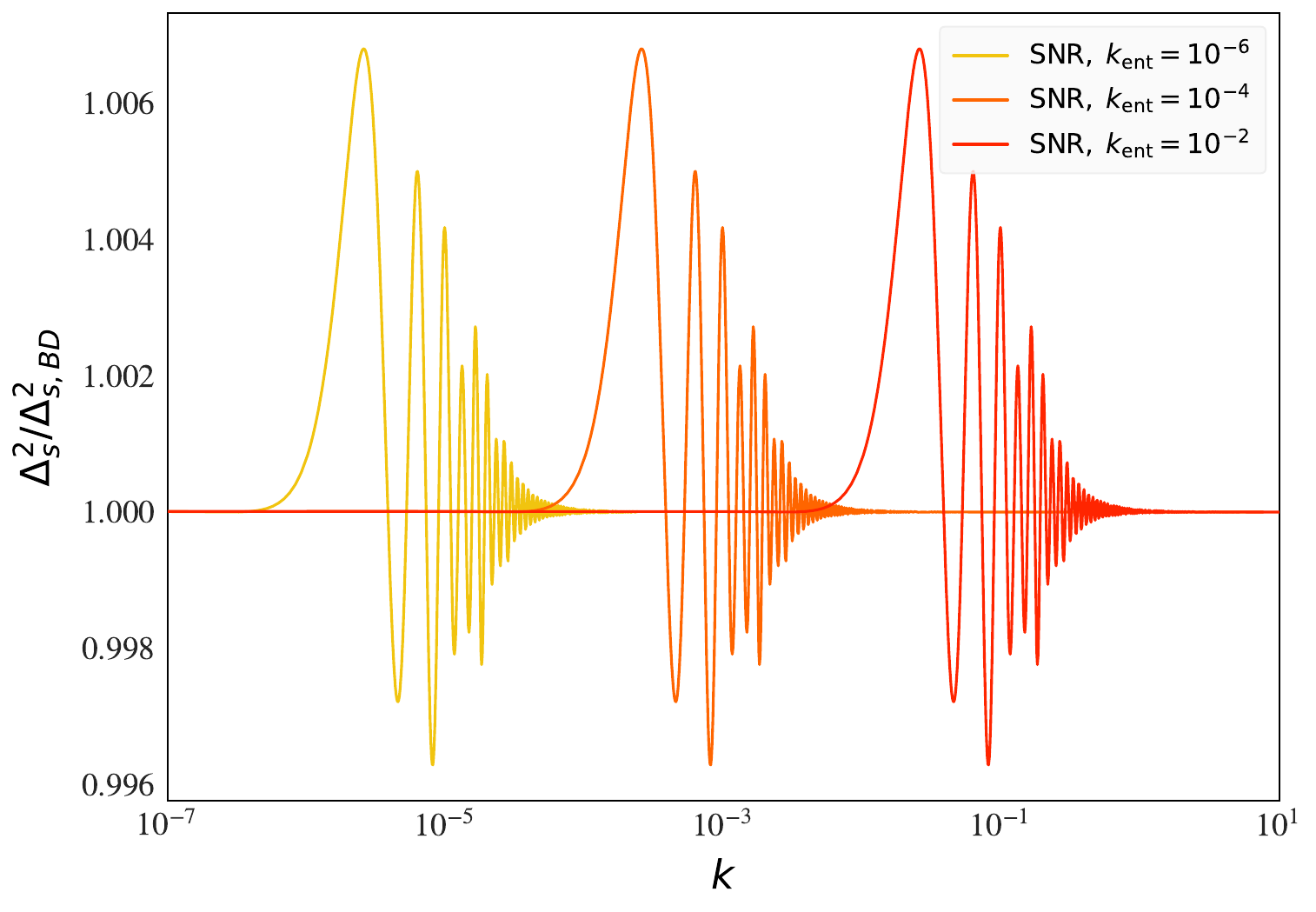}
    \vspace{-0.3cm}
    \caption{Fractional corrections to the scalar primordial power spectrum due to entanglement, $\frac{\Delta_s^{2}}{\Delta_{s,BD}^{2}}$, for the SNR Higgs-like spectator, given three different values of $k_{\mathrm{ent}}$. $k_{\mathrm{ent}} = 10^{-6}$ corresponds to the SNR spectra plotted in figure~\ref{fig:ImposterPS} and figure~\ref{fig:s8}.}
    \label{fig:kent}
\end{figure}

To compare the primordial power spectra results in the previous section with CMB data, I use the form of the scalar power spectrum defined in eq.~\eqref{eq:psCMB} with $A_s$ and $n_s$ taken from best fit Planck values~\cite{2020Planck}. I also include the effects of varying $k_{\mathrm{ent}}$. Varying $k_{\rm ent}$ corresponds to shifting the scale that leaves the inflationary horizon at the onset of entanglement, and therefore the time at which entangled evolution begins (as discussed in section~\ref{sec:entReview}). As 
investigated previously~\cite{Baunach:2021yvu}, many modifications to the primordial power spectrum render CMB observables largely unchanged when the onset of entanglement corresponds to the largest observable scale, $k=10^{-6} \mathrm{Mpc}^{-1}$. By shifting $k_{\mathrm{ent}}$ to smaller scales, the data becomes much more constraining. Figure~\ref{fig:kent} shows the SNR Higgs-like spectra from figure~\ref{fig:ImposterPS} for a few different values of $k_{\mathrm{ent}}$.
\vspace{-0.5cm}
\vspace{-0.2cm}

\begin{figure}[h!]
    \centering
   \includegraphics[width=0.68\textwidth]{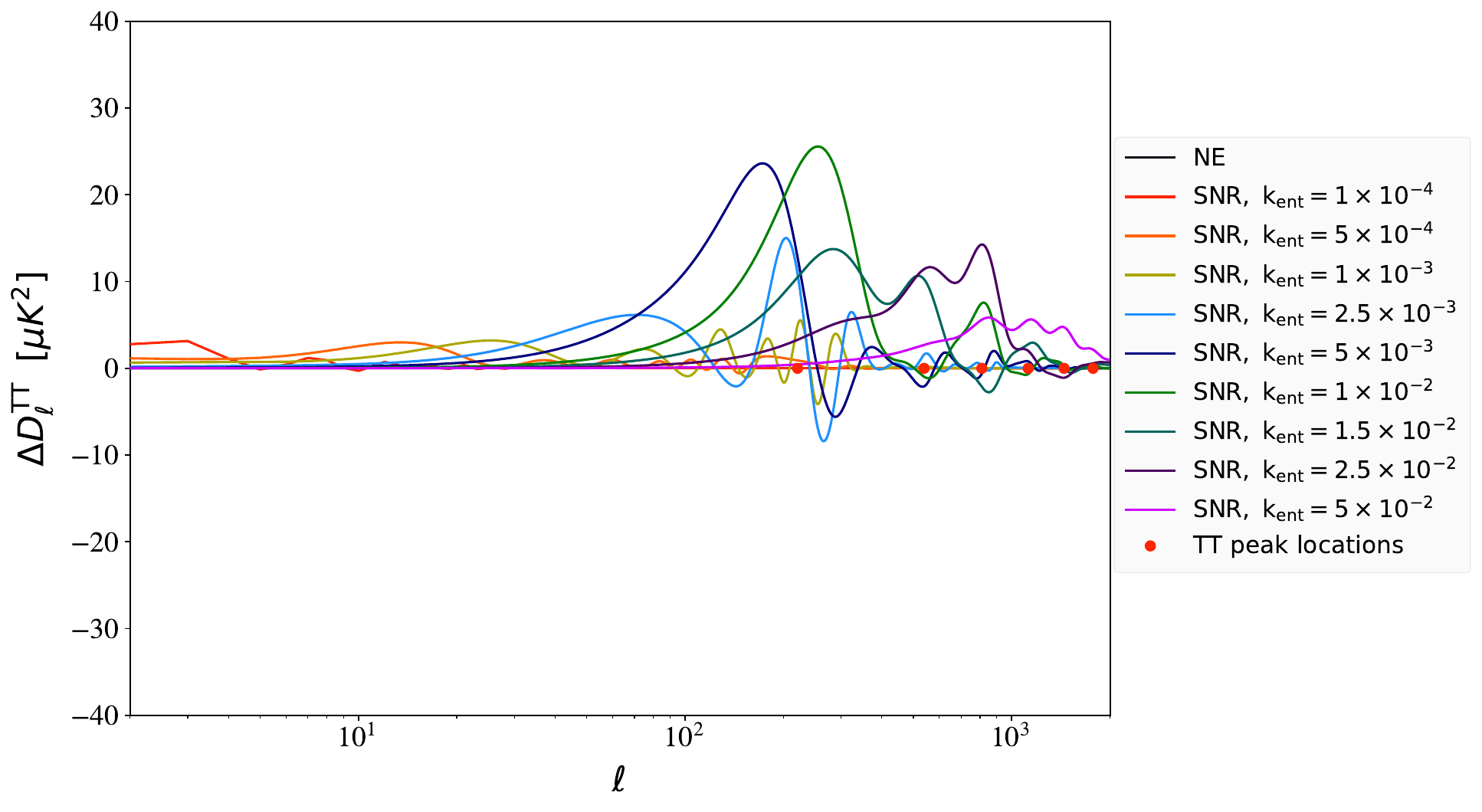} \hfill
    \includegraphics[width=0.68\textwidth]{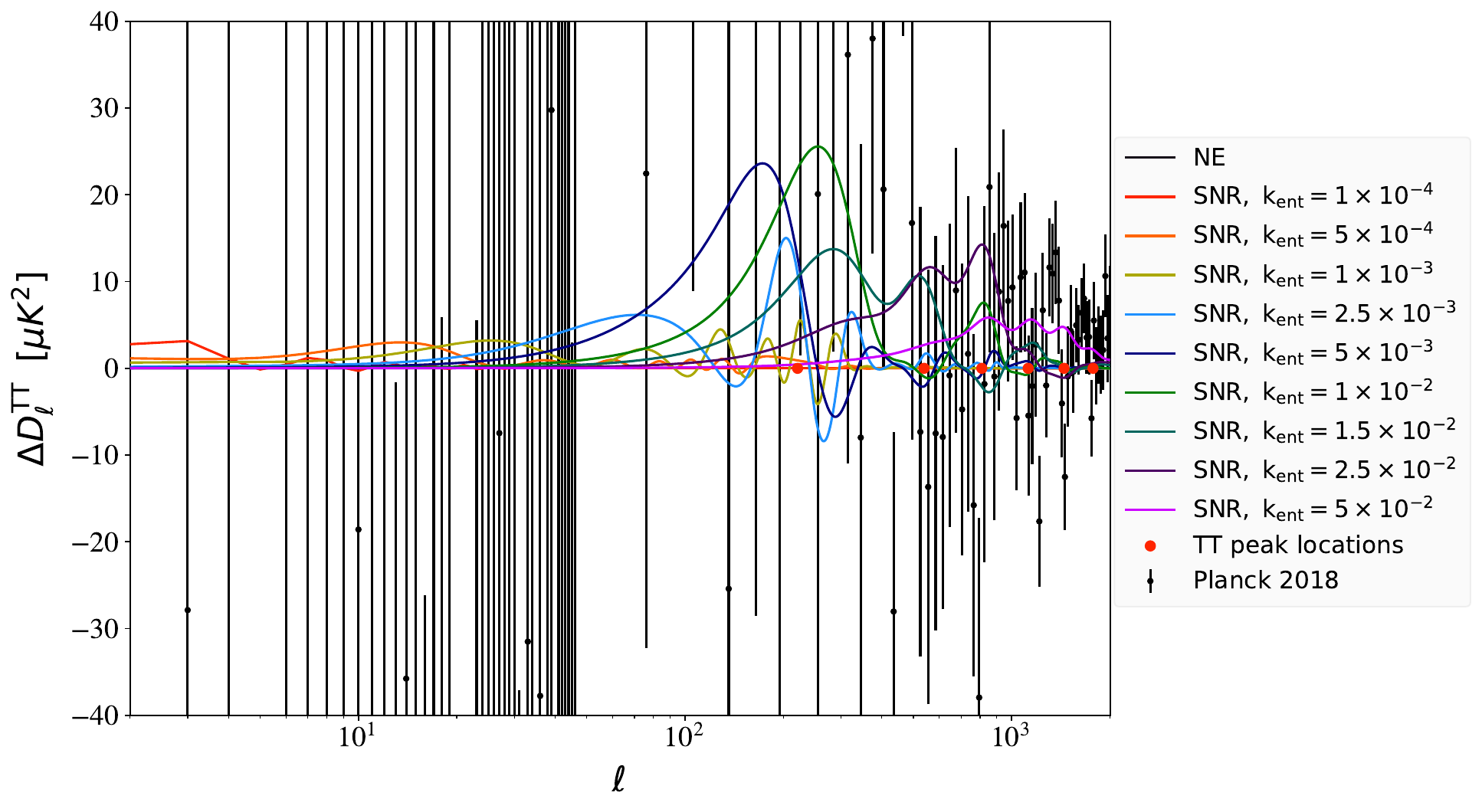} 
    \vspace{-0.2cm}
    \caption{\label{fig:SNR_cl} TT power spectrum residuals given the primordial power spectrum corrections in figure~\ref{fig:s8} for the SNR Higgs-like spectator. $D_{\ell}^{TT} = \frac{\ell (\ell + 1)}{2 \pi} C_{\ell}^{TT}$.  Residuals are calculated with respect to the non-entangled (NE) Bunch-Davies result. $k_{\rm ent}$ is varied as labeled in the caption. Locations of the TT peaks are also plotted to guide the eye. Data from the Planck 2018 data release.}
\end{figure}

Figure~\ref{fig:SNR_cl} shows the corresponding $C_{\ell}^{TT}$ residuals for the SNR Higgs-like spectator---computed using CLASS~\cite{Blas_2011}---both with and without error-bars from Planck data~\cite{2020Planck}. The residuals are with respect to the standard Bunch-Davies non-entangled solution (plotted as NE in the figure). I found spectra with $k_{\mathrm{ent}}$ in the range $10^{-4} \leq k_{\mathrm{ent}} \leq 10^{-2}$ were most highly constrained by the data, so I have displayed a selection of solutions within that range in the figure.


 The amplitude of the $C_{\ell}$ residuals in figure~\ref{fig:SNR_cl} are small but not insignificant. In particular, the region near the first three peaks of the TT-spectra has the most potential for constraining power, especially if future experiments are able to shrink some of the errorbars in that region. The signals are also clearly oscillatory---if one compares the results in figure~\ref{fig:SNR_cl} to the TT peak locations (plotted in red), the residuals shown clearly have unique oscillatory patterns (rather than simply shifting the amplitude of the TT peaks, for example). Furthermore, note the fact that the largest oscillatory features in the corresponding primordial power spectrum of figure~\ref{fig:s8} occur within 1-2 orders of magnitude (k decades) directly following $k_{ \rm ent}$. From figure~\ref{fig:SNR_cl}, it appears that values of $k_{\rm ent}$ close to $k_{piv}$---specifically $ 0.05 \; k_{piv}\leq k_{\rm ent} \leq k_{piv}$---are the most observationally relevant signals from the perspective of the TT spectra. For all of those signals, their values of $k_{\rm ent}$ would place the largest oscillatory features in the corresponding primordial spectrum to dominate the scales around $k_{piv}$.
\vspace{-0.3cm}

\vspace{-0.3cm}

\begin{figure}[h!]
    \centering
    \subfloat[125 GeV free massive scalar (eq.~\ref{eq:FMSHiggs})]{\label{fig:a}\includegraphics[width=0.68\textwidth]{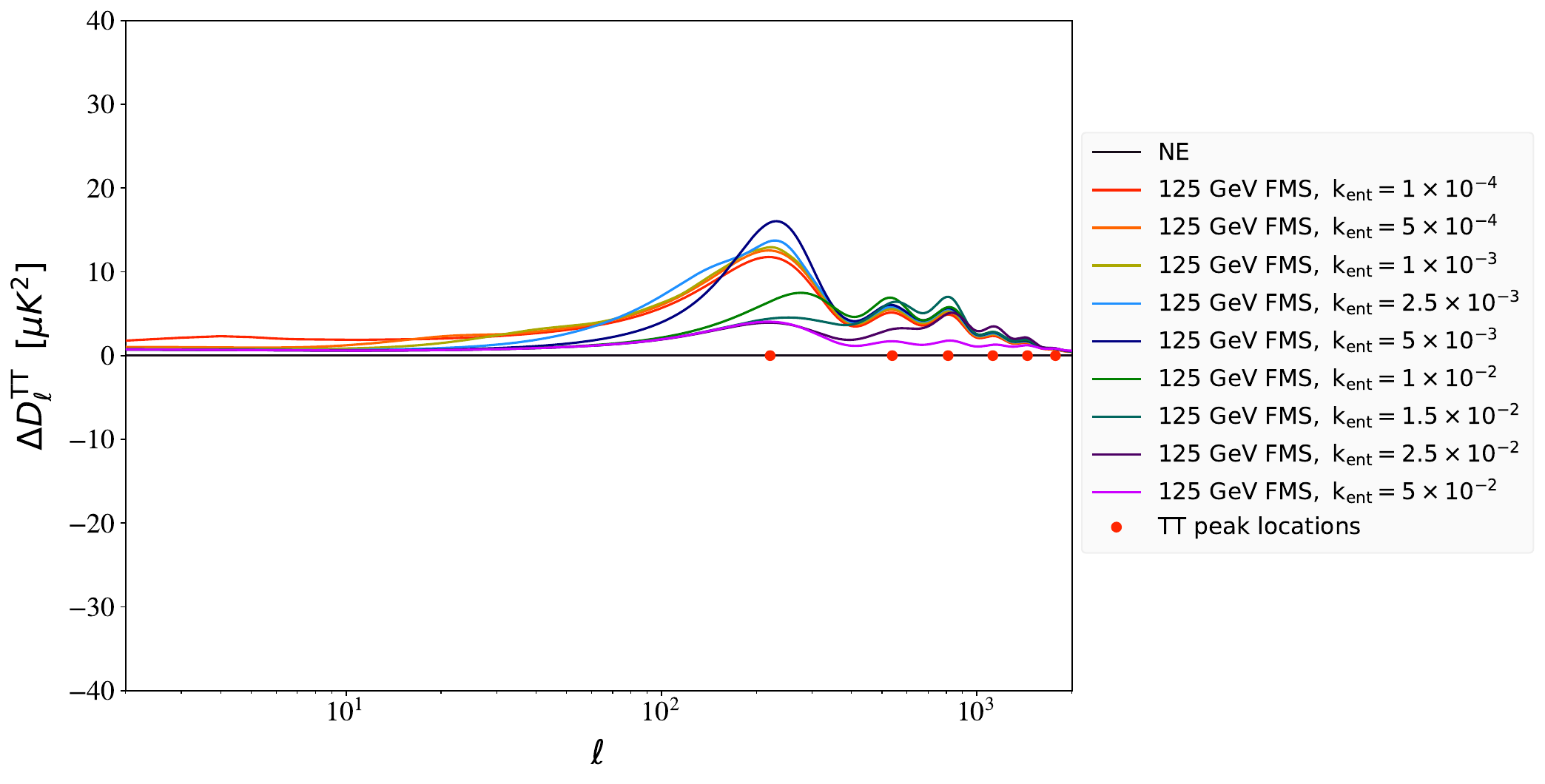}} \\
    \subfloat[11 TeV free massive scalar (eq.~\ref{eq:11TeV})]{\label{fig:b}\includegraphics[width=0.68\textwidth]{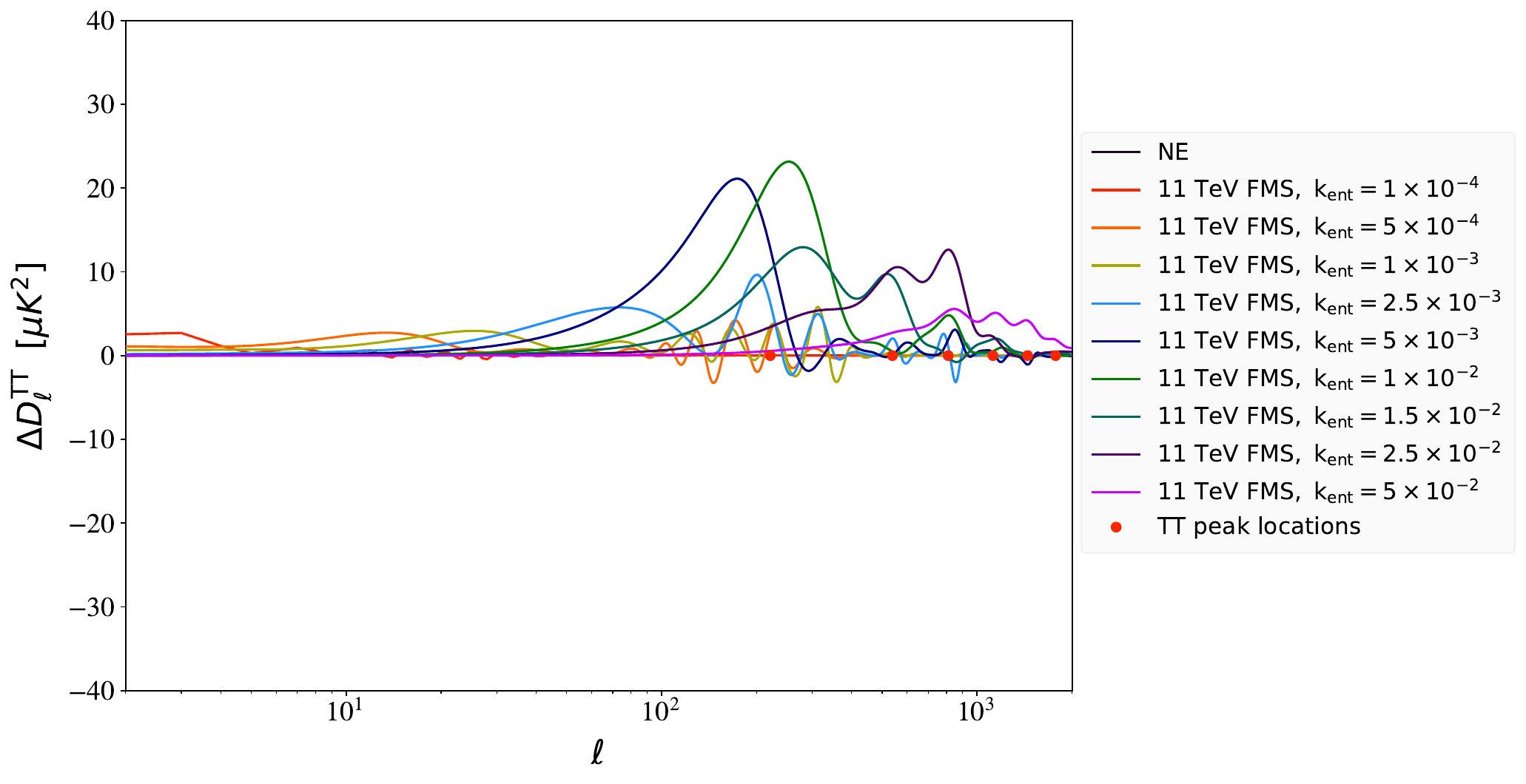}} \\
    \subfloat[SSNR Higgs-like spectator (eq.~\ref{eq:SSNR})]{\label{fig:c}\includegraphics[width=0.68\textwidth]{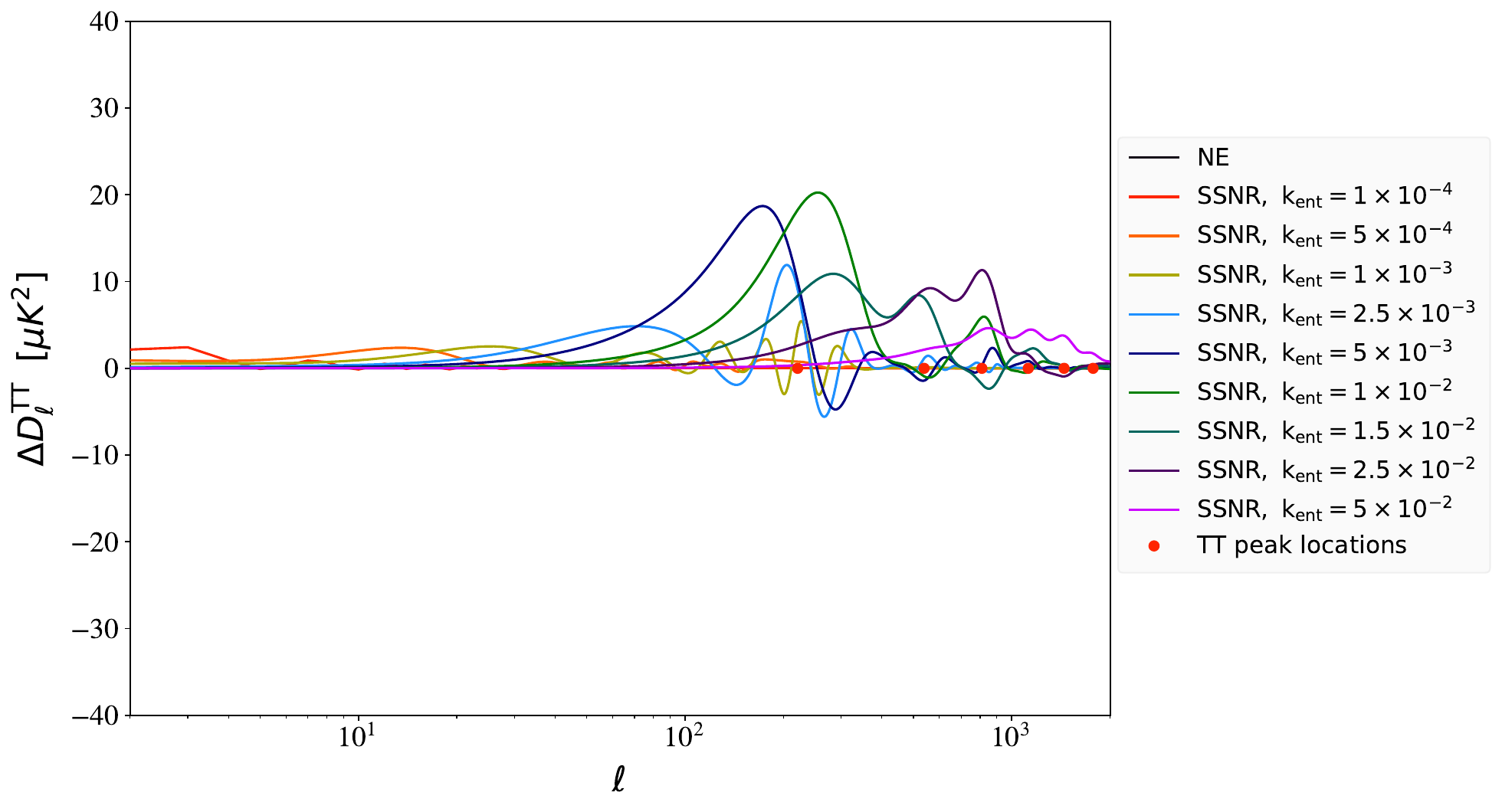}}
    \vspace{-0.2cm}
    \caption{\label{fig:others} TT power spectrum residuals given the primordial power spectrum corrections in figure~\ref{fig:ImposterPS} for the spectators scalars with potentials defined in eqs.~\eqref{eq:FMSHiggs}, \eqref{eq:11TeV}, and~\eqref{eq:SSNR}. $D_{\ell}^{TT} = \frac{\ell (\ell + 1)}{2 \pi} C_{\ell}^{TT}$.  Residuals are calculated with respect to the non-entangled (NE) Bunch-Davies result. $k_{\rm ent}$ is varied as labeled in the caption. Locations of the TT peaks are also plotted to guide the eye.}
\end{figure}

 One can repeat the same experiments for the other cases plotted in figure~\ref{fig:ImposterPS}. Figure~\ref{fig:others} shows the $C_{\ell}^{TT}$ residuals for the 125 GeV free massive scalar (figure~\ref{fig:a}), the 11 TeV free massive scalar (figure~\ref{fig:b}) and the SSNR Higgs-like spectator (figure~\ref{fig:c}), given the potentials defined in section~\ref{sec:Imposter}. 

 The results in figure~\ref{fig:a} are similar to what 
 was seen in previous work~\cite{Adil:2022rgt}, in that lower mass (i.e. $M_{eff} \leq H_{ds}$) free massive scalars typically cannot impart many oscillatory features due to entanglement at the level of CMB anisotropies. If one compares the residuals in figure~\ref{fig:a} to the locations of the TT peaks, one can see the main net effect of entanglement is to increase the amplitude of the TT peaks. Contrast this to the results in figures~\ref{fig:b} and~\ref{fig:c}, where there are clearly oscillatory features of the type in figure~\ref{fig:SNR_cl}.~\footnote{The spectator that generates the results in figure~\ref{fig:b} is also a free massive scalar, however its mass is larger than the Hubble scale, $H_{ds}$, which generates further oscillatory features (as shown in figure~\ref{fig:ImposterPS}). As discussed in section~\ref{sec:PowSpec}, this is a mass region
 the analysis in~\cite{Adil:2022rgt} did not investigate due to technical limitations, but that I was able to investigate for this work given a (non-ideal) workaround in my code.} 

\begin{figure}[h!]
    \centering
    \subfloat[$k_{\rm ent} = 2.5 \times 10^{-3}$ comparison]{\label{fig:compare_a}\includegraphics[width=0.72\textwidth]{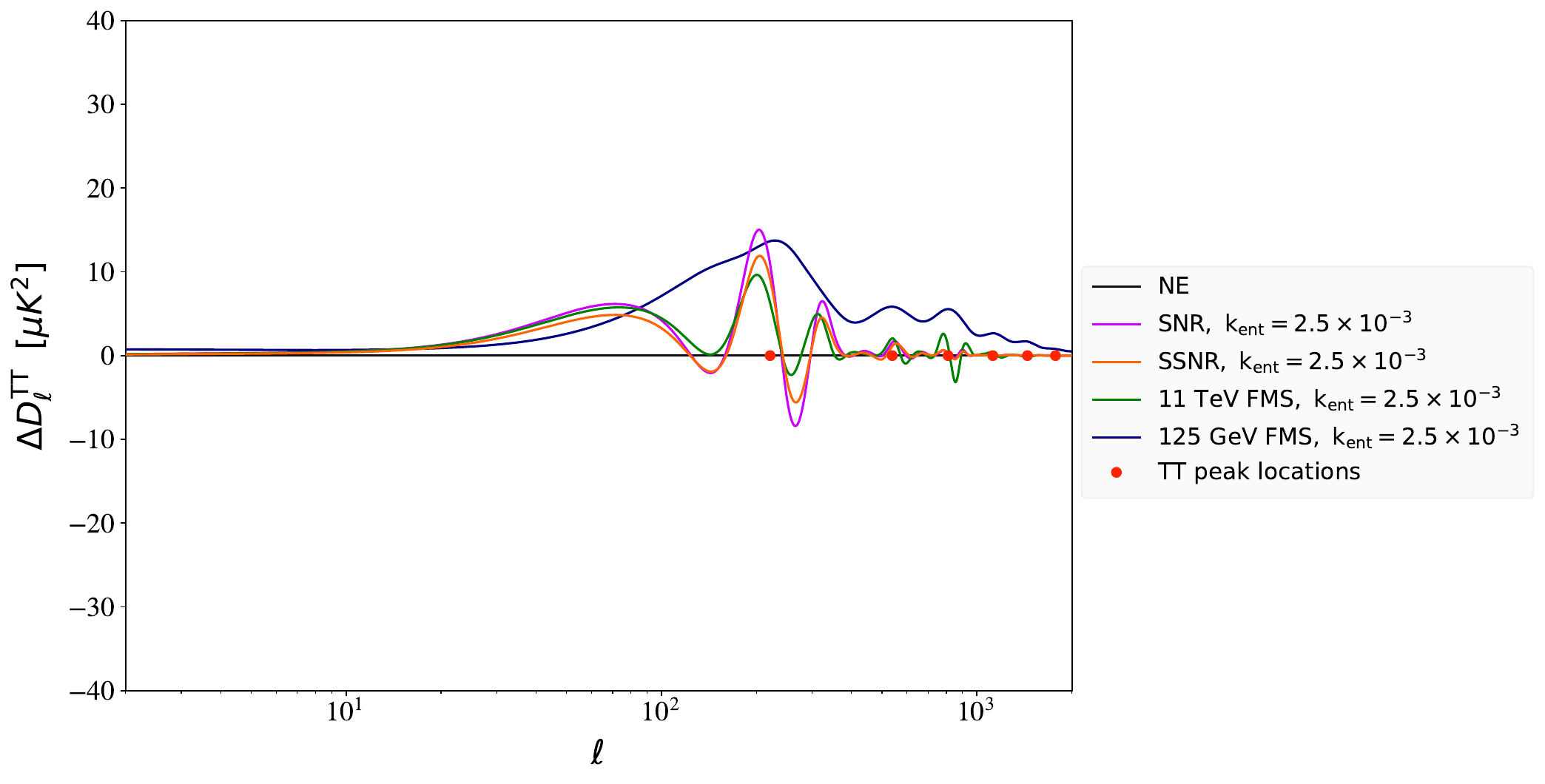}} \\
    \subfloat[$k_{\rm ent} = 5 \times 10^{-3}$ comparison]{\label{fig:compare_b}\includegraphics[width=0.72\textwidth]{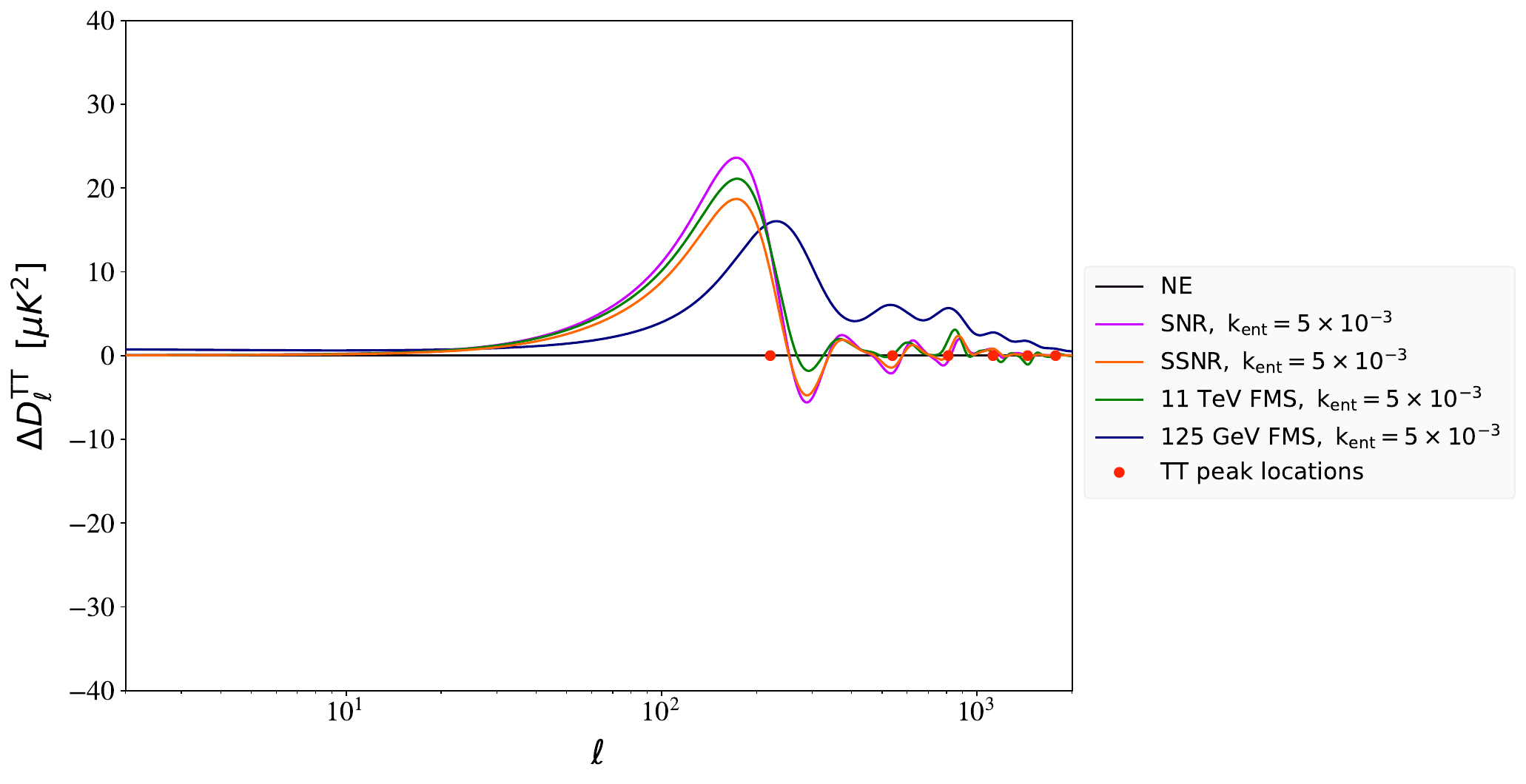}} 
    \vspace{-0.2cm}
    \caption{\label{fig:compare} Comparison of TT power spectrum residuals, given the primordial power spectrum corrections in figure~\ref{fig:ImposterPS}. $D_{\ell}^{TT} = \frac{\ell (\ell + 1)}{2 \pi} C_{\ell}^{TT}$.  Residuals are calculated with respect to the non-entangled (NE) Bunch-Davies result. $k_{\rm ent}$ is fixed for each subfigure. Locations of the TT peaks are also plotted to guide the eye.}
\end{figure}

Having shown that the primordial spectra in figure~\ref{fig:ImposterPS} yield CMB residuals that are not completely insignificant, I conclude this section with some direct comparisons. As a representative example, figure~\ref{fig:compare} contrasts the $C_{\ell}^{TT}$ residuals for the four spectator scalars in question, for two different values of $k_{\rm ent}$. If one compares these results to the primordial comparison of figure~\ref{fig:ImposterPS}, one can see many of the visual `tells' discussed with the primordial results still exist at the level of CMB anisotropies.
\vspace{-0.2cm}

\vspace{-0.3cm}
For example, in both figure~\ref{fig:compare_a} and figure~\ref{fig:compare_b}, the 125 GeV free massive scalar has a visually distinct impact compared to the other alternatives, just as its primordial signature in figure~\ref{fig:ImposterPS} is also distinct. Differences between the other three spectra are more suble, but still exist by eye. The oscillations in $C_{\ell}^{TT}$ residuals for the 11 TeV free massive scalar are phase shifted compared to the Higgs-like and SSNR Higgs-like spectators for several $\ell$ values in both plots.  There is also evidence of a secondary oscillatory feature at higher $\ell$ (clearest in figure~\ref{fig:compare_a}) that the other three spectra do not have, most likely due to the secondary feature that also exists in the corresponding primordial power spectrum (see figure~\ref{fig:ImposterPS}).\footnote{For example, using the approximation $\ell \approx (14000 \; \mathrm{Mpc})k$ and taking the $k_{\mathrm{ent}}$ shift in figure~\ref{fig:compare_a} into account, the secondary oscillatory feature at $\ell = 8 \times 10^{2} - 1 \times 10^{3}$ corresponds to $k = 2.3 - 2.9 \times 10^{-5} \mathrm{Mpc^{-1}}$, which is the approximate location of the secondary feature in figure~\ref{fig:ImposterPS} for the 11 TeV free massive scalar.} Even the Higgs-like and SSNR Higgs-like spectators show clear visual differentiation in the amplitudes of their residuals. 

Since this work is an exploratory case study, my goal is not to prove every minutia of my guiding questions, but instead explore what might be possible. In this section I have shown it \textit{is} possible to distinguish entanglement with a Higgs-like spectator versus other spectator scalars, at the level of CMB residuals. In fact, I have done one step better, as two of the contrasting spectators had features that were highly similar to the Higgs-like spectator at a primordial level. And since this coarse grained distinction appears feasible, I turn to more fine grained questions in the next section---namely whether one can determine if the Higgs-like potential is symmetry broken or symmetry restored, and to what degree such signatures are unique to a given inflationary energy scale.

\section{\label{sec:Sensitivity}Sensitivity to phase transitions and the inflationary energy scale}

Given the simple model for a Higgs-like spectator laid out in section~\ref{sec:Higgs}, questions about phase transitions and the inflationary energy scale are inherently linked. In my model the inflationary energy scale, $H_{ds}$, is linked to the first slow roll parameter, $\epsilon$, and $T_{GH}$---which plays the role of `temperature' in this work---as discussed in section~\ref{sec:EnergyScale}. Thus, if one were able to detect a signal in the data consistent with entanglement with a SNR Higgs-like spectator, then one would also implicitly know the inflationary energy scale. Similarly, if a signal consistent with a Higgs-like spectator at a given inflationary energy scale was detected, then one would know by definition if the potential was symmetry restored or not. In this section I make a first pass investigating how unique such signals are, or if degeneracies exist to impede such an analysis.

\begin{figure}[h]
    \centering
    \includegraphics[width=0.53\textwidth]{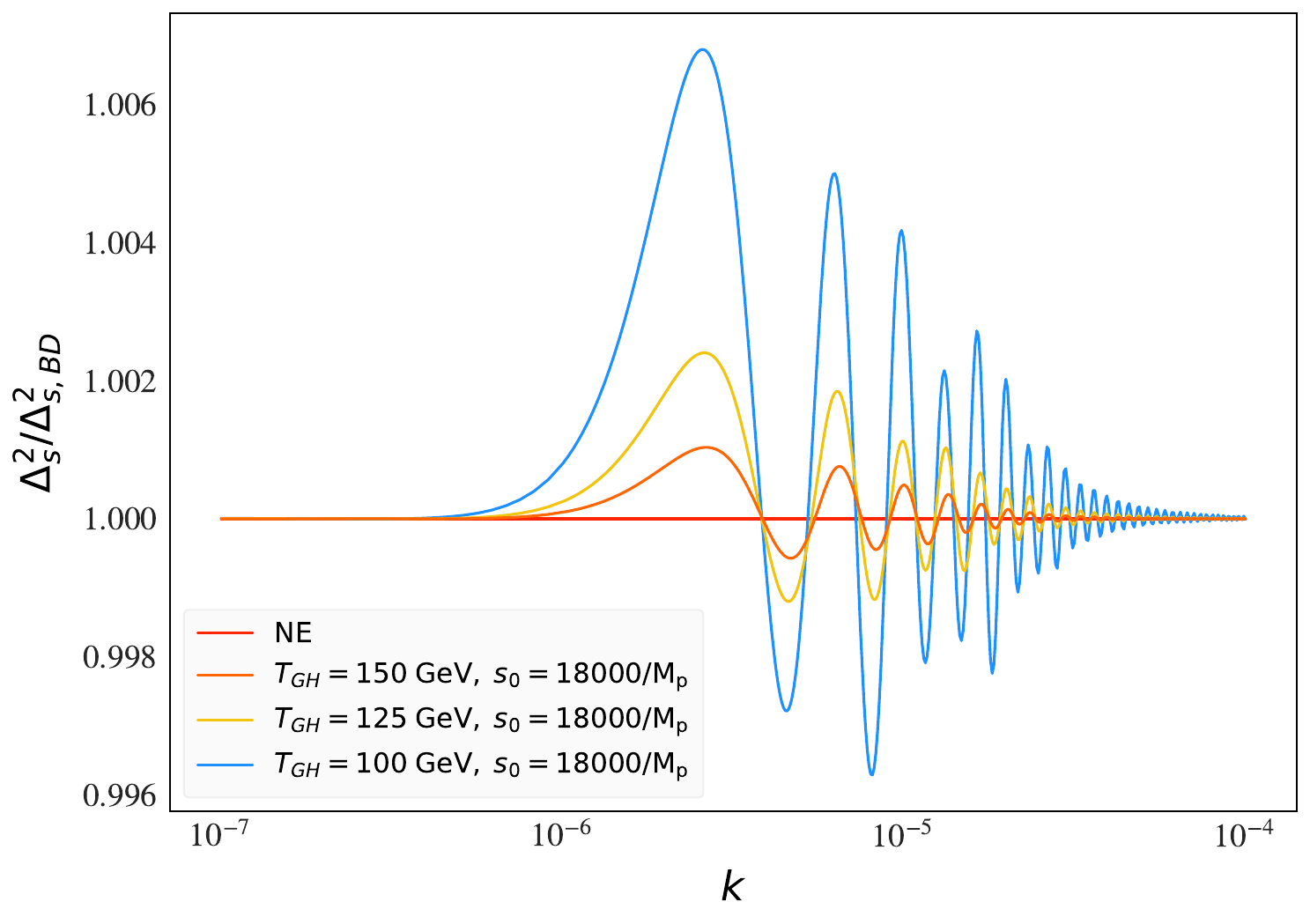}
    \caption{\label{fig:SNR_Tcompare} Fractional corrections to the scalar primordial power spectrum due to entanglement, $\frac{\Delta_s^{2}}{\Delta_{s,BD}^{2}}$, for the Higgs-like spectator. Initial conditions for the zero mode are held fixed---with $s_0 = 18000/M_p$ and $v_0 = 0$---but $T=T_{GH}$ (and therefore $\epsilon$ and $H_{ds}$) is varied. The non entangled (NE) case corresponds to $\frac{\Delta_s^{2}}{\Delta_{s,BD}^{2}} = 1$. $k_{\mathrm{ent}} = 10^{-6}$ (see eq.~\ref{eq:kent}).}
\end{figure}

Consider again the signal in figure~\ref{fig:s8}. The underlying spectator potential is the symmetry non-restored Higgs-like potential of eq.~\eqref{eq:HiggspotDimless}. Figure~\ref{fig:SNR_Tcompare} shows what happens when $T = T_{GH}$ (and therefore therefore $\epsilon$ and $H_{ds}$) are varied, keeping the rest of the spectator initial conditions fixed. 


Clearly, increasing $T$ causes a decrease in amplitude, as well as subtle shifts in the character of the oscillations. The values of $T$ shown in figure~\ref{fig:SNR_Tcompare} are all within $T < T_{c}$ for my simplified model, and so all correspond to SNR potentials, but I found the trend continues even when $T > T_{c}$ in my numerical experiments (in that $\frac{\Delta_s^{2}}{\Delta_{s,BD}^{2}}$ continues to decrease as $T$ is increased, if the spectator zero mode initial conditions are fixed). From this evidence one might be tempted to say, yes, a signal is unique to a particular inflationary energy scale---which would also enable one to answer questions about phase transitions by definition, as discussed above. However, this is unfortunately not the case.

\begin{figure}[h!]
    \centering
     \subfloat[]{\label{fig:deg_a}\includegraphics[width=0.48\textwidth]{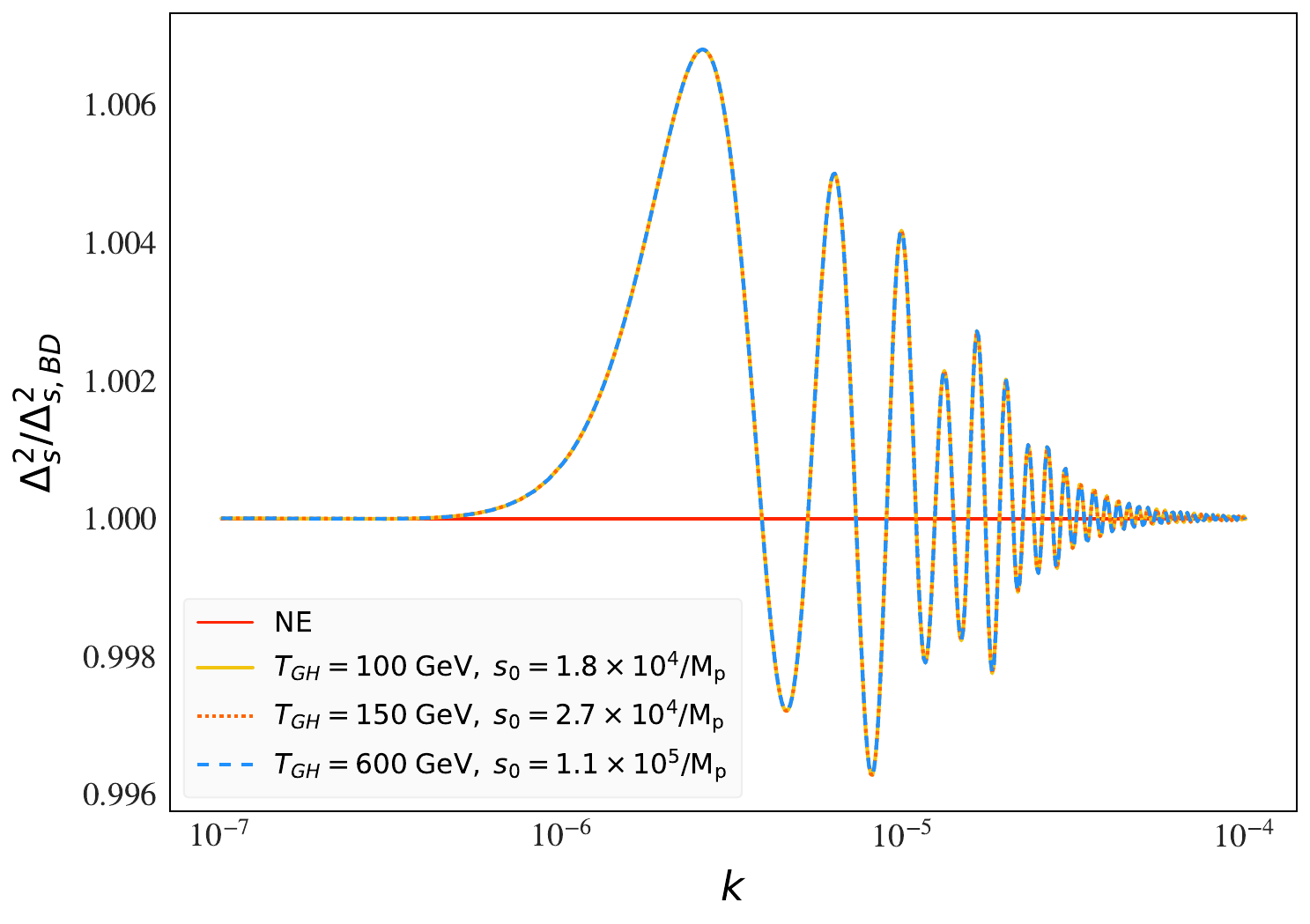}} \quad
     \subfloat[]{\label{fig:deg_b}\includegraphics[width=0.48\textwidth]{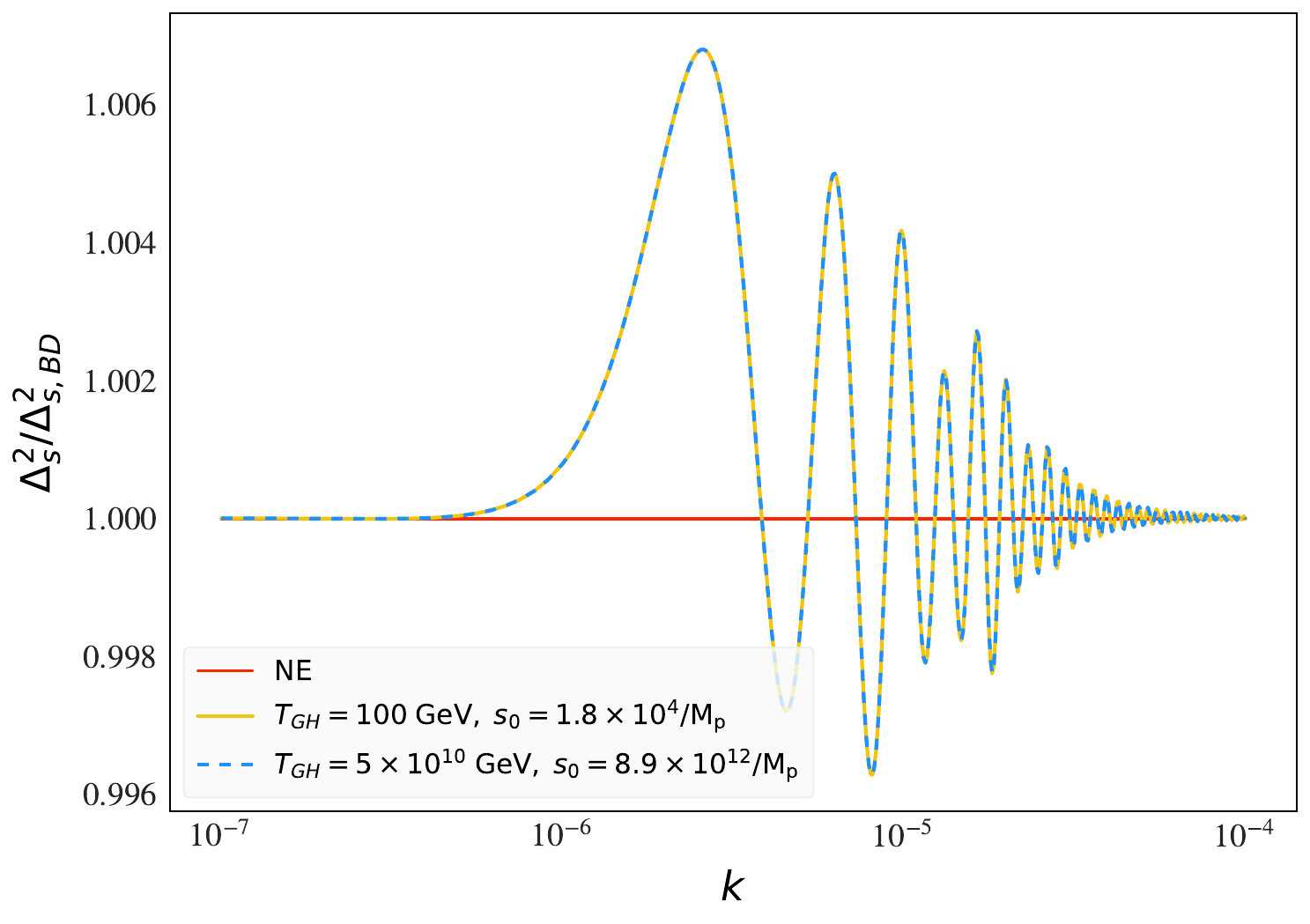}}
    \caption{Fractional corrections to the scalar primordial power spectrum due to entanglement, $\frac{\Delta_s^{2}}{\Delta_{s,BD}^{2}}$, for the Higgs-like spectator. Both the initial position of the zero mode and $T=T_{GH}$ (and therefore $\epsilon$ and $H_{ds}$) are varied. $v_0 = 0$ for all cases, as in figure~\ref{fig:SNR_Tcompare}. The non entangled (NE) case corresponds to $\frac{\Delta_s^{2}}{\Delta_{s,BD}^{2}} = 1$. $k_{\mathrm{ent}} = 10^{-6}$ for both plots (see eq.~\ref{eq:kent}).}
    \label{fig:SNR_degeneracy}
\end{figure}

Figure~\ref{fig:SNR_degeneracy} demonstrates a degeneracy that exists within the technical framework.  The trend identified previously is still true, i.e. that increasing T (and therefore $\epsilon$ and $H_{ds}$) causes $\frac{\Delta_s^{2}}{\Delta_{s,BD}^{2}}$ to decrease.  However, it is possible to vary the initial position of the zero mode to compensate for that effect, as shown in figure~\ref{fig:deg_a} for a few values of T above and below $T_{c}$. More precisely, the near exact degeneracy in figure~\ref{fig:deg_a} was obtained by selecting a value of the initial position of the zero mode such that $\nu_{g,i}$, (eq.~\ref{eq:Hankelnug} at the initial time, $\tau_0$), was the same for both potentials despite the change in T ($\epsilon$, $H_{ds}$). (This then caused the initial conditions in eq.~\eqref{eq:realimA0B0} to be the same for both cases, since I set the initial velocity of the spectator to be zero.) This degeneracy also persists for very high temperatures, and therefore high scale inflation, as shown in figure~\ref{fig:deg_b}.  And if such a degeneracy already exists at the primordial power spectrum level, it is highly unlikely CMB data will be able to differentiate anything.
\vspace{-0.3cm}
\vspace{-0.4cm}

So given the evidence in figures~\ref{fig:SNR_Tcompare} and~\ref{fig:SNR_degeneracy}, one might conclude it is difficult to determine the energy scale of inflation using entanglement for a spectator scalar field with a Higgs-like potential unless one also knows the initial conditions of the spectator zero mode.  If the initial conditions of the zero mode as well as the energy scale of inflation are free parameters in a Monte Carlo analysis, such an analysis would be set up for failure due to the degeneracy demonstrated in figure~\ref{fig:SNR_degeneracy}.  However, if a model had some theoretical motivation for the values of the initial conditions of the spectator field zero mode---and if such values produced large enough corrections to the primordial power spectrum due to entanglement to be observationally viable, as discussed in section~\ref{sec:PowSpec}---then one might be able to say something about the energy scale of inflation using CMB data. 

\subsection{\label{sec:Further}Further analysis}

\begin{figure}[h!]
    \centering
    \includegraphics[width=0.6\textwidth]{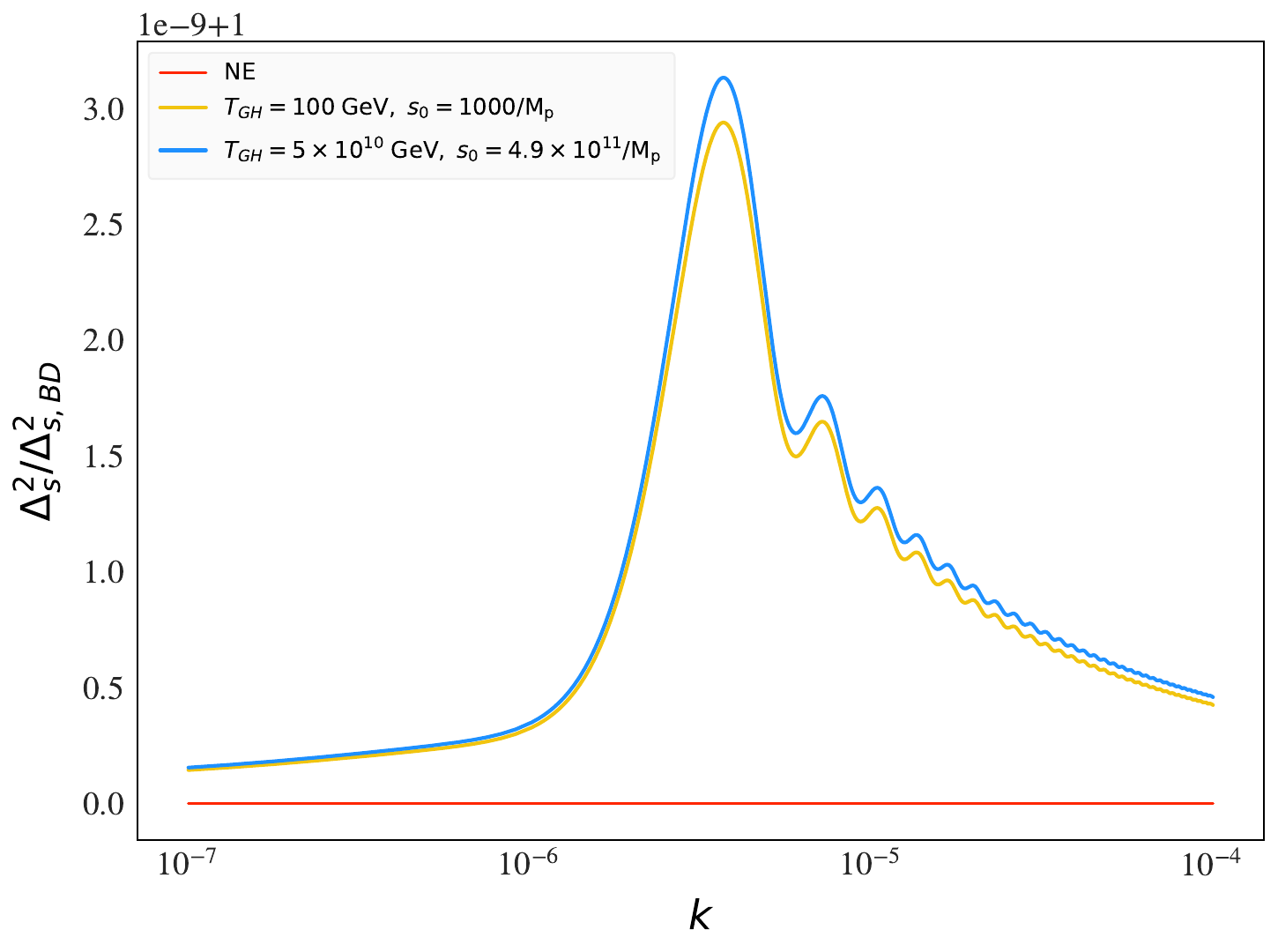}
    \caption{\label{fig:SNR_Tcompare_lows0} Fractional corrections to the scalar primordial power spectrum due to entanglement, $\frac{\Delta_s^{2}}{\Delta_{s,BD}^{2}}$, for the Higgs-like spectator. Both the initial position of the zero mode and $T=T_{GH}$ (and therefore $\epsilon$ and $H_{ds}$) are varied (see discussion in the text), but with a smaller initial position of the zero mode compared to figure~\ref{fig:SNR_degeneracy}. The non entangled (NE) case corresponds to $\frac{\Delta_s^{2}}{\Delta_{s,BD}^{2}} = 1$. $k_{\mathrm{ent}} = 10^{-6}$ (see eq.~\ref{eq:kent}).}
\end{figure}

There are some additional caveats to the discussion above.  First, in figures~\ref{fig:SNR_Tcompare} and~\ref{fig:SNR_degeneracy} I have restricted my attention to primordial power spectra that have large enough corrections due to entanglement to be potentially observable in CMB data. Figure~\ref{fig:SNR_Tcompare_lows0} shows what happens if I attempt to set up the degeneracy in figure~\ref{fig:SNR_degeneracy}---i.e. pick the initial position of the zero mode such that $\nu_{g,i}$ is the same for both potentials despite the change in T ($\epsilon$, $H_{ds}$)---for a case where the initial position of the zero mode is nearer the origin. In contrast to figure~\ref{fig:SNR_degeneracy}, there is a clear difference between the two power spectra. 

\begin{figure}[h!]
    \centering
    \includegraphics[width=0.6\textwidth]{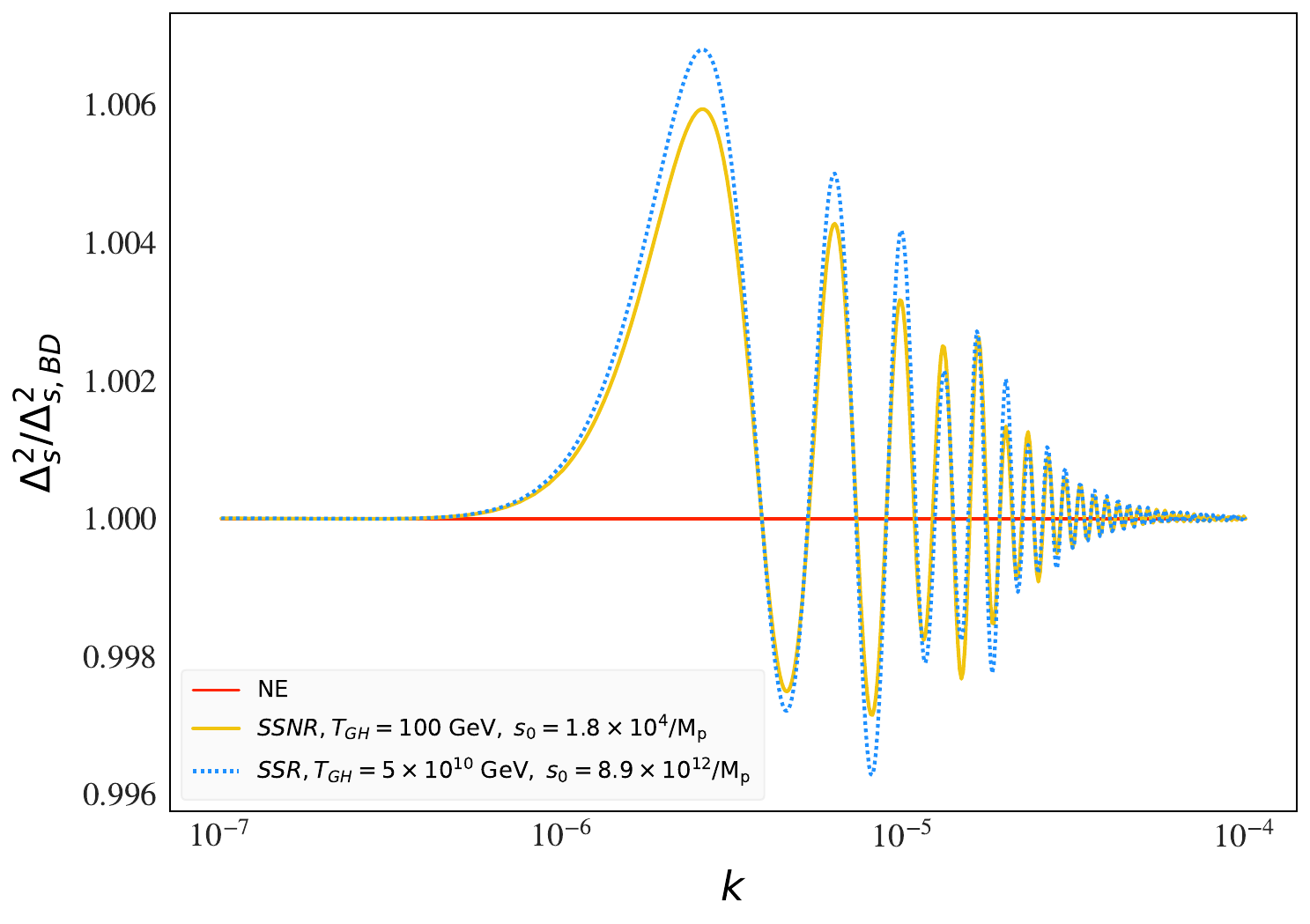}
    \caption{\label{fig:SSNR_Tcompare} Fractional corrections to the scalar primordial power spectrum due to entanglement, $\frac{\Delta_s^{2}}{\Delta_{s,BD}^{2}}$, for the SSNR/SSR Higgs-like spectator. Both the initial position of the zero mode and $T=T_{GH}$ (and therefore $\epsilon$ and $H_{ds}$) are varied. The non entangled (NE) case corresponds to $\frac{\Delta_s^{2}}{\Delta_{s,BD}^{2}} = 1$. $k_{\mathrm{ent}} = 10^{-6}$ (see eq.~\ref{eq:kent}).}
\end{figure}

As another piece of evidence, consider the SSNR Higgs-like spectator from section~\ref{sec:Imposter}. The only different between this spectator's potential and the standard Higgs-like potential I have been considering so far in this section is the fact that its mass is 25 times heavier and the location of its minima is 25 times further from the origin (these parameters were chosen so that the quartic and temperature dependant parts of each potential would stay the same).  Figure~\ref{fig:SSNR_Tcompare} shows the result of performing the equivalent experiment to figure~\ref{fig:deg_b} for this potential. 
Despite setting up the $\nu_{g,i}$ `degeneracy conditions' described previously, and choosing an initial value of the spectator zero mode position that corresponds to an observationally viable amount of entanglement, the two spectra are visually different. Unlike with the more standard Higgs-like spectator, one appears to be able to distinguish differences in the inflationary energy scale at the primordial level, even though the initial position of the zero mode was artificially varied to make this distinction as difficult as possible.

In both figure~\ref{fig:deg_b} and~\ref{fig:SSNR_Tcompare}, the comparisons are set up to have identical initial conditions for the entangled evolution equations (kernel equations) between the low and high inflationary energy scale results. Both results correspond to near percent level corrections due to entanglement to the primordial power spectrum. However, the results in figure~\ref{fig:deg_b} show evidence of a degeneracy, but those in figure~\ref{fig:SSNR_Tcompare} do not. This suggests that some difference in the entangled evolution during inflation due to the difference between the two potentials must be sourcing the difference in these results. 

\begin{figure}[h!]
    \centering
    \subfloat[$\sigma = s M_{p}$, Higgs-like potential]{\label{fig:sHiggs}\includegraphics[width=0.48\textwidth]{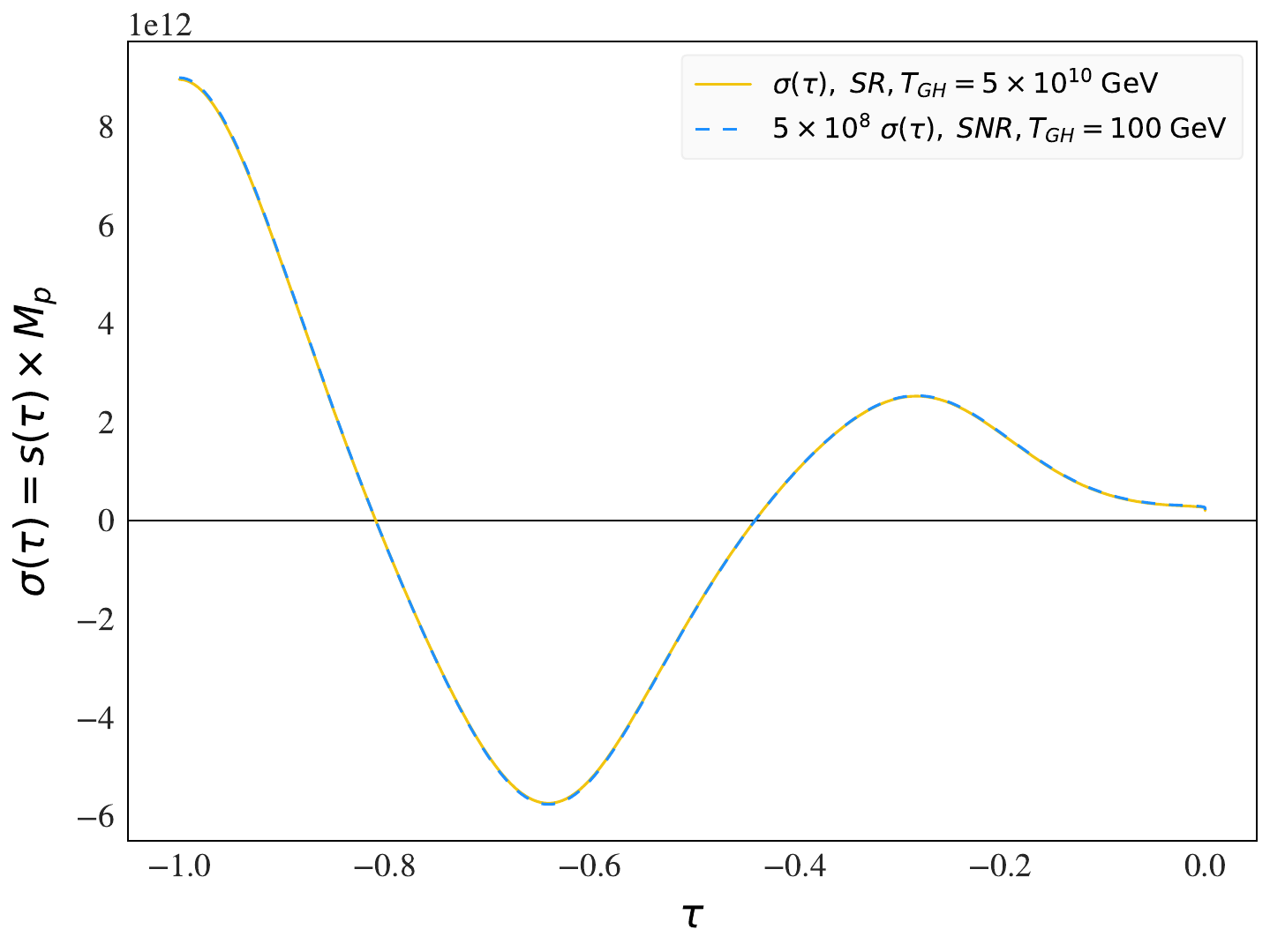}}\quad
\subfloat[$\sigma= s M_{p}$, SSNR/SSR Higgs-like potential]{\label{fig:sSSNR}\includegraphics[width=0.48\textwidth]{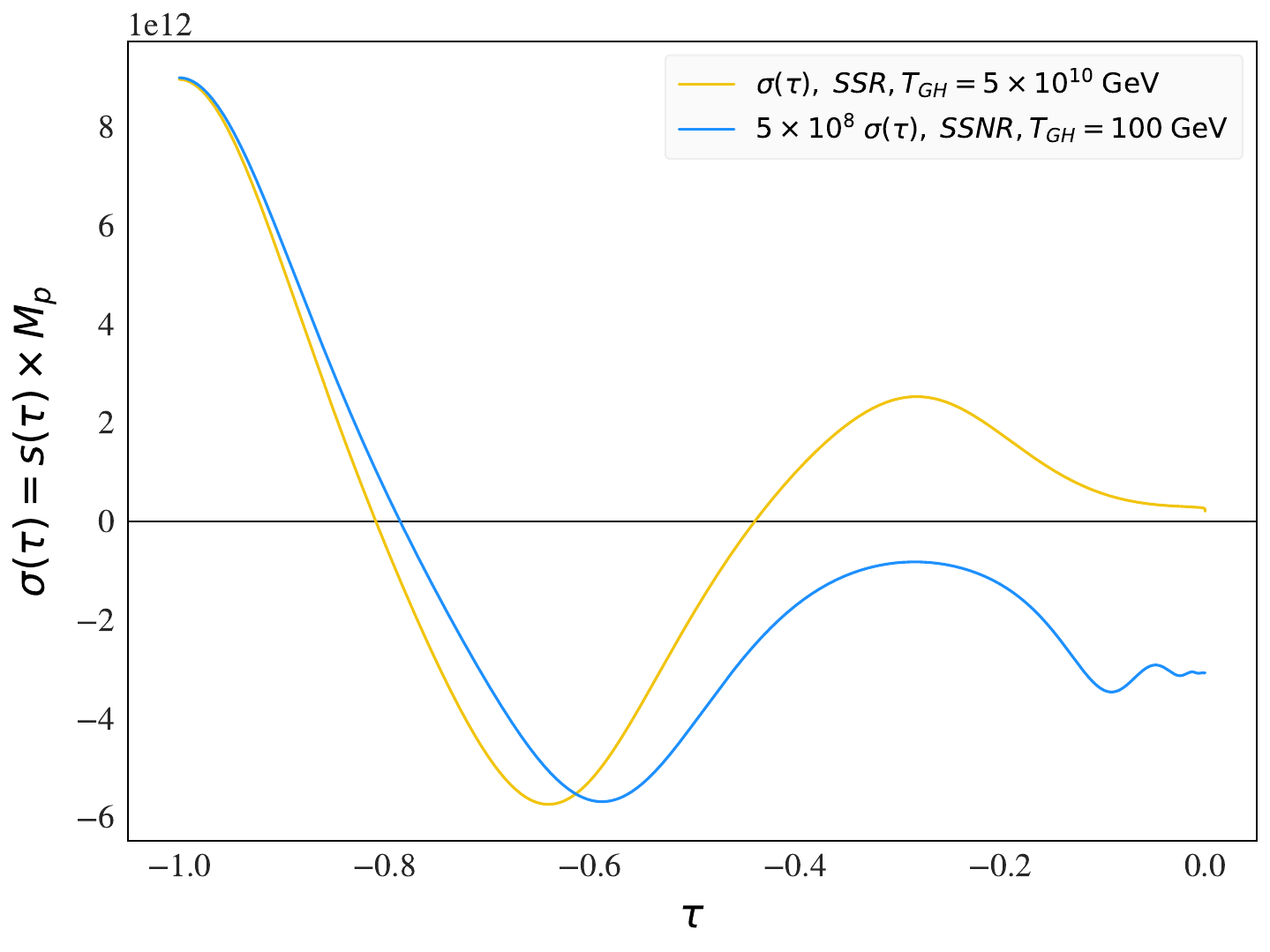}}\\
\caption{\label{fig:s_compare} Location of the zero mode, $\sigma = s M_{p}$, on the Higgs-like spectator potential of eq.~\eqref{eq:HiggspotDimless} and the SSNR/SSR Higgs-like potential of eq.~\eqref{eq:SSNR} during inflation. I qualitatively compare low and high scale inflation by scaling the zero mode location for the low scale results ($T_{GH} = 100 \; GeV$ and $s_0 = 1.8 \times 10^{4}/M_{p}$) by $5 \times 10^{8}$, to more easily compare with the high scale results ($T_{GH}= 5 \times 10^{10} \; GeV$ and $s_0 = 8.9 \times 10^{12} /M_p$). (Dimensionless conformal time $\tau$ is defined in eq.~\eqref{eq:dimless}.)}
\end{figure}

To support this claim, figure~\ref{fig:s_compare} investigates the behavior of the zero mode during inflation for the two cases shown in figure~\ref{fig:deg_b} and figure~\ref{fig:SSNR_Tcompare}. After applying a vertical scaling for the low-scale inflation results, one can see that the zero mode behaves remarkably similarly for both energy scales for the Higgs-like potential. However this is not the case for the SSNR potential. The high scale, or SSR, results are similar to those in figure~\ref{fig:sHiggs},\footnote{This suggests it may not be quite so easy to distinguish the Higgs-like spectator from the SSR Higgs-like spectator for high scale inflation, in contrast to the low scale results shown in section~\ref{sec:Clres}.} however the low-scale results are quite different (i.e. showing evidence of vacuum selection before inflation even finishes).
And since the evolution of the zero mode feeds directly into all the equations governing entangled evolution (see section~\ref{sec:entReview}), any differences or similarities will have an effect on the primordial power spectrum.

\subsubsection{\label{sec:phi4}$\phi^{4}$ diagnostics}
The analysis in the previous section suggests the following interpretation of the $\sigma - H_{ds}$ degeneracies. Both initial conditions and evolution during inflation play a role in determining entangled corrections to the primordial power spectrum.  A degeneracy in the initial conditions for the entangled equations (the kernel equations in section~\ref{sec:entReview}) exists between $\epsilon$ and $\sigma$ ($s_0$ in dimensionless units), such that the initial position of the zero mode can be adjusted to exactly compensate for a change in $\epsilon$ ($H_{ds}, T_{GH}$). As shown in figure~\ref{fig:SNR_degeneracy}, this means that a prospective primordial power spectrum result for a Higgs-like spectator could correspond to a variety of energy-scale and initial position pairs in parameter space, if the initial position of the zero mode is a free parameter. However, even if such a degeneracy in initial conditions is assumed (a `worst case scenario'), the degeneracy can still be broken if the evolution of the zero mode position---and therefore the evolution of the entangled wavefunction parameters---is significantly different for the different energy scales (as investigated in figure~\ref{fig:s_compare} and the surrounding text). 

\begin{figure}[h!]
    \centering
    \includegraphics[width=0.6\textwidth]{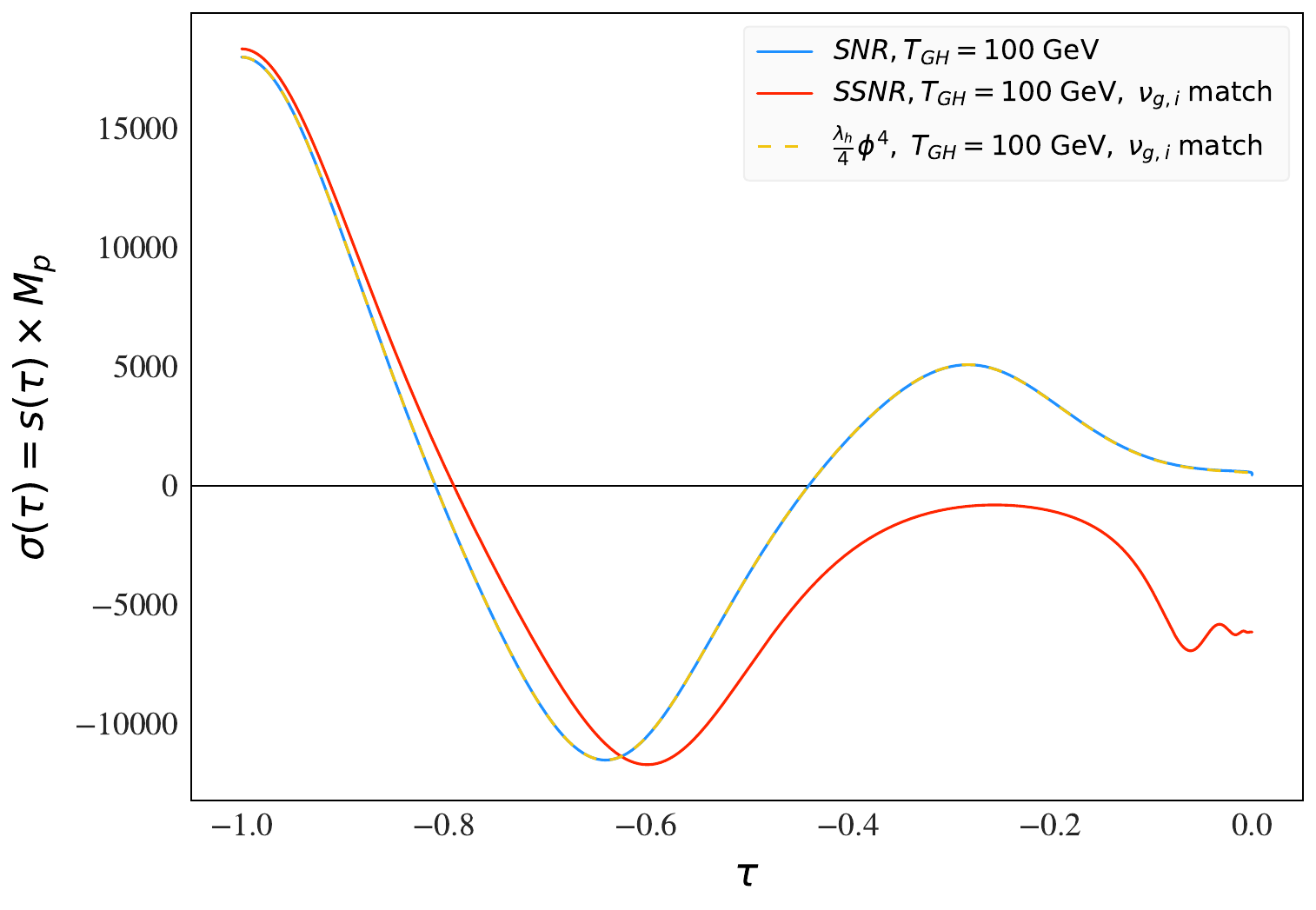}
    \caption{\label{fig:s_compare_phi4} Location of the zero mode, $\sigma = s M_{p}$, on the Higgs-like spectator potential of eq.~\eqref{eq:HiggspotDimless}, the SSNR Higgs-like potential of eq.~\eqref{eq:SSNR}, and $V = \frac{\lambda_h}{4}\phi^{4}$ during inflation. For all three potentials $T_{GH} = 100 \; GeV$, but the initial position of the zero mode has been varied slightly so that $\nu_{g,i}$  is the same for all three cases (to investigate the degeneracy discussed in the text).}
\end{figure}

Additional insight into how evolution during inflation can break these degeneracies can be obtained by comparing how similar a given result is to one obtained from a spectator scalar field with the equivalent $\phi^{4}$ potential (i.e. $V = \frac{\lambda_h}{4}\phi^{4}$).
\begin{figure}[hptb!]
    \centering
    \subfloat[SNR Higgs-like spectator vs $\frac{\lambda_h}{4}\phi^{4}$, compare with figure~\ref{fig:SNR_Tcompare_lows0}]{\label{fig:sphi_1}\includegraphics[width=0.52\textwidth]{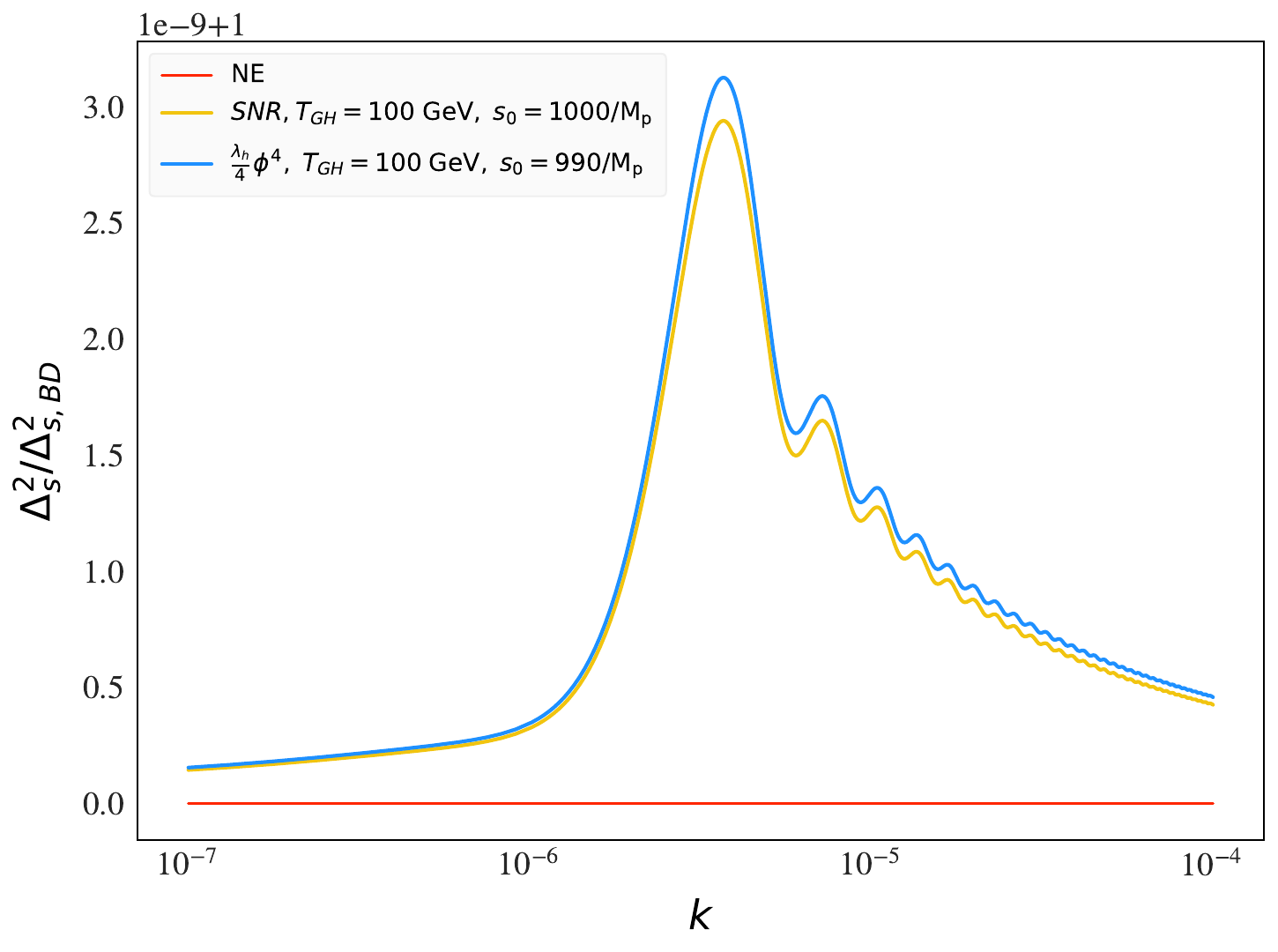}}\hfill \\
   \subfloat[SNR Higgs-like spectator vs $\frac{\lambda_h}{4}\phi^{4}$, compare with figure~\ref{fig:deg_b}]{\label{fig:sphi_2}\includegraphics[width=0.52\textwidth]{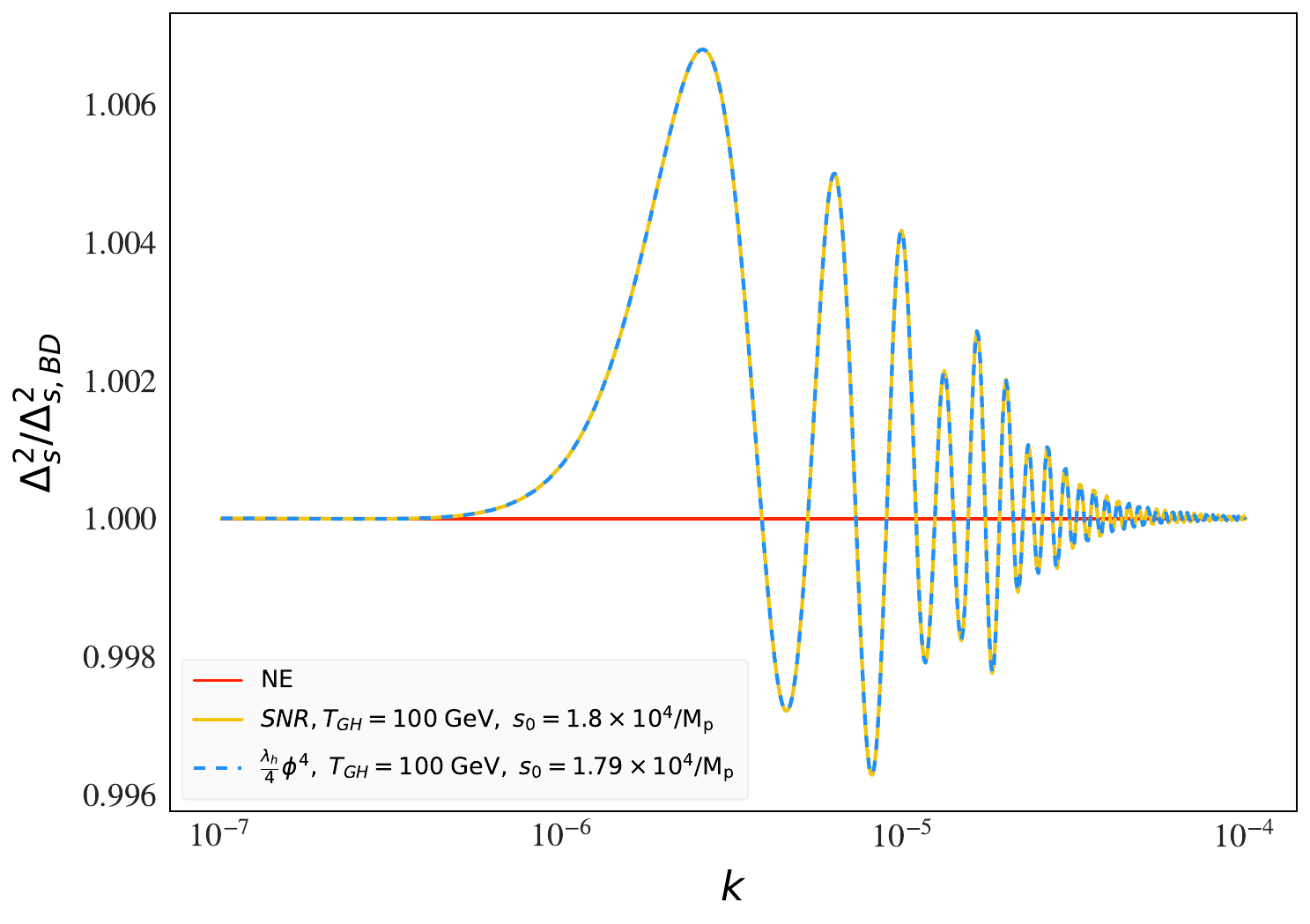}}\hfill \\
   \subfloat[SSNR Higgs-like spectator vs $\frac{\lambda_h}{4}\phi^{4}$, compare with figure~\ref{fig:SSNR_Tcompare}]{\label{fig:sphi_3}\includegraphics[width=0.52\textwidth]{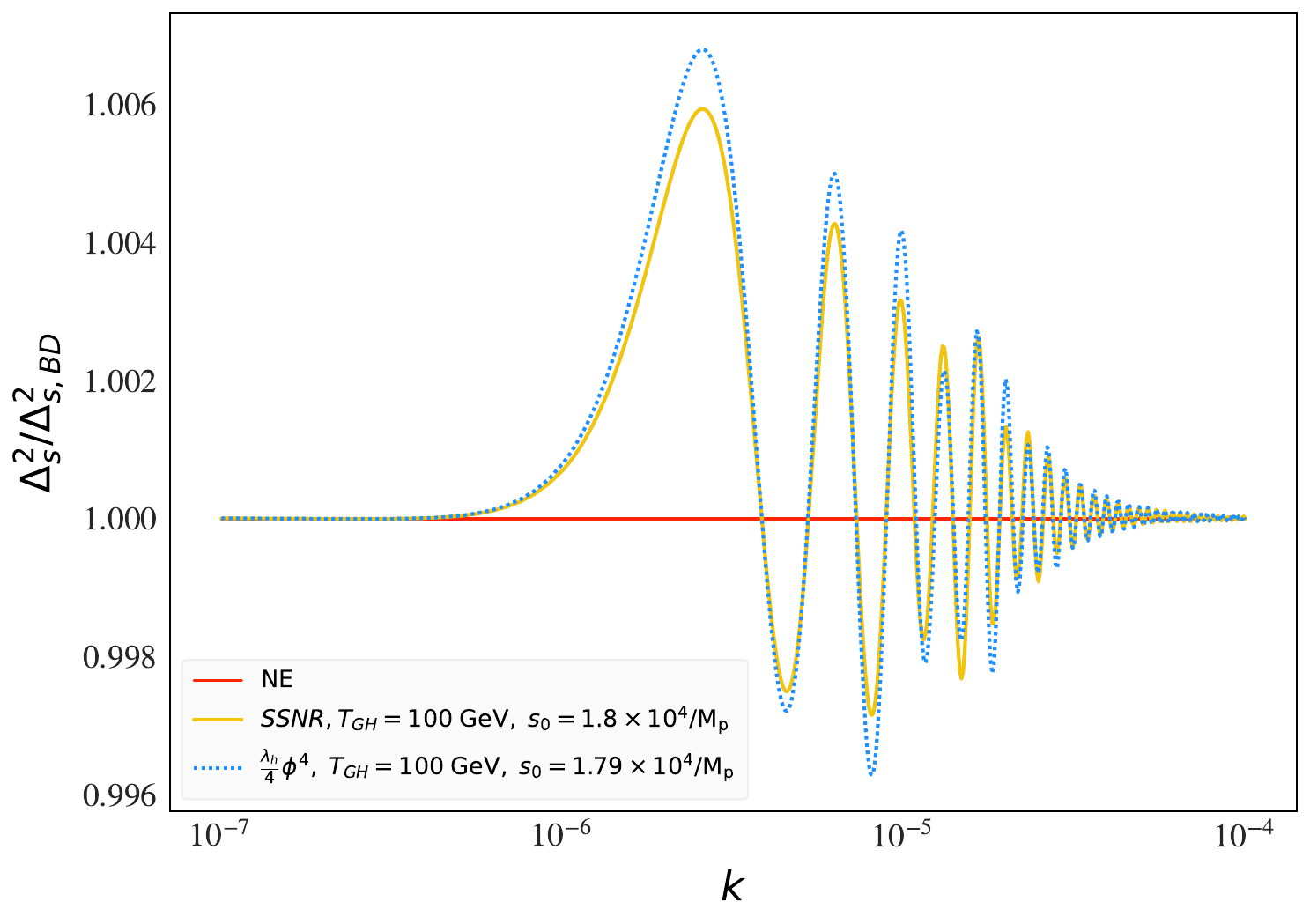}}
   \vspace{-0.2cm}
    \caption{\label{fig:SNR_vphi4} Fractional corrections to the scalar primordial power spectrum due to entanglement, $\frac{\Delta_s^{2}}{\Delta_{s,BD}^{2}}$, for the Higgs-like and SSNR Higgs-like spectators and for a spectator scalar with the equivalent $\phi^{4}$ potential, $V = \frac{\lambda_h}{4}\phi^{4}$. The initial position of the zero mode is varied (such that $s_0$ causes $\nu_{g,i}$ to be the same for each set of comparisons, as discussed in the text), but $T=T_{GH}$ (and therefore $\epsilon$ and $H_{ds}$) is held fixed. The non entangled (NE) case corresponds to $\frac{\Delta_s^{2}}{\Delta_{s,BD}^{2}} = 1$. $k_{\mathrm{ent}} = 10^{-6}$ for all plots (see eq.~\ref{eq:kent}).}
\end{figure} 
Figure~\ref{fig:SNR_vphi4} demonstrates that the degeneracies (or lack thereof) show in figures~\ref{fig:deg_b},~\ref{fig:SNR_Tcompare_lows0}, and~\ref{fig:SSNR_Tcompare} between high and low inflationary energy scale results can be mimicked by a comparison with the equivalent $\phi^{4}$ spectator at low scale. And figure~\ref{fig:s_compare_phi4} does the same comparison for the evolution of the zero mode for the results in figure~\ref{fig:sphi_2} and~\ref{fig:sphi_3}.



Examining the results of figures~\ref{fig:SNR_vphi4} and~\ref{fig:s_compare_phi4}, it appears that there is a threshold above which the quadratic terms in the Higgs-like potential---which control the symmetry breaking features through the presence or absence of multiple minima---become subdominant, such that the zero mode evolution is phenomenologically equivalent to $\phi^{4}$. (This similarity in zero mode evolution then feeds into the primordial spectra result, as discussed above.) For the SSNR Higgs-like potential the degeneracy threshold is higher since the field is more massive.  This allows one to choose larger initial positions for a given energy scale and still see noticeable differences with $\phi^{4}$ (see figure~\ref{fig:sphi_3}), or between energy scales comparing SSR vs SSNR behavior (see figure~\ref{fig:SSNR_Tcompare}). For the standard Higgs-like potential this threshold\footnote{Further work would be needed to identify exactly where this degeneracy line is in the space of possible initial conditions for a given Higgs-like potential. The result would depend in part on resolution at the $C_{\ell}$ level between different cases. It would also depend on whether the initial position was being treated as a totally free or partially free parameter in the analysis.} is lower, which is why the results in figures~\ref{fig:SNR_Tcompare_lows0} and~\ref{fig:sphi_1} are visually distinct, but the results in figures~\ref{fig:SNR_degeneracy} and~\ref{fig:sphi_2} are not.

Interpreting further still, the results in figures~\ref{fig:SNR_degeneracy} -~\ref{fig:s_compare_phi4} appear to indicate that high-scale inflation results may be degenerate with $\phi^{4}$ for some spectator initial conditions with these types of potentials. Further study would be needed to confirm this, but this is not a total red flag for my motivating questions. For example, take the SSNR Higgs-like spectator results. If one had a theoretical reason to put a prior on the initial conditions of that spectator at a level corresponding to potentially observable entanglement, then the results in figures~\ref{fig:SSNR_Tcompare} and~\ref{fig:sphi_3}---supported by the work in section~\ref{sec:Clres}---demonstrate one could potentially answer whether symmetry breaking has occurred during inflation for that spectator potential. If the data preferred a high-scale result (which may be degenerate with $\phi^{4}$), one could still potentially say with some certainty that symmetry breaking did not occur during inflation, because the low-scale results are visually distinct from both the high-scale and/or $\phi^{4}$ options. In contrast, it appears the SM Higgs may be too light to be able to answer questions about symmetry breaking from the perspective of CMB data with entanglement, since it generates visually indistinct primordial spectra for different energy scales for observationally viable initial conditions.

\section{\label{sec:Conclude}Discussion and conclusions}
In this paper I have made a case study analysis to investigate whether the technical framework for dynamically generated entangled states developed in~\cite{Baunach:2021yvu, Adil:2022rgt} can be used to answer questions about early universe phase transitions and/or help determine the inflationary energy scale using CMB data. To investigate this, I used a simple model of a spectator scalar field with a Higgs-like potential, and considered entanglement between its perturbations and those of the inflaton. In section~\ref{sec:PowSpec} I demonstrated the rich phenomenology of corrections to the primordial power spectrum this type of spectator can induce, and noted that only some of these results would produce large enough corrections to be observationally viable. Restricting my focus to observationally viable results, in section~\ref{sec:CMB} I then demonstrated how distinct entangled results from a Higgs-like spectator might be compared to other similar scalar spectators, at the level of $C_{\ell}^{TT}$ residuals. After demonstrating that such coarse grained distinctions would be possible, I then turned to the related questions of phase transitions and the inflationary energy scale in section~\ref{sec:Sensitivity}. My analysis in section~\ref{sec:Sensitivity} demonstrates that there are some initial values of the spectator zero mode that inhibit distinguishing spectra at different inflationary energy scales---due to a combined degeneracy in initial conditions and dynamical evolution of entangled parameters---and some values that do not. In particular, it appears the SM Higgs (at least in the form of the simple Higgs-like potential I considered here) may be too light to break degeneracies between different energy scales for observationally viable results, whereas a heavier Higgs-like spectator---perhaps a `dark Higgs'---may be able to give some insight into such questions with current data.

The case study format of this work means that not every question was answered in full generality.  Still, I hope these results will be a useful signpost for future researchers interested in questions of our Universe's early beginnings. For example, given a narrative of symmetry breaking in the early universe with particular model parameters, this work will give one a rough idea of whether corrections due to entanglement and their potential impact on CMB observables should be considered or not. Also, as discussed in~\cite{Adil:2022rgt}, it remains an open question how such corrections due to entanglement might impact the bi-spectrum. Such a calculation might reveal further opportunities to differentiate narratives of the early universe using entanglement.

\acknowledgments
I thank A.~Adil, A.~Albrecht, N.~Bolis, R.~Holman,  L.~Knox, R.~Ribeiro, and B.~Richard for useful discussions. This work was supported
in part by the U.S. Department of Energy, Office of Science, Office of High Energy Physics
QuantISED program under Contract No. KA2401032. I was also supported by a UC Davis Graduate Research Award.

\appendix

\section{\label{app:Review} Supplementary equations and derivations}

Here I collect additional results that the main text in section~\ref{sec:entReview} makes reference to, including some details of the technical formalism previously presented in~\cite{Adil:2022rgt}.

\subsection{\label{app:Kernel}Dimensionful kernel equations}

I re-print the kernel equations (as derived in~\cite{Adil:2022rgt}) obtained from solving the functional Schr\"odinger equation:
\begin{equation} \label{eq:Schrodinger1}
i \frac{\partial \Psi_{k}}{\partial \eta} = \mathcal{H}_{k} \Psi_{k}
\end{equation}
given a wavefunctional of the form:
\begin{align}
\Psi_{\vec{k}} \left[ v_{\vec{k}} ,  \theta_{\vec{k}} ; \eta \right] & = \mathcal{N}_k(\eta)\  \mathrm{\exp}
\bigg[
-\dfrac{1}{2} \bigg( A_{k}(\eta)v_{\vec{k}}v_{-\vec{k}} + B_{k}(\eta)\theta_{\vec{k}}\theta_{-\vec{k}} 
 + C_{k}(\eta) \big(v_{\vec{k}}\theta_{-\vec{k}} + \theta_{\vec{k}}v_{-\vec{k}} \big) \bigg]\ ,
\label{eq:wavefunctional1}
\end{align}
and Hamiltonian defined in eq.~\eqref{eq:kspacehamiltonian2}. One obtains:
\begin{flalign}
i\partial_{\eta} A_k  &= A_k^2 - \left(k^2 - \frac{z^{\prime\prime}}{z}\right) \nonumber \\
 &+ \bigggl\{\sinh \alpha\ A_k + \cosh\alpha\ C_k   + \frac{i}{2} \left[\left(3 - \epsilon + \frac{\eta_{\rm sl}}{2} \right){\mathcal H}\sinh\alpha + \frac{a^2\partial_{\sigma}V}{{\mathcal H}^2\sqrt{2 M_{\rm Pl}^2 \epsilon}}{\mathcal H}\cosh\alpha\right]\bigggr\}^2  \label{eq:kernelEqsA} 
 \end{flalign}
 \begin{flalign}
i\partial_{\eta} B_k  &= B_k^2 - \left(k^2 - \frac{a^{\prime\prime}}{a} + a^2\partial^2_{\sigma}V \right) \nonumber \\ 
&+ \bigggl\{\sinh\alpha\ B_k + \cosh\alpha\ C_k 
- \frac{i}{2} \left[\left(3 - \epsilon + \frac{\eta_{\rm sl}}{2} \right){\mathcal H}\sinh\alpha + \frac{a^2\partial_{\sigma}V}{{\mathcal H}^2\sqrt{2 M_{\rm Pl}^2 \epsilon}}{\mathcal H}\cosh\alpha\right]\bigggr\}^2 \nonumber \\
 &+ 2\epsilon {\mathcal H}^2\left[\left(\epsilon - 3\right)\tanh^2\alpha - 2\tanh\alpha\left(\frac{a^2\partial_{\sigma}V}{{\mathcal H}^2\sqrt{2 M_{\rm Pl}^2 \epsilon}}\right)\right] \label{eq:kernelEqsB} 
 \end{flalign}
 \begin{flalign}
i\partial_{\eta} C_k &= \cosh^2\alpha\ C_k\Big(A_k + B_k\Big) \nonumber\\
&+ \frac{\sinh 2\alpha}{2}\bigggl\{C_k^2 + A_k B_k \nonumber \\
&\hspace{6em}+ i \frac{\mathcal{H}}{2} \left[\left(3 - \epsilon + \frac{\eta_{\rm sl}}{2} \right) + \coth\alpha\frac{a^2\partial_{\sigma}V}{{\mathcal H}^2\sqrt{2 M_{\rm Pl}^2 \epsilon}}\right]\Big(B_k - A_k \Big) \nonumber \\
&\hspace{6em}+  \frac{\mathcal{H}^2}{4} \left[\left(3 - \epsilon + \frac{\eta_{\rm sl}}{2} \right)\tanh\alpha + \frac{a^2\partial_{\sigma}V}{{\mathcal H}^2\sqrt{2 M_{\rm Pl}^2 \epsilon}}\right]^2 \biggg\} \nonumber \\
&+ \tanh\alpha \biggg[k^2 + \frac{1}{2}a^2\partial^2_{\sigma}V + {\mathcal{H}}^2\left(1 + \frac{5}{4}\eta_{\rm sl}\right)\biggg] + 
{\mathcal{H}}^2 \left(1 + \epsilon + \frac{\eta_{\rm sl}}{2} \right)\frac{a^2\partial_{\sigma}V}{{\mathcal H}^2\sqrt{2 M_{\rm Pl}^2 \epsilon}} .  \label{eq:kernelEqsC}
\end{flalign}
where $\alpha$ is defined through eq.~\eqref{eq:alpha_def}.

\subsection{\label{app:ICs}Initial conditions for the dimensionless kernel equations}
In this section I summarize the derivation of the initial conditions for the dimensionless kernel equations---eqs.~\eqref{eq:A0real} -~\eqref{eq:A2im}---defined in section~\ref{sec:Perturb}. This material previously appeared in~\cite{Adil:2022rgt}, and is included here as an additional resource for the reader interested in the details of the technical formalism.

As previously discussed in~\cite{Adil:2022rgt}, initial conditions for the kernel equations are constructed so that the corresponding modes are the standard Bunch--Davies ones at the initial time $\eta_{0}$ (corresponding to $\tau = -1$) at which entanglement begins to be dynamically generated. One does this via a Riccati transform. Given a kernel equation of the form
\begin{equation}
\label{eq:Ricatti1}
i K^{\prime}(\tau)=\alpha_2(\tau) K^2(\tau)+\alpha_1(\tau) K(\tau)+\alpha_0(\tau).
\end{equation}
one can transform it into a mode equation of the form
\begin{equation}
\label{eq:RicattiToModeFinal1}
f''(\tau)+\Omega^2 f(\tau)=0,\ \ \textrm{with} \ \  \Omega^2=\frac{1}{4}\alpha_1^2-\alpha_0 \alpha_2-\frac{i}{2} \alpha_1^{\prime}+\frac{i \alpha_1 \alpha_2^{\prime}}{2 \alpha_2}-\frac{3}{4} \left(\frac{\alpha_2^{\prime}}{\alpha_2}\right)^2+\frac{\alpha_2^{\prime \prime}}{2 \alpha_2}.
\end{equation}
by choosing
\begin{equation}
\label{eq:Ricatti_trans1}
i K(\tau) = \frac{1}{\alpha_2(\tau)}\left(\frac{f'(\tau)}{f(\tau)}-\Delta(\tau)\right),
\end{equation}
with $2\Delta=i\alpha_1-\alpha_2^{\prime}\slash \alpha_2$. One can then use eq.~\eqref{eq:Ricatti_trans1} to set the initial conditions for the real and imaginary parts of the kernel equations, respectively given by \cite{Adil:2022rgt,Baunach:2021yvu}:
\begin{subequations}\label{eq:realimKernelsIC1}
\begin{align}
\label{eq:realinKernelsIC:real1}
K_R(\tau_0) =& \frac{1}{2 \alpha_2(\tau_0)}\left(\frac{1}{\left| f(\tau_0)\right|^2}-\alpha_{1 R}(\tau_0)\right)\\
\label{eq:realinKernelsIC:imag1}
K_I(\tau_0) =& -\frac{1}{2 \alpha_2(\tau_0)}\left(\left . \partial_{\tau}\ln\left | f\right |^2\right |_{\tau=\tau_0}+\alpha_{1 I}(\tau_0)+\left . \frac{\alpha_2^{\prime}}{\alpha_2}\right |_{\tau=\tau_0}\right) \ .
\end{align}
\end{subequations}
One only needs to use eqs.~\eqref{eq:realimKernelsIC1} to set the initial conditions for $A_{q}^{(0)}$, $B_{q}^{(0)}$ and $A_{q}^{(2)}$---the initial condition for $C_{q}^{(1)}$ is $C_{q}^{(1)} (\tau_0 = -1) = 0$ in this work. For $A_{q}^{(0)}$ and $B_{q}^{(0)}$ the results are straightforward to compute, and one obtains:
\begin{subequations}\label{eq:realimA0B01}
\begin{align}
\label{eq:A0real1}
A_{q R}^{(0)}(\tau_0 = -1) =& \frac{2}{ \pi \left |H^{(2)}_{\nu_{f}}(\frac{q}{1-\epsilon})\right|^2} \\
\label{eq:A0im1}
A_{q I}^{(0)}(\tau_0 = -1) =& \frac{1}{2} \left[ 1 - 2 \nu_{f} + x \left . \left[ \frac{H^{(1)}_{\nu_{f} -1}(x)}{H^{(1)}_{\nu_{f}}(x)} + \frac{H^{(2)}_{\nu_{f}-1} (x)}{H^{(2)}_{\nu_{f}}(x)} \right] \right|_ {x=\frac{q}{(1-\epsilon)}} \right] \\
\label{eq:B0real1}
B_{q R}^{(0)}(\tau_0 = -1) =& \frac{2}{ \pi \left |H^{(2)}_{\nu_{g}}(\frac{q}{1-\epsilon})\right|^2} \\
\label{eq:B0im1}
B_{q I}^{(0)}(\tau_0 = -1) =& \frac{1}{2} \left[ 1 - 2 \nu_{g} + x \left . \left[ \frac{H^{(1)}_{\nu_{g} -1}(x)}{H^{(1)}_{\nu_{g}}(x)} + \frac{H^{(2)}_{\nu_{g}-1} (x)}{H^{(2)}_{\nu_{g}}(x)} \right] \right |_ {x=\frac{q}{(1-\epsilon)}} \right]  
\end{align}
\end{subequations}
as discussed in section~\ref{sec:Perturb}.

For $A_{q}^{(2)}$ the situation is slightly more obtuse. The easiest thing to do is to start with the dimensionless, but unexpanded, $A_{q}$ equation, i.e.,
\begin{align}
\label{eq:Aqeqn1}
i \partial_{\tau} A_{q} & = A_{q}^{2} - \left[ \left(\frac{q}{1-\epsilon}\right)^2 - \frac{\left(\nu_{f}^2 -\frac{1}{4}\right)}{\tau^2} \right]  \notag \\
& {} {} \quad \quad + \frac{1}{1 - \lambda_{1}^{2}} \left[ \lambda_1 A_{q} + C_{q} -\frac{i}{2(1-\epsilon)\tau} \left[ \left( 3 - \epsilon +\frac{\eta_{sl}}{2} \right)\lambda_1 + \lambda_2 \right] \right]^{2},
\end{align}
identify the $\alpha$ coefficients (of the type in eq.~\eqref{eq:Ricatti1}), use those along with the $\lambda$ expansions for $A_{q}$ and $C_{q}$ in eqs.~\eqref{eq:realinKernelsIC:real1} and \eqref{eq:realinKernelsIC:imag1}, and then collect the second order in $\lambda$ terms that remain.  Finally, after noting that $C_{q}(\tau_0 = -1) = 0$, one obtains:
\begin{subequations}\label{eq:realimA21}
\begin{align}
\label{eq:A2real1}
A_{q R}^{(2)}(\tau_0 = -1) & = -\tilde{\lambda}_{1,0}^{2}A_{q R}^{(0)}(\tau_0 = -1) \\
\label{eq:A2im1}
A_{q I}^{(2)}(\tau_0 = -1) & = -\tilde{\lambda}_{1,0}^{2}A_{q I}^{(0)}(\tau_0 = -1) -\frac{1}{2(1-\epsilon)} \left[\left(3 - \epsilon + \frac{\eta_{sl}}{2} \right) \tilde{\lambda}_{1,0}^{2} + \tilde{\lambda}_{1,0}\tilde{\lambda}_{2,0} \right] \notag \\
& {} {} \quad \quad +\left[ \frac{\eta_{sl}\tilde{\lambda}_{1,0}^{2}}{2(1-\epsilon)} + \tilde{\lambda}_{1,0}^{2} + \frac{2\tilde{\lambda}_{1,0}^{2}}{(1-\epsilon)} +\frac{\tilde{\lambda}_{1,0}\tilde{\lambda}_{2,0}}{(1-\epsilon)} \right]
\end{align}
\end{subequations}
where $\tilde{\lambda}_{1,0}$ denotes that $\tilde{\lambda}_{1}$ should be evaluated at $\tau_0 = -1$, and a term $\mathcal{O}(\eta_{sl}\epsilon)$ has been dropped from $A_{q I}^{(2)}$. From glancing at eqs.~\eqref{eq:realimA21}, one can see that $A_{q}^{(2)}$ will be zero initially, unless there is a non-zero initial velocity in the spectator zero mode (which causes $\tilde{\lambda}_{1,0}$ to be non-zero). 

To explain why there are no terms first order in $\lambda$ in eq.~\eqref{eq:kernels_expansion}, as discussed in section~\ref{sec:Perturb}, consider eq.~\eqref{eq:Aqeqn1}.  If we add a term $\mathcal{O}(\lambda)$ to $A_{q}$ and expand (keeping $C_{q} = \lambda C_{q}^{(1)}$), the first order result will be
\begin{equation}
\label{eq:A1q1}
i \partial_{\tau}A_{q}^{(1)} = 2 A_{q}^{(0)}A_{q}^{(1)},
\end{equation}
since the terms in the second line of eq.~\eqref{eq:Aqeqn1} will always be $\mathcal{O}(\lambda^{2})$ at lowest order. The solution to this equation is $A_{q}^{(1)}(\tau) = \frac{D}{f_{q}^{2}(\tau)}$ where $D$ is an integration constant and $f_{q}(\tau)$ is the dimensionless BD mode function given by $f_{v}(\tau)= \frac{\sqrt{-\pi \tau}}{2} H_{\nu_{f}}^{(2)}\left(\frac{-q \tau}{(1-\epsilon)}\right) $, since $A_{q}^{(0)} = -i \frac{f'}{f}$ by definition. 

Then, however, if one consults the initial conditions for $A_{q}^{(1)}$, using the same method to obtain them as was described for $A_{q}^{(2)}$ above, one would find the integration constant $D$ must vanish. This is because the only term that can be $\mathcal{O}(\lambda)$ in the initial conditions is proportional to $C_{q}^{(1)}$, and for this work I have the condition that $C_{q}(\tau_0 = -1) = 0$.  The exact same procedure holds for the $B_{q}$ equation. Thus, unless one considers some initial entanglement---which is not part of this analysis---there are no first-order terms in $A_{q}$ and $B_{q}$ for the $\lambda$ expansion because there are no non-zero initial conditions to source them.

\newpage

\bibliographystyle{unsrt}
\bibliography{HiggsEnt.bib}

\begin{thebibliography}{10}

\bibitem{Adil:2022rgt}
Arsalan Adil, Andreas Albrecht, Rose Baunach, R.~Holman, Raquel~H. Ribeiro, and Benoit~J. Richard.
\newblock {Entanglement masquerading in the CMB}.
\newblock {\em JCAP}, 06:024, 2023.

\bibitem{Baunach:2021yvu}
Rose Baunach, Nadia Bolis, R.~Holman, Stacie Moltner, and Benoit~J. Richard.
\newblock {Does Planck actually \textquotedblleft{}see\textquotedblright{} the Bunch-Davies state?}
\newblock {\em JCAP}, 07:050, 2021.

\bibitem{Guth:1980zm}
Alan~H. Guth.
\newblock {The Inflationary Universe: A Possible Solution to the Horizon and Flatness Problems}.
\newblock {\em Adv. Ser. Astrophys. Cosmol.}, 3:139--148, 1987.

\bibitem{Kazanas:1980tx}
D.~Kazanas.
\newblock {Dynamics of the Universe and Spontaneous Symmetry Breaking}.
\newblock {\em Astrophys. J. Lett.}, 241:L59--L63, 1980.

\bibitem{Linde:1981mu}
Andrei~D. Linde.
\newblock {A New Inflationary Universe Scenario: A Possible Solution of the Horizon, Flatness, Homogeneity, Isotropy and Primordial Monopole Problems}.
\newblock {\em Phys. Lett. B}, 108:389--393, 1982.

\bibitem{Albrecht:1982wi}
Andreas Albrecht and Paul~J. Steinhardt.
\newblock {Cosmology for Grand Unified Theories with Radiatively Induced Symmetry Breaking}.
\newblock {\em Phys. Rev. Lett.}, 48:1220--1223, 1982.

\bibitem{Guth:1982ec}
Alan~H. Guth and S.~Y. Pi.
\newblock {Fluctuations in the New Inflationary Universe}.
\newblock {\em Phys. Rev. Lett.}, 49:1110--1113, 1982.

\bibitem{Bardeen:1983qw}
James~M. Bardeen, Paul~J. Steinhardt, and Michael~S. Turner.
\newblock {Spontaneous Creation of Almost Scale - Free Density Perturbations in an Inflationary Universe}.
\newblock {\em Phys. Rev. D}, 28:679, 1983.

\bibitem{Bunch:1978yq}
T.~S. Bunch and P.~C.~W. Davies.
\newblock {Quantum Field Theory in de Sitter Space: Renormalization by Point Splitting}.
\newblock {\em Proc. Roy. Soc. Lond. A}, 360:117--134, 1978.

\bibitem{Tristram:2021tvh}
M.~Tristram et~al.
\newblock {Improved limits on the tensor-to-scalar ratio using BICEP and Planck data}.
\newblock {\em Phys. Rev. D}, 105(8):083524, 2022.

\bibitem{PhysRevD.93.025003}
Michela D'Onofrio and Kari Rummukainen.
\newblock Standard model cross-over on the lattice.
\newblock {\em Phys. Rev. D}, 93:025003, Jan 2016.

\bibitem{Mukhanov:2005sc}
V.~Mukhanov.
\newblock {\em {Physical Foundations of Cosmology}}.
\newblock Cambridge University Press, Oxford, 2005.

\bibitem{Freese:2017ace}
Katherine Freese, Evangelos~I. Sfakianakis, Patrick Stengel, and Luca Visinelli.
\newblock {The Higgs Boson can delay Reheating after Inflation}.
\newblock {\em JCAP}, 05:067, 2018.

\bibitem{Arnowitt:1962hi}
Richard~L. Arnowitt, Stanley Deser, and Charles~W. Misner.
\newblock {The Dynamics of general relativity}.
\newblock {\em Gen. Rel. Grav.}, 40:1997--2027, 2008.

\bibitem{Albrecht:2014aga}
Andreas Albrecht, Nadia Bolis, and R.~Holman.
\newblock {Cosmological Consequences of Initial State Entanglement}.
\newblock {\em JHEP}, 11:093, 2014.

\bibitem{Bolis:2016vas}
Nadia Bolis, Andreas Albrecht, and Rich Holman.
\newblock {Modifications to Cosmological Power Spectra from Scalar-Tensor Entanglement and their Observational Consequences}.
\newblock {\em JCAP}, 12:011, 2016.
\newblock [Erratum: JCAP 08, E01 (2017)].

\bibitem{Albrecht:2018hoh}
Andreas Albrecht, Nadia Bolis, and R.~Holman.
\newblock {Cosmic Inflation: The Most Powerful Microscope in the Universe}.
\newblock 6 2018, arXiv:1806.00392.

\bibitem{2020Planck}
N.~Aghanim, Y.~Akrami, M.~Ashdown, J.~Aumont, C.~Baccigalupi, M.~Ballardini, A.~J. Banday, R.~B. Barreiro, N.~Bartolo, and et~al.
\newblock Planck 2018 results.
\newblock {\em Astronomy $\&$ Astrophysics}, 641:A6, Sep 2020.

\bibitem{Kolb:1990vq}
Edward~W. Kolb and Michael~S. Turner.
\newblock {\em {The Early Universe}}, volume~69.
\newblock 1990.

\bibitem{Quiros:1999jp}
Mariano Quiros.
\newblock {Finite temperature field theory and phase transitions}.
\newblock In {\em {ICTP Summer School in High-Energy Physics and Cosmology}}, pages 187--259, 1 1999.

\bibitem{Carena:2021onl}
Marcela Carena, Claudius Krause, Zhen Liu, and Yikun Wang.
\newblock {New approach to electroweak symmetry nonrestoration}.
\newblock {\em Phys. Rev. D}, 104(5):055016, 2021.

\bibitem{Meade:2018saz}
Patrick Meade and Harikrishnan Ramani.
\newblock {Unrestored Electroweak Symmetry}.
\newblock {\em Phys. Rev. Lett.}, 122(4):041802, 2019.

\bibitem{Aguirre:2009ug}
Anthony Aguirre and Matthew~C. Johnson.
\newblock {A Status report on the observability of cosmic bubble collisions}.
\newblock {\em Rept. Prog. Phys.}, 74:074901, 2011.

\bibitem{Linde:2007fr}
Andrei~D. Linde.
\newblock {Inflationary Cosmology}.
\newblock {\em Lect. Notes Phys.}, 738:1--54, 2008.

\bibitem{Guth:2007ng}
Alan~H. Guth.
\newblock {Eternal inflation and its implications}.
\newblock {\em J. Phys. A}, 40:6811--6826, 2007.

\bibitem{Blas_2011}
Diego Blas, Julien Lesgourgues, and Thomas Tram.
\newblock The {C}osmic {L}inear {A}nisotropy {S}olving {S}ystem ({CLASS}). {P}art {II}: Approximation schemes.
\newblock {\em Journal of Cosmology and Astroparticle Physics}, 2011(07):034–034, Jul 2011.

\end{thebibliography}







\end{document}